\documentclass[fleqn,usenatbib,useAMS]{mnras}

\usepackage{newtxtext,newtxmath}
\usepackage[T1]{fontenc}
\usepackage{ae,aecompl}

\DeclareRobustCommand{\VAN}[3]{#2}
\let\VANthebibliography\thebibliography
\def\thebibliography{\DeclareRobustCommand{\VAN}[3]{##3}\VANthebibliography}

\usepackage{graphicx}	% Including figure files
\usepackage{amsmath}	% Advanced maths commands

\usepackage{amssymb}	% Extra maths symbols
\usepackage{array}
\usepackage{multirow}
\usepackage{tabularray}
\usepackage{subcaption} 
\usepackage{hyperref}
\usepackage{multicol}        % Multi-column entries in tables
\usepackage{bm}		% Bold maths symbols, including upright Greek
\usepackage{pdflscape}	% Landscape pages
\usepackage{siunitx}
\usepackage{threeparttable}
\usepackage{lscape}
\usepackage{rotating}
\usepackage{xcolor}
\usepackage{enumitem}
\usepackage{tabularx}

\newcommand{\angstrom}{\textup{\AA}}

\defcitealias{Onorato2025}{O25}
\def\citeO25{\citetalias{Onorato2025}}

\title[Quasar proximity zones]{Homogeneous measurements of proximity zone sizes for 59 quasars in the Epoch of Reionization}

\author[Onorato et al.]{
Silvia Onorato$^{1,2}$\thanks{E-mail: onorato@strw.leidenuniv.nl; silvia.onorato@noirlab.edu},
Joseph F. Hennawi$^{1,3}$, Elia Pizzati$^{1,4}$, Bram P. Venemans$^{1}$, and Anna-Christina Eilers$^{5}$
\\
$^{1}$Leiden Observatory, Leiden University, P.O. Box 9513, NL-2300 RA Leiden, the Netherlands\\
$^{2}$International Gemini Observatory, NSF NOIRLab, 670 N A’ohoku Place, HI-96720, Hilo, USA\\
$^{3}$Department of Physics, Broida Hall, University of California, Santa Barbara, Santa Barbara, CA 93106-9530, USA\\
$^{4}$Center for Astrophysics | Harvard \& Smithsonian, 60 Garden St., Cambridge, MA 02138, USA\\
$^{5}$MIT Kavli Institute for Astrophysics and Space Research, Massachusetts Institute of Technology, Cambridge, MA 02139, USA
}

\date{Accepted 2026 February 20. Received 2026 February 20; in original form 2025 May 14.}

\pubyear{2026}

\begin{document}
\label{firstpage}
\pagerange{\pageref{firstpage}--\pageref{lastpage}}
\maketitle

\begin{abstract}
The overionized regions surrounding high-redshift quasars, known as proximity zones, provide a window into the interaction between supermassive black holes (SMBHs) and the intergalactic medium (IGM) during the epoch of reionization (EoR). We present new homogeneous measurements of proximity zone sizes ($R_{\mathrm{p}}$) for a sample of $59$ quasars spanning redshifts $5.77 \leq z \leq 7.54$ (median $z = 6.59$). For $15$ of these sources, we measure $R_{\mathrm{p}}$ for the first time. The quasars in our catalog have absolute magnitudes at rest-frame $1450$ {\angstrom} in the range $-29.13 \leq M_{1450} \leq -25.20$ (median $M_{1450} \simeq -26.49$), providing one of the most extensive data sets for exploring $R_{\mathrm{p}}$ at these epochs. The distribution of $R_{\mathrm{p}}$ values shows a large scatter at fixed redshift and luminosity, likely reflecting variations in quasar lifetimes ($t_{\mathrm{Q}}$), IGM density fluctuations, and IGM neutral fraction. We fit a bivariate power-law model to a large sample of $100$ objects to study the dependence of $R_{\mathrm{p}}$ with both $M_{1450}$ and $z$: we find that the evolution of $R_{\mathrm{p}}$ with luminosity is in agreement with the models ($R_{\mathrm{p}} \propto 10^{-0.4 M_{1450}/2.87}$), while the evolution of $R_{\mathrm{p}}$ with $z$ is steeper than previous works ($R_{\mathrm{p}} \propto (1+z)^{-2.44}$). We identify $13$ quasars with small proximity zone size, defined using the residuals of our fit. In all cases, except for J2211$-$6320, we rule out the presence of associated dense absorbers that prematurely truncate $R_{\mathrm{p}}$, and suggest a short $t_{\mathrm{Q}}$ ($\lesssim 10^4$ yr) as a possible explanation for their small proximity zone sizes.
\end{abstract}


%%%%%%%%%%%%%%%%%%%%%%%%%%%%%%%%%%%%%%%%%%%%%%%%%%

%%%%%%%%%%%%%%%%% BODY OF PAPER %%%%%%%%%%%%%%%%%%
\section{Introduction}\label{sec:intro}
One of the fundamental goals of modern observational cosmology is to study the epoch of reionization (EoR), during which the intergalactic medium (IGM) transitioned from a predominantly neutral state to the highly ionized state observed today. While considerable progress has been made in constraining the timeline and evolution of this process, critical questions remain regarding the morphology of reionization and the driving physical mechanisms. A principal observational probe of this epoch is the study of Ly$\alpha$ absorption in the spectra of high-redshift quasars. The rapid increase in Ly$\alpha$ optical depth beyond $z \gtrsim 5.5$ and the associated scatter in absorption measurements provide strong evidence for substantial changes in the IGM, consistent with the final stages of reionization \citep{Fan2006, Becker2015, Eilers2018, Yang2020b, Bosman2022}. The absence of extended Gunn-Peterson troughs \citep{Gunn1965} at $z \lesssim 5.5$ and the study of dark gaps in $z\simeq 6$ quasars \citep{Zhu2022}, further support the conclusion that hydrogen reionization was largely complete by this time \citep{McGreer2015,Bosman2022}.

At $z\gtrsim6$, constraints on the reionization history are still relatively weak. One of the main ways to probe reionization at these redshifts is by analyzing quasar proximity zones (or near zones), regions of enhanced transmission in quasar spectra immediately blueward of the Ly$\alpha$ emission line. These zones arise from the intense ionizing radiation of the quasar, which produces a local ionized bubble within the surrounding IGM \citep{Cen2000, Haiman2001, Bolton2007b, Bolton2007a}. The proper size of this ionized region ($R_{\mathrm{ion}}$) depends on the quasar's ionizing photon emission rate ($\dot{N}_{\gamma}$), its lifetime ($t_{\mathrm{Q}}$), and the ambient neutral hydrogen fraction ($x_{\mathrm{HI}}$; \citealt{Haiman2001}). In the simplest scenario of a homogeneous, predominantly neutral IGM with a uniform neutral fraction throughout and negligible recombination on quasar timescales, $R_{\mathrm{ion}}$ is given by:
%% JFH This equation is only correct if the IGM is very neutral. If the IGM is highly ionized, then you don't want to use this equation. I suggest omitting this description entirely. If the IGM is neutral, there is an ionized bubble around the quasars; if the IGM is ionized, there is an overionized region. Neither of these is R_prox, although obviously they can be related. But I don't think you need to use this equation here, which only applies to the neutral case. If you prefer to include this equation, then you need to state the regime in which it applies (a homogenous predominantly neutral IGM with uniform neutral fraction throughout)
%% SO Done
%
\begin{equation}\label{eq:prox_zone}
    R_{\mathrm{ion}} \approx \left( \frac{3 \dot{N}_\gamma t_{\mathrm{Q}}}{4 \pi n_{\mathrm{H}} x_{\mathrm{HI}}} \right)^{1/3} ,
\end{equation}
where $n_{\mathrm{H}}$ is the hydrogen number density.
For a highly ionized IGM, instead, the proximity zone reflects the region where the IGM has had time to respond to the quasar’s radiation field. In this regime, the size of the proximity zone is governed by the equilibration timescale of the IGM ($t_{\mathrm{eq}}\simeq 3 \cdot 10^4$ yr), and the proximity zone becomes sensitive to $t_{\mathrm{Q}}$ only when the quasar age is shorter than this timescale \citep{Davies2020}.
On top of that, Equation \ref{eq:prox_zone} does not fully account for the complexity of the IGM, including pre-existing ionization effects by local galaxies, small-scale density fluctuations, and the presence of overlapping ionized regions \citep{Bolton2007b, Lidz2007, Khrykin2016}.
Furthermore, the measured proximity zone size in the spectrum ($R_{\mathrm{p}}$; \citealt{Fan2006}) may differ from the physical size $R_{\mathrm{ion}}$ due to the presence of residual neutral hydrogen within the ionized bubble, which can produce significant absorption before the light reaches the ionization front \citep{Bolton2007b}. Accounting for most of these effects is possible using semi-analytical models and radiative transfer simulations \citep{Davies2016, Davies2018a, Davies2020}. 
%% JFH You can remove Hennawi24 from the RT sims references. It should be Davies16, Davies18 (DW paper), and possibly also Davies20
%% SO Done
These theoretical efforts offer a way to model and disentangle the effects of the average neutral hydrogen fraction, the quasar luminosity, and the quasar lifetime.

The first systematic measurements of proximity zones were carried out by \citet{Fan2006}, who analyzed $16$ quasars at $z\sim6$ and found that luminosity-scaled proximity zone sizes decrease with increasing redshift, providing evidence for an evolving IGM neutral fraction. Subsequent studies refined these measurements using different, but comparable, resolution 
%% JFH I don't think it is true that these were high-resolution spectra. I think the resolution of all these studies varies but all are comparable to Fan
%% SO Done
spectra \citep{Willott2007, Carilli2010, Mortlock2011, Venemans2015, Reed2015, Reed2017, Eilers2017, Eilers2020, Mazzucchelli2017, Banados2018, Davies2018a, Ishimoto2020, Greig2022, Satyavolu2023, Bigwood2024}. These works confirmed a general decline in the proximity zone size with $z$, but also reported an intrinsically large scatter, indicating a diversity in quasar properties and environments and highlighting the need for large statistical samples.

In addition to proximity zones, damping wings \citep{Miralda-Escude1998} offer a complementary probe of reionization from quasar spectra. They arise due to the presence of neutral regions at $z \gtrsim  6.5$, which imprint a specific off-resonant absorption feature on the red side of the Ly$\alpha$ emission line. As a result, the strength of the damping wing features in $z \gtrsim  6.5$ quasar spectra can provide a direct constraint on the neutral fraction of hydrogen in the IGM \citep{Banados2018, Davies2018a, Greig2022, Greig2024}. Jointly analyzing proximity zone and damping wing effects in high-$z$ quasar spectra offers a promising future avenue to map out the entire evolution of the average hydrogen neutral fraction \citep{Durovcikova2024, Hennawi2024}.

%% JFH You need to revisit this paragraph it is not really accurate. The main thing you are missing is that the dependence of the proximity zone on the lifetime is different in a highly-ionized IGM than it is in a neutral IGM. This is the basis of my objection to your usage of equation 1. It only handles the highly neutral case. In an ionized IGM, one is sensitive mostly to the time it takes for the IGM to equilibrate to a new radiation field. In a neutral IGM something like equation 1 applies.
%% SO Done
Aside from reionization, proximity zones and damping wings also offer an indirect probe of past quasar activity.
For a simple lightbulb model (in which quasars shine continuously for a time $t_\mathrm{Q}$), and in the case of a predominantly neutral IGM, the size of the ionized region created by a quasar is directly related to the quasar lifetime $t_\mathrm{Q}$, as described by Equation \ref{eq:prox_zone}. In a highly ionized IGM, the proximity zone size becomes sensitive to quasar lifetimes mainly for $t_\mathrm{Q} \lesssim t_{\mathrm{eq}}\simeq 3 \cdot 10^4$ yr, as described above.
%% JFH It is unclear what you mean here, but I think you mean flickering light-curves
%% SO Taken out
% Accounting for recombination effects weakens the dependence of $R_\mathrm{p}$ on $t_\mathrm{Q}$ for timescales $\gtrsim 10^5$ yr \citep{Davies2020}.
Nonetheless, several studies have revealed that some $z \gtrsim 6$ quasars exhibit unexpectedly small proximity zones, which may indicate short quasar lifetimes of $t_{\mathrm{Q}} \lesssim 10^4 - 10^5$ yr \citep{Eilers2017, Eilers2020}.  A global average of a larger set of high-$z$ quasars suggests quasar activity timescales of $t_\mathrm{Q}\approx10^6\,\mathrm{yr}$ \citep{Eilers2021a}.

These findings have major consequences for our understanding of supermassive black holes (SMBH) growth: current theories (e.g., \citealt{Fan2023}) suggest that high-$z$ black holes grow from smaller seeds via accretion of material, and this accretion is directly related to quasar activity via a radiative efficiency parameter ($\varepsilon \simeq 0.1$). A short quasar lifetime $t_\mathrm{Q} \lesssim 10^6-10^7\,\mathrm{yr}$ implies that the accretion timescale of SMBHs is not long enough to explain their large $\gtrsim 10^9\,\mathrm{M}_\odot$ masses inferred from quasar spectra \citep{Davies2019,Eilers2021a}. This conclusion assumes a simple lightbulb scenario for the radiative history of quasars. Whether more complex quasar lightcurves can account for the small observed proximity zone sizes while being compatible with the constraints coming from SMBH growth is currently unknown \citep{Satyavolu2023b}. 

Interestingly, quasar clustering measurements can offer independent constraints on the \textit{total} time quasars are active, on average, across cosmic time (or, equivalently, on the \textit{duty cycle} of quasars, $t_\mathrm{dc}$). Measurements of quasar clustering at low to intermediate-$z$ generally predict $t_{\mathrm{dc}} \sim 10^7 - 10^9$ yr \citep{Haiman2001a, Martini2001, Martini2004, Shen2007, Shen2009, White2012, Pizzati2024a}. Recent measurements of the quasar-galaxy cross-correlation function at $z \gtrsim 6$, however, suggest a lower timescale for quasar activity of $t_{\mathrm{dc}} \sim 10^6-10^7$ yr \citep{Eilers2024, Pizzati2024b}, broadly compatible with constraints from proximity zone sizes and in tension with the timescale required for SMBH growth in a standard, Eddington-limited scenario \citep{Eilers2024}. 

In summary, the study of proximity zones in high-$z$ quasar spectra offers key insight into the reionization process and the early evolution of SMBHs. However, current measurements are limited by small sample sizes and heterogeneous assumptions between different data sets. In this work, we overcome these limitations by providing uniform proximity zone size measurements for a sample of $59$ quasars at $5.77 \leq z \leq 7.54$, the largest to date. We combine three different existing data sets, providing updated measurements of proximity zone sizes for most quasars and adding $15$ new measurements, which expand the number of proximity zone sizes available.
We then fit a bivariate power-law model to our results and the literature to explore the dependence of $R_{\mathrm{p}}$ on quasar luminosity and redshift.
% Our analysis reveals a population of quasars with systematically smaller $R_{\mathrm{p}}$ than expected from model predictions, for which we rule out associated dense absorbers as a primary cause. Instead, we suggest a short $t_{\mathrm{Q}}$ as a plausible explanation.

This paper is organized as follows: in Section \ref{sec:dataset}, we describe our data set and its properties; in Section \ref{sec:methods}, we explain the method used for the quasar continuum normalization and proximity zone size measurements; in Section \ref{sec:results}, we detail our measurements, compare them to the literature, fit our data and the literature with a bivariate power-law model to study the evolution of $R_{\mathrm{p}}$ with $M_{1450}$ and $z$, and investigate the cases of small $R_{\mathrm{p}}$; in Section \ref{sec:summary}, we present our summary and main conclusions.

\section{The Data Set}\label{sec:dataset}
We assemble our high-$z$ quasar data set by selecting a sub-sample of objects from three existing catalogs (\citealt{Onorato2025}, \citealt{D'Odorico2023}, \citealt{Durovcikova2024}), each of which is described in detail in the following sections.

\subsection{The ENIGMA sub-sample}\label{sec:sample}
This work involves a sub-sample of $35$ out of the $45$ quasars analyzed in \citealt{Onorato2025} (hereafter, \citeO25) that we will refer to as the ENIGMA sub-sample. \citeO25 contains $12$ Broad Absorption Lines (BAL) quasars (sources with absorption lines with FWHM $\gtrsim$ $2000$ km s$^{-1}$) which in principle we want to exclude from this study since their features could make it difficult to determine the continuum level and cannot be disentangled from IGM absorption \citep{Eilers2020}.
% contaminate the proximity zones with unseen strong \ion{N}{V} associated
%% JFH unseen is not the right word. The point is it cannot be disentangled from IGM absorption
%% SO Done
%% JFH \ion{N}{5} for NV I think. Also, note that other elements can contaminate depending on where the BAL absorption is, i.e. there could be SiIV for a highly blueshifted BAL trough
%% SO Cool, thanks
However, for two of these BALs (J2348$-$3054 and J1526$-$2050; flagged in Table \ref{tab:sample}) the proximity zone sizes were already measured from other works from the literature (see \citealt{Mazzucchelli2017} and \citealt{Bigwood2024}), so we keep them in the sub-sample to make a comparison with previous measurements, but we exclude them from our bivariate power-law fit (see Section \ref{subsec:doublePL}). 
%% JFH I am not sure this makes sense, i.e. you can keep them for comparison but I would exclude them from your fit. If they are contaminated, what is the motivation to keep them. You can also look at the specifics of the BAL and figure out if it could be contaminated by looking to see whether bluer transitions than CIV clobber the proximity zone. 
%% SO Done
Thus, this newly defined sample contains $35$ quasars, where $33$ are non-BAL sources and two are classified as BALs in the literature. It spans the redshift range $6.50 < z \leq 7.54$ ($z_{\mathrm{median}}=6.70$) and a range of absolute magnitudes at $1450$ {\angstrom} of $-27.19 \leq M_{1450} \leq -25.20$ ($M_{1450,\mathrm{median}} \simeq -26.0$, see Fig. \ref{fig:M1450} and Table \ref{tab:sample}).
We include two out of the three unpublished high-$z$ quasars reported in \citeO25; where J0410$-$0139 is now published in \citealt{Banados2025}, and J1917$+$5003 is included in \citet{Belladitta2025}. We exclude the one classified as a BAL (J0430$-$1445, \citealt{Belladitta2025}).

The main facilities used to collect the data are 
%% JFH I would say optical-near-IR since Xshooter is both
%% SO Done
optical and near-infrared (VIS and NIR) echelle spectrographs, such as Gemini/GNIRS (Gemini Near-Infrared Spectrograph; \citealt{Elias2006a,Elias2006b}), Keck/NIRES (Near-Infrared Echellette Spectrometer; \citealt{Wilson2004}), and VLT/X-Shooter \citep{Vernet2011}, both VIS and NIR arms.
However, for some quasars, the Ly$\alpha$ region was not adequately covered by these instruments. To address this, we supplemented our data set with observations from additional long-slit instruments, including Gemini/GMOS (Gemini Multi-Object Spectrographs; \citealt{Hook2004}), Keck/LRIS (Low-Resolution Imaging Spectrometer; \citealt{Oke1995, Rockosi2010}), and LBT/MODS (Multi-Object Double Spectrographs; \citealt{Pogge2010}).

The spectra are consistently reduced with the open-source Python-based spectroscopic data reduction pipeline \texttt{PypeIt}\footnote{\url{https://github.com/pypeit/PypeIt}}, using versions between 1.7.1 and 1.14.1 \citep{Prochaska2020}.
More details on the data reduction and quasar properties can be found in \citeO25, which presents the largest medium-to-moderate resolution sample of $z>6.5$ quasar spectra from ground-based instruments.

\subsection{The E-XQR-30 sub-sample}\label{sec:exqr30}
An analysis of quasar proximity zones on a sub-sample of the enlarged XQR-30 (E-XQR-30) data has already been performed and described in \cite{Satyavolu2023}. However, all the spectra are publicly available from \cite{D'Odorico2023}, so we perform our own analysis including two more objects than \cite{Satyavolu2023}, and compare it with their results. We use coadded, flux-calibrated spectra with J-band photometry, which are rebinned to $50$ km s$^{-1}$. They are hosted in the "Flux-calibrated-spectra" directory of the project's GitHub repository\footnote{\url{https://github.com/XQR-30}}. We refer to \cite{D'Odorico2023} for a full description of the data reduction and quasar details, but we summarize here the main information relevant to this work. From the $42$ quasars in the E-XQR-30 sample, only $21$ are considered, as four sources are already included in the ENIGMA sub-sample, and we exclude all those labeled as BALs, proximate Damped Ly$\alpha$ systems (pDLAs), reddened, and some of the mini BALs in Table 1 from \cite{Greig2024}. The mini BALs included in our sub-sample and bivariate power-law fit are indicated in Table \ref{tab:sample} (i.e., J0842+1218, J0408$-$5632, and J1213$-$1246). The $21$ analyzed quasars span the redshift range $5.76<z<6.34$ ($z_{\mathrm{median}}=6.03$), and the magnitude range $-29.13 < M_{1450} < -26.40$ ($M_{1450,\mathrm{median}} \simeq -26.77$, see Fig. \ref{fig:M1450} and Table \ref{tab:sample}), and all are
%% JFH are all --> all are 
%% SO Done
collected with VLT/X-Shooter.
%% JFH So I have 21 XQR-30 objects in my IGM DW sample qso_fitting/jobs/targfiles/targfile_full_sample_dv500. However, I always used the XQR30 data instead of the ENIGMA reductions, because typically the XQR30 reductions looked better. I think you may be excluding too many sources from Greig's table. Can you check your XQR-30 targets against mine? I also provide notes in that file on associated absorbers.
%% SO For example, I exclude ULASJ0148+0600 as it is a DLA with also a strong BAL feature.

\subsection{The FIRE sub-sample}\label{sec:fire}
Taking advantage of the public data release from \cite{Durovcikova2024} at doi: \href{https://doi.org/10.5281/zenodo.11402934}{10.5281/zenodo.11402934}, we make use of their quasars to enlarge our collection of proximity zone size measurements.
We refer to \cite{Durovcikova2024} for more details, while we summarize here the relevant information to this analysis. From the $18$ quasar spectra they provide, we exclude those already present in the ENIGMA and the E-XQR-30 samples described above, leading to a final sub-sample (FIRE) of three objects in the redshift range $6.37<z<6.74$ 
%% JFH this redshift range is incorrect. The range you actually use is 6.38 < Z< 6.74. Also I used 5 FIRE objects, but I think that is because in a few cases I thought they looked better than the ENIGMA reduction, sothat is fine if you use ENIGMA
%% SO Corrected. And yes, I am favoring the ENIGMA reductions always.
and magnitude range $-27.65 < M_{1450} < -26.31$ (see Fig. \ref{fig:M1450} and Table \ref{tab:sample}). Their spectra come from Magellan/FIRE (Folded port InfraRed Echellette; \citealt{Simcoe2013}), VLT/X-Shooter, Keck/MOSFIRE (Multi-Object Spectrograph for Infrared Exploration; \citealt{McLean2010,McLean2012}) and Keck/ESI (Echellette Spectrograph and Imager; \citealt{Sheinis2002}).

\subsection{Final catalog}\label{subsec:final_subsample}
\begin{figure*}
    \includegraphics[width=\linewidth]{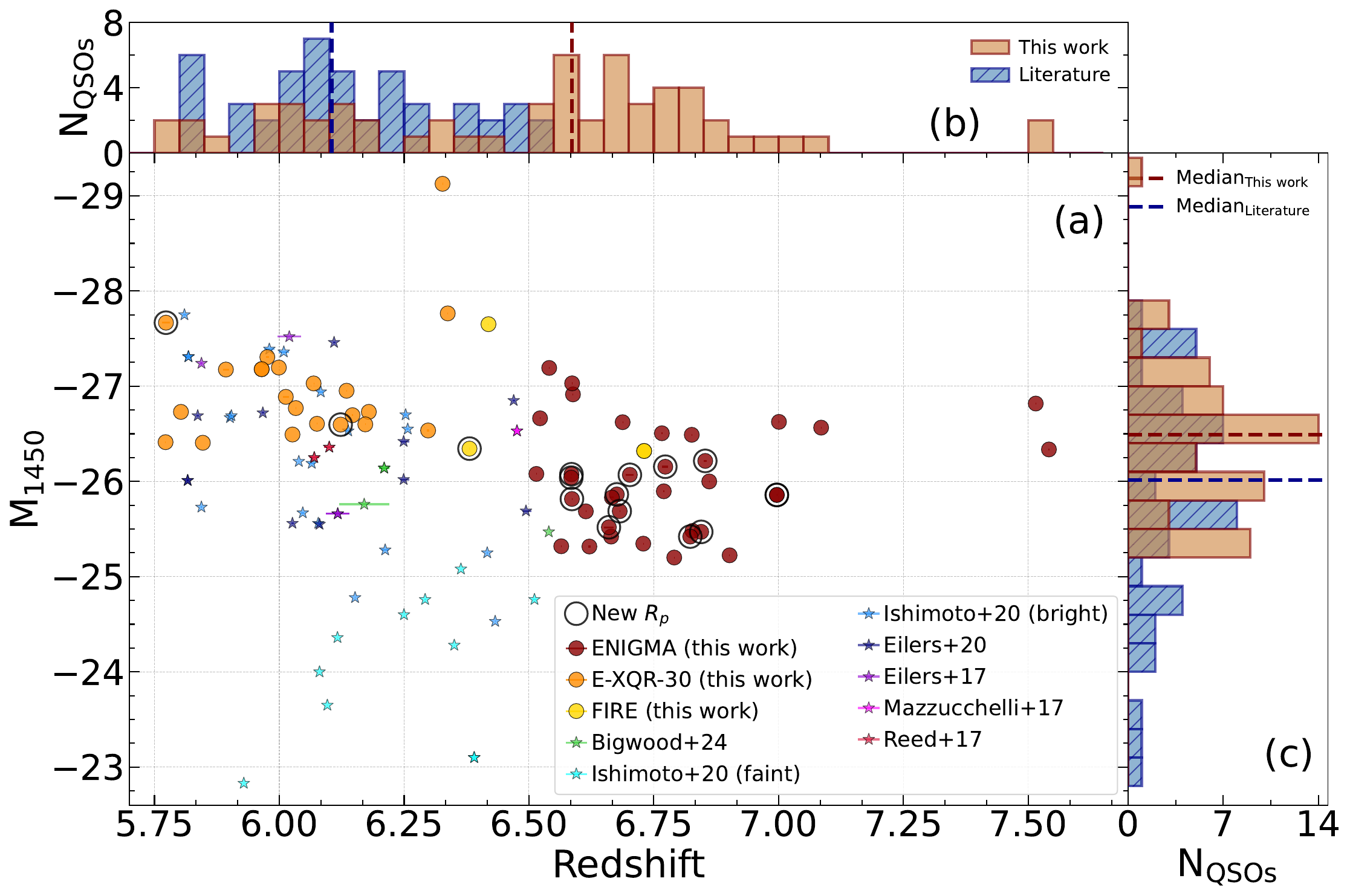}
    \caption{\textit{Panel (a)}: Distribution of $M_{1450}$ as a function of $z$ for the sub-samples of quasars considered in this study ($35$ from ENIGMA in red, $21$ from E-XQR-30 in orange, and three from FIRE in yellow), and for others from the literature in which the proximity zone sizes have also been measured, i.e., \citealt{Bigwood2024} (green), \citealt{Ishimoto2020} (both faint and bright samples, respectively in cyan and light blue), \citealt{Eilers2020} (dark blue), \citealt{Eilers2017} (purple), \citealt{Mazzucchelli2017} (magenta), and \citealt{Reed2015,Reed2017} (crimson). The error bars show the uncertainties on $z$. All the quasars in common between this work and the literature or among themselves have been included only once, always prioritizing this work in the first case and the most recent work in the second one. The big black circles surrounding some points indicate the quasars from this work that did not have previous proximity zone size measurements from the literature. \textit{Panel (b)}: Histogram of the redshift distribution of the quasars considered in this study (light brown) and of the quasars from the literature (blue), with bins of size $0.05$. The dashed lines represent the median redshift for this sample ($z_{\mathrm{This\ work}}=6.59$, in dark red) and the literature ($z_{\mathrm{Literature}}=6.10$, in blue). \textit{Panel (c)}: Histogram of the $M_{1450}$ distribution of the quasars considered in this study (light brown) and of the quasars from the literature (blue), with bins of size $0.3$. The dashed lines represent the median $M_{1450}$ for this work ($M_{1450,\mathrm{This\ work}}\simeq-26.49$, in dark red) and the literature ($M_{1450,\mathrm{Literature}}\simeq-26.02$, in blue).}
    \label{fig:M1450}
\end{figure*}

To summarize, our final catalog consists of $59$ objects coming from the three samples described above (ENIGMA, E-XQR-30, and FIRE). It spans the redshift range $5.77 \leq z \leq 7.54$ and the magnitude range $-29.13 \leq M_{1450} \leq -25.20$ (with $z_{\mathrm{median}}=6.59$ and $M_{1450,\mathrm{median}} \simeq -26.49$, see Fig. \ref{fig:M1450}).
Many of these quasars already have documented analyses of proximity zone sizes in the literature (i.e., \citealt{Bigwood2024,Satyavolu2023,Greig2022,Ishimoto2020, Eilers2020,Eilers2017,Banados2018,Davies2018a,Mazzucchelli2017,Venemans2015}). For those in common with other works, we perform new measurements and then compare the different results (see Section \ref{subsec:comp}).
In addition, there are $15$ objects that do not have previous proximity zone measurements. 
Here we provide for the first time measurements of their proximity zones. They are flagged with an asterisk in Table \ref{tab:sample}, where we report all the sources grouped by sample (first ENIGMA, then E-XQR-30, and finally FIRE), and sorted by decreasing redshift. The properties of the quasars are listed in the table as follows: the number of the quasar, the short version of its name, its redshift, the method adopted to determine the redshift, the reference for this measurement, its magnitude ($M_{1450}$), the reference for the discovery of each quasar, the proximity zone size measurement with associated uncertainties in proper Mpc (pMpc), and notes. The references are listed at the bottom of the table.

We interpret our results along with those coming from several other samples \citep{Bigwood2024,Ishimoto2020,Eilers2020,Eilers2017,Mazzucchelli2017,Reed2017,Reed2015}, described in Section \ref{subsec:comp}. We do not re-analyze these data, but rather we use the published literature measurements, which adopt a methodology similar to this work.
For these sources, we quote the median values of redshift and magnitude which are, respectively, $z_{\mathrm{Literature}}=6.10$ and $M_{1450,\mathrm{Literature}}\simeq-26.02$ (see Fig. \ref{fig:M1450}).
%% JFH Make it more clear you don't reanalyze this data like the samples above, but rather you use the published literature measurements, which use similar methodology. Right now the distinction between literature and your measurements could be more clear. 
%% SO Done

\subsection{Redshift uncertainties}\label{sec:redshift}
As already discussed in \citeO25, we address uncertainties in the systemic redshift of each quasar depending on the emission line used for the redshift determination. Systemic redshifts are challenging to obtain due to factors like the broad widths of emission lines, Gunn-Peterson absorption \citep{Gunn1965}, and discrepancies between different ionization lines (\citealt{Gaskell1982}; \citealt{Tytler1992}; \citealt{VandenBerk2001}; \citealt{Richards2002}; \citealt{Shen2016}). Additionally, because of the winds and significant internal motions that many quasars exhibit, the emission lines deviate substantially from the systemic redshift of the host galaxy.

Thus, following \cite{Eilers2017}, we 
%% JFH conservatively assign
%% SO Done
conservatively assign a redshift error of $\Delta v = 100$ km s$^{-1}$ for the most accurate determinations based on emission lines from the host galaxy's atomic gas reservoir ([\ion{C}{II}] lines). 
However, we state clearly that obtaining this measurement 
%% JFH this measurement error
%% SO Done
error is not trivial: the typical velocity dispersion ($\sigma_v$) measured in quasar host galaxies is approximately $300$ km s$^{-1}$ for detections with a signal-to-noise ratio (SNR) greater than $5$. The statistical uncertainty on the mean redshift is given by $\sigma_v/\mathrm{SNR} \approx 60$ km s$^{-1}$. Consequently, our adopted uncertainty of $100$ km s$^{-1}$ represents a conservative estimate. This assumption relies on the expectation that the quasar is embedded within the gravitational potential traced by the cool interstellar medium.
We adopt the same uncertainty of $\Delta v = 100$ km s$^{-1}$ also for measurements inferred from the narrow component of the [\ion{O}{III}] line $\lambda 5008.24$ {\angstrom}
%% JFH convention is to state line then \lambda 5008 etc.
%% SO Done
observed with mid-IR JWST WFSS (\textit{James Webb Space Telescope} Wide Field Slitless Spectroscopy; 
%% JFH grism data is unclear and vague. I think you want to say mid-IR JWST WFSS of the [OIII] line
%% SO Done
\citealt{Kashino2023a,Wange2023,Yang2023}).

For quasars with redshift measurements from low-ionization lines like \ion{Mg}{II}, we assume a redshift error of $\Delta v = 390$ km s$^{-1}$ to accommodate the discrepancy between the \ion{Mg}{II} line redshift and that of the host galaxy, measured by [\ion{C}{II}] line (e.g., \citealt{Schindler2020, Mazzucchelli2017}). This value is consistent with previous findings and is further validated by our analysis of a sub-sample of quasars with both \ion{Mg}{II} and [\ion{C}{II}] redshifts, for which we measure a median velocity offset of $\Delta v \simeq -388$ km s$^{-1}$, in excellent agreement with \cite{Schindler2020}.
%% JFH maybe make it more clear this is by comparing MgII to CII zs
%% SO Done
In Table \ref{tab:sample} we report redshifts, methods, and references for all the quasars.
We use these $\Delta v$ as uncertainties on the redshift, and consequently, on the position of the Ly$\alpha$ line in the analysis of the proximity zone sizes (see Section \ref{subsub:err}). Estimates of $\Delta z$ for every quasar are also available from the literature (see $z_{\mathrm{Ref}}$ in Table \ref{tab:sample}), but we do not report them in this work. 
%% JFH I'm confused, are you using the literature values or the values you just discussed. And why not just add the column to the table? You have space. 
%% SO We use \delta v as uncertainties. The values of \delta z are only used for Figure 1 and reported in the table on GitHub. This is the approach we adopted for Paper I, and I am using it here as well.
They are used in Fig. \ref{fig:M1450} as uncertainties on $z$, and available online in the GitHub repository\footnote{\url{https://github.com/enigma-igm/onorato24_hiz_qsos}} referenced in \citeO25.

\subsection{Absolute magnitudes at 1450 {\angstrom}}\label{sec:M1450}
We want to carefully determine the quasars' absolute magnitudes at $1450$ {\angstrom} ($M_{1450}$) as they will be crucial for evaluating the evolution of the proximity zone sizes with luminosity (see Section \ref{subsec:doublePL}). The values of $M_{1450}$ are calculated from the flux-scaled spectra of the quasars with J, Y, or $\mathrm{K_{p}}$-band photometry, as already described in \citeO25.
Briefly, we compute the rest-frame wavelengths for every spectrum based on the redshift estimates. From each median flux between $1445$ and $1455$ {\angstrom} converted to Jansky ($f_{\nu,1450}$), we determine the luminosity per unit frequency ($L_{\nu}$) at $1450$ {\angstrom} as:
\begin{equation}\label{eq:luminosity}
L_{\nu} = \frac{4\pi {d_{L}}^{2} \cdot f_{\nu,1450}}{1 + z} ,
\end{equation}
where $d_{L}$ is the luminosity distance to the object at a redshift $z$.
Finally, we find $M_{1450}$ from the luminosity per unit frequency $L_{\nu}$ via:
\begin{equation}\label{eq:M1450}
M_{1450} = -2.5 \cdot \log_{10} \left( \frac{L_{\nu}}{4 \pi {d_{0}}^{2} \cdot 3631 \text{Jy}} \right) ,
\end{equation}
where $d_0$ is the reference distance ($10$ pc), and $3631$ Jy is the zero-point flux density in the AB system.
We show the absolute magnitude values $M_{1450}$ as a function of redshift $z$ for this sample in panel (a) of Fig. \ref{fig:M1450}, with a comparison to other literature works in which the study of proximity zone size measurements has also been performed. To avoid repetitions among the different samples, all quasars in common between this work and the literature, or among themselves, have been considered only once, always prioritizing this work in the first case and the most recent work in the second one.
We use big black circles 
%% JFH circumfrences --> circles
%% SO Done
surrounding some points 
%% JFH circles --> points
%% SO Done
to indicate quasars from this work that did not have previous proximity zone size measurements from the literature.
In panel (b) of Fig. \ref{fig:M1450} we plot the histogram of the redshift distribution of the quasars for both this work and the quasars from the literature, with bins of size $0.05$. In panel (c) of Fig. \ref{fig:M1450} we plot the histogram of the $M_{1450}$ distribution for the quasars in this work and for the quasars considered from the literature, with bins of size $0.3$.

%% JFH I'm not sure it make sense to quote 3 decimal place in M_1450. I don't think we know M_1450 this well. Specifically, the spectrophotometric errors that these are largely based are surely not 0.001. Typically what one does in this case is quote say two decimal places, but then in the table you publish put in the full precision of the number.
%% SO This is what I have done in Paper I. I basically copied and pasted the magnitudes from there, so I prefer leaving them as they are.
\begin{table*}
    \centering
    \begin{threeparttable}
    \caption{Information on the $59$ quasars in this work, sorted by sub-sample and decreasing redshift. The details on the columns are explained in Section \ref{subsec:final_subsample}.}
    \label{tab:sample}
    \begin{tabularx}{\textwidth}{l|lccccc|XXX|l}
    % \begin{tabular}{l|lccccc|ccc|l}
    \hline
    N$_{\mathrm{QSO}}$ & Name & $z$ & $z_{\mathrm{method}}$ & $z_{\mathrm{Ref}}$ & $M_{1450}$ & Discovery & $R_{\mathrm{p}}$ [pMpc] & $\sigma_{\mathrm{up}}$ & $\sigma_{\mathrm{low}}$ & Notes\\
    \hline
        1 & J1342$+$0928 & 7.5413 & [\ion{C}{II}] & 1 & $-$26.336 & 2 & 1.41 & 0.11 & 0.11 & \\
        2 & J1007$+$2115 & 7.5149 & [\ion{C}{II}] & 3 & $-$26.818 & 3 & 1.40 & 0.11 & 0.11 & \\
        3 & J1120$+$0641 & 7.0851 & [\ion{C}{II}] & 1 & $-$26.565 & 4 & 1.41 & 0.17 & 0.12 & \\
        4 & J0252$-$0503 & 7.0006 & [\ion{C}{II}] & 5 & $-$26.625 & 6 & 1.47 & 0.12 & 0.12 & \\
        5* & J0410$-$0139 & 6.9964 & [\ion{C}{II}] & 23 & $-$25.858 & 23 & 2.17 & 0.12 & 0.12 & \\
        6$^{\mathrm{a}}$ & J2348$-$3054 & 6.9018 & [\ion{C}{II}] & 9 & $-$25.224 & 10 & 2.26 & 0.13 & 0.13 & BAL quasar. \\
        7 & J0020$-$3653 & 6.861 & [\ion{C}{II}] & 32 & $-$25.999 & 11 & 2.23 & 0.12 & 0.13 \\
        8* & J1917$+$5003 & 6.853 & \ion{Mg}{II} & 24 & $-$26.208 & 24 & 1.28 & 0.47 & 0.47 & \\
        9* & J2211$-$6320 & 6.8449 & [\ion{C}{II}] & 5 & $-$25.470 & 6 & 0.19 & 0.12 & 0.12 & \\
        10 & J0319$-$1008 & 6.8275 & [\ion{C}{II}] & 5 & $-$25.480 & 6 & 1.77 & 0.13 & 0.12 & \\
        11 & J0411$-$0907 & 6.8260 & [\ion{C}{II}] & 5 & $-$26.490 & 8 & 3.27 & 1.18 & 0.13 \\
        12* & J1129$+$1846 & 6.823 & \ion{Mg}{II} & 7 & $-$25.421 & 12 & 4.25 & 0.47 & 0.47 \\
        13 & J0109$-$3047 & 6.7909 & [\ion{C}{II}] & 9 & $-$25.200 & 10 & 1.33 & 0.13 & 0.13 & \\
        14* & J0829$+$4117 & 6.773 & \ion{Mg}{II} & 7 & $-$26.154 & 8 & 3.36 & 0.48 & 0.48 \\
        15 & J0218$+$0007 & 6.7700 & [\ion{C}{II}] & 5 & $-$25.896 & 7,13 & 1.01 & 0.12 & 0.12 & \\
        16 & J1104$+$2134 & 6.7662 & [\ion{C}{II}] & 5 & $-$26.506 & 8 & 4.03 & 0.12 & 0.12 \\
        17 & J0910$+$1656 & 6.7289 & [\ion{C}{II}] & 5 & $-$25.346 & 8 & 0.49 & 0.12 & 0.13 & \\
        18* & J0837$+$4929 & 6.702 & \ion{Mg}{II} & 7 & $-$26.069 & 8 & 3.10 & 0.48 & 0.48 \\
        19 & J2002$-$3013 & 6.6876 & [\ion{C}{II}] & 5 & $-$26.622 & 7 & 2.19 & 0.12 & 0.13 \\
        20* & J0923$+$0753 & 6.6817 & [\ion{C}{II}] & 5 & $-$25.687 & 7 & 1.87 & 0.12 & 0.13 & \\
        21* & J1048$-$0109 & 6.6759 & [\ion{C}{II}] & 14 & $-$25.864 & 15 & 2.23 & 0.12 & 0.12 \\
        22 & J2232$+$2930 & 6.666 & [\ion{C}{II}] & 25 & $-$25.831 & 16 & 5.02 & 0.13 & 2.98 \\
        23 & J2102$-$1458 & 6.6645 & [\ion{C}{II}] & 5 & $-$25.421 & 8 & 2.39 & 0.12 & 0.13 \\
        24* & J1216$+$4519 & 6.66 & \ion{Mg}{II} & 7 & $-$25.518 & 8 & 1.35 & 0.49 & 0.49 & \\
        25 & J0024$+$3913 & 6.6210 & [\ion{C}{II}] & 18 & $-$25.316 & 19 & 3.26 & 0.13 & 0.13 \\
        26 & J0305$-$3150 & 6.6139 & [\ion{C}{II}] & 26 & $-$25.690 & 10 & 3.13 & 0.13 & 0.33 \\
        27 & J2132$+$1217 & 6.5881 & [\ion{C}{II}] & 18 & $-$26.914 & 18 & 6.11 & 0.13 & 0.13 \\
        28$^{\mathrm{a}}$ & J1526$-$2050 & 6.5864 & [\ion{C}{II}] & 14 & $-$27.030 & 18 & 3.98 & 0.13 & 0.13 & BAL quasar. \\
        29* & J2338$+$2143 & 6.586 & [\ion{C}{II}] & 32 & $-$25.816 & 7 & 0.53 & 0.13 & 0.13 & \\
        30* & J1135$+$5011 & 6.5851 & [\ion{C}{II}] & 5 & $-$26.075 & 8 & 4.18 & 0.13 & 0.13 \\
        31* & J1058$+$2930 & 6.5846 & [\ion{C}{II}] & 22 & $-$26.039 & 7 & 1.94 & 0.13 & 0.13 & \\
        32 & J0921$+$0007 & 6.5646 & [\ion{C}{II}] & 5 & $-$25.319 & 17 & 2.09 & 0.13 & 0.13 \\
        33 & J0226$+$0302 & 6.5405 & [\ion{C}{II}] & 26 & $-$27.192 & 16 & 3.61 & 0.13 & 0.13 \\
        34 & J0224$-$4711 & 6.5222 & [\ion{C}{II}] & 5 & $-$26.663 & 21 & 5.97 & 0.22 & 0.19 \\
        35 & J1110$-$1329 & 6.5148 & [\ion{C}{II}] & 14 & $-$26.079 & 16 & 1.40 & 0.13 & 0.16 & \\ \hline
        36 & J0142$-$3327 & 6.3373 & [\ion{C}{II}] & 26 & $-$27.764 & 30 & 7.81 & 0.13 & 0.14 \\
        37 & J0100$+$2802 & 6.3269 & [\ion{C}{II}] & 26 & $-$29.126 & 31 & 6.99 & 0.16 & 0.14 \\
        38 & J1030$+$0524 & 6.298 & [\ion{O}{III}] & 33 & $-$26.538 & 34 & 5.27 & 0.14 & 0.19 \\
        39 & J0402$+$2451 & 6.1793 & [\ion{C}{II}] & 35 & $-$26.730 & 29 & 4.13 & 0.15 & 0.15 \\
        40 & J2356$-$0622 & 6.1722 & [\ion{C}{II}] & 36 & $-$26.599 & 29 & 2.73 & 0.17 & 0.15 \\
        41 & J1428$-$1602 & 6.1466 & [\ion{C}{II}] & 14 & $-$26.695 & 29 & 2.69 & 0.17 & 0.17 \\
        42 & J1319$+$0950 & 6.1347 & [\ion{C}{II}] & 26 & $-$26.952 & 37 & 4.96 & 0.15 & 0.14 \\
        43* & J1509$-$1749 & 6.1228 & [\ion{C}{II}] & 35 & $-$26.596 & 38 & 4.49 & 0.14 & 0.15 \\
        44 & J0842$+$1218 & 6.0754 & [\ion{C}{II}] & 26 & $-$26.605 & 39 & 6.76 & 0.15 & 0.15 & Mini BAL quasar (see 46). \\
        45 & J1034$-$1425 & 6.0687 & [\ion{C}{II}] & 36 & $-$27.029 & 40,41 & 1.92 & 0.14 & 0.15 & \\
        46 & J1306$+$0356 & 6.033 & [\ion{C}{II}] & 26 & $-$26.770 & 34 & 6.33 & 0.15 & 0.14 \\
        47 & J0408$-$5632 & 6.0264 & [\ion{C}{II}] & 35 & $-$26.493 & 21 & 2.58 & 0.15 & 0.15 & Mini BAL quasar (see 46). \\
        48 & J0159$-$3633 & 6.013 & \ion{Mg}{II} & 42 & $-$26.888 & 30 & 3.79 & 0.56 & 0.55 \\
        49 & J0818$+$1722 & 5.9991 & [\ion{C}{II}] & 35 & $-$27.196 & 43 & 5.10 & 0.14 & 0.15 \\
        50 & J0158$-$2905 & 5.976 & \ion{Mg}{II} & 42 & $-$27.305 & 29 & 4.38 & 0.56 & 0.56 \\
        51 & J0713$+$0855 & 5.9647 & [\ion{C}{II}] & 35 & $-$27.179 & 29 & 2.51 & 0.47 & 0.14 & \\
        52 & J1213$-$1246 & 5.893 & \ion{Mg}{II} & 42 & $-$27.175 & 44 & 1.53 & 0.57 & 0.57 & Mini BAL quasar (see 46). \\
        53 & J1609$-$1258 & 5.8468 & [\ion{C}{II}] & 35 & $-$26.406 & 29 & 5.93 & 0.15 & 0.15 \\
        54 & J2033$-$2738 & 5.803 & [\ion{C}{II}] & 35 & $-$26.730 & 29 & 3.23 & 2.43 & 0.16 \\
        55* & J0836$+$0054 & 5.773 & \ion{Mg}{II} & 42 & $-$27.667 & 34 & 2.76 & 0.59 & 0.58 \\
        56 & J0927$+$2001 & 5.7722 & [\ion{C}{II}] & 45 & $-$26.412 & 43 & 4.66 & 0.16 & 0.15 \\ \hline
        57 & J0244$-$5008 & 6.7306 & [\ion{C}{II}] & 25 & $-$26.320 & 11 & 2.28 & 0.40 & 0.13 \\
        58 & J1148$+$5251 & 6.4189 & [\ion{C}{II}] & 27 & $-$27.651 & 28 & 4.47 & 0.14 & 0.14 \\
        59* & J1036$-$0232 & 6.3809 & [\ion{C}{II}] & 14 & $-$26.344 & 29 & 4.83 & 0.13 & 0.14 \\
    \hline
    \end{tabularx}
    \begin{tablenotes}
    \scriptsize
    \item Ref: 1 - \cite{Venemans2017}; 2 - \cite{Banados2018}; 3 - \cite{Yang2020a}; 4 - \cite{Mortlock2011}; 5 - \cite{Wang2021b}; 6 - \cite{Yang2019}; 7 - \cite{Yang2021}; 8 - \cite{Wang2019}; 9 - \cite{Venemans2016}; 10 - \cite{Venemans2013}; 11 - \cite{Reed2019}; 12 - \cite{Banados2021}; 13 - \cite{Matsuoka2022}; 14 - \cite{Decarli2018}; 15 - \cite{Wang2017}; 16 - \cite{Venemans2015}; 17 - \cite{Matsuoka2018}; 18 - \cite{Mazzucchelli2017}; 19 - \cite{Tang2017}; 20 - \cite{Banados2015}; 21 - \cite{Reed2017}; 22 - \cite{Wang2024}; 23 - \cite{Banados2025}; 24 - \cite{Belladitta2025}; 25 - \cite{Yang2023}; 26 - \cite{Venemans2020}; 27 - \cite{Maiolino2005}; 28 - \cite{Fan2003}; 29 - \cite{Banados2016}; 30 - \cite{Carnall2015}; 31 - \cite{Wu2015}; 32 - Bouwens et al., in prep; 33 - \cite{Kashino2023a}; 34 - \cite{Fan2001}; 35 - Private comm. with Sarah Bosman; 36 - \cite{Eilers2021}; 37 - \cite{Mortlock2009}; 38 - \cite{Willott2007}; 39 - \cite{Jiang2015}; 40 - \cite{Banados2023}; 41 - \cite{Chehade2018}; 42 - \cite{Bischetti2022,D'Odorico2023}; 43 - \cite{Fan2006a}; 44 - \cite{Banados2014}; 45 - \cite{Carilli2007}; 46 - \cite{Greig2024}.
    \item[*] New $R_{\mathrm{p}}$ measurement.
    \item[a] Excluded from the bivariate power-law fit.
    \end{tablenotes}
    \end{threeparttable}
\end{table*}

\section{Methods}\label{sec:methods}

\begin{figure*}
    \includegraphics[width=\linewidth]{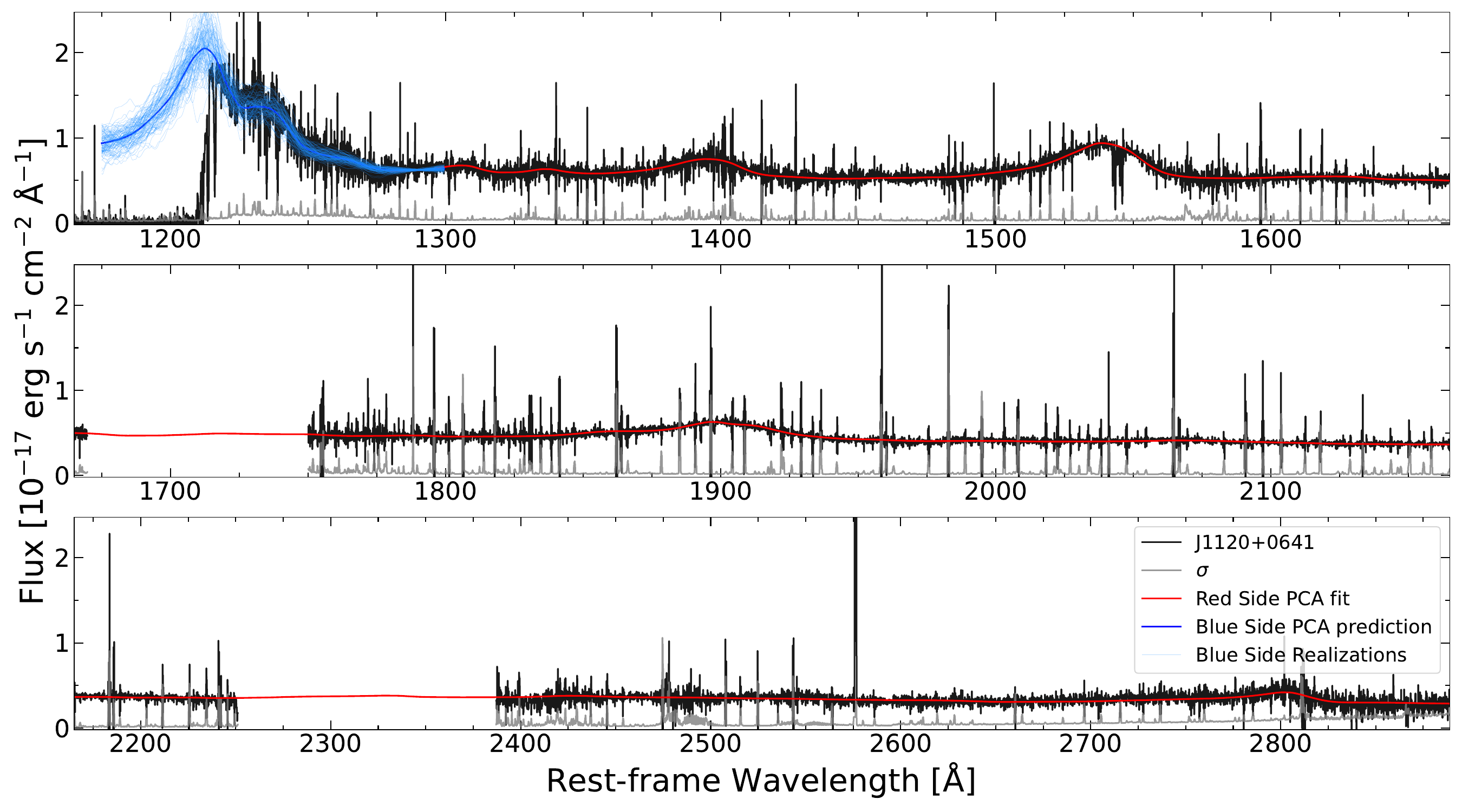}
    \caption{Example case for the quasar spectrum of J1120$+$0641 (black), its associated noise vector $\sigma$ (grey), and its best-fitting continuum model determined using principal component analysis (PCA). We estimate the continuum across the full spectral range by fitting $10$ red principal components to the region $1280  < \lambda_{\mathrm{rest}}$ [{\angstrom}] $< 2890$ (red curve), and projecting onto a set of $6$ blue principal components covering $1175  < \lambda_{\mathrm{rest}}$ [{\angstrom}] $< 1280$ (blue curve). The $100$ realizations of the interpolated PCA projection errors for the blue side are shown in light blue. The spectrum is smoothed for visual purposes, and we have masked regions of strong telluric absorption. The $3$-panel plot lets us evaluate the fit at every wavelength in detail. The continuum model reconstructions for all the quasars included in this sample are shown in Fig. \ref{fig:pca_plot_enigma}, \ref{fig:pca_plot_exqr30}, and \ref{fig:pca_plot_fire}.}
    \label{fig:pca}
\end{figure*}

\subsection{Quasar continuum estimates using PCA}\label{subsec:pca}
To obtain proximity zone size measurements, we first require a prediction for the unabsorbed quasar continuum in the Ly$\alpha$ forest. The detailed procedure is described in \cite{Davies2018b}, but we summarize here the key aspects of the process. For estimating the quasar continuum, we adopt \textit{principal component analysis} (PCA), a statistical technique used to simplify a data set by reducing its dimensionality while retaining most of the variation in the data \citep{Suzuki2005,Paris2011,Davies2018b}. To construct the PCA model, we process $12764$ quasar spectra with $2.09 < z < 2.51$ and SNR $>7$ from the BOSS DR14 database.
The logarithm of each quasar spectrum ($F_{\lambda}$) can be described as:
\begin{equation}\label{eq:pca}
\log F_{\lambda} \approx \langle \log F_{\lambda} \rangle + \sum_{i=0}^{N_{\mathrm{PCA}}} a_i A_i,
\end{equation}
where $\langle \log F_{\lambda} \rangle$ is the mean of the natural logarithm of the spectrum, 
%% JFH mean of the log of the spectrum. Is it log10 or ln?? Be explicit.
%% SO It's natural. I now wrote it explicitly.
$N_{\mathrm{PCA}}$ is the number of PCA components, 
%% JFH vectors and components refer to the same thing, choose one and use consistently
%% SO Done, I'll stick with "components".
$A_i$ are the PCA components, and $a_i$ the coefficients weighting them.
This logarithmic space is useful because it accounts for variations between power-law quasar continua.
We adopt the same parameter choices as in \cite{Davies2018b}, decomposing the red side ($1280<\lambda_{\mathrm{rest}}$ [{\angstrom}] $< 2890$) and the blue side ($1175<\lambda_{\mathrm{rest}}$ [{\angstrom}] $< 1280$) of the spectra independently, keeping $10$ red ($R_i$) and $6$ blue ($B_i$) PCA components. We compute the best estimate for the red PCA coefficients ($r_i$) by fitting them to the red side of the quasar spectra. Finally, the best estimated $r_i$ are projected onto a set of blue coefficients ($b_j$) for the blue PCA components using a \textit{projection matrix} ($P_{ij}$) determined from the training spectra, following:
\begin{equation}\label{eq:proj}
b_j = \sum_{i=1}^{N_{\mathrm{PCA}},r} r_i P_{ij}.
\end{equation}
The only difference in our approach is the use of iterative sigma clipping and outlier rejection during the fitting process to mask absorption lines and bad pixels.
The errors on the continuum prediction are derived by generating multiple realizations of the PCA model using the covariance matrix of the PCA components.

An example case of a quasar spectrum and its PCA continuum model is shown in Fig. \ref{fig:pca} with J1120$+$0641. The spectra and estimated continuum models for all the sources in this work are reported in Fig. \ref{fig:pca_plot_enigma}, \ref{fig:pca_plot_exqr30}, and \ref{fig:pca_plot_fire}. We see that the predicted continua match the data with an overall good agreement.
While we could have used other continuum fitting methods, the differences are shown to have a negligible impact on the proximity zone size measurements \citep{Eilers2017,Greig2024a}.

\subsection{Proximity zone sizes}\label{subsec:proxzonesize}
\subsubsection{Measurements}\label{subsub:Rp}
To determine the proximity zone sizes of the quasars in this work, we adopt the definition introduced by \cite{Fan2006} and widely used in the literature referenced above. First, we normalize the quasar spectra by dividing the observed flux by the estimated continuum reconstruction obtained as described in Section \ref{subsec:pca}. We then smooth the continuum-normalized flux using a boxcar function with a $20$ {\angstrom} width in the observed frame.

The size of the proximity zone ($R_{\mathrm{p}}$) is defined as the proper distance at which the smoothed normalized flux first drops below $10\%$ of the continuum level. 
Previous work defined that the smoothed normalized flux had to remain below the $10\%$ threshold for at least three consecutive pixels for a proper measurement.
However, such a requirement is strongly dependent on the pixel size adopted, which changes from instrument to instrument.

We generalize this condition by consistently re-binning all the spectra to a new wavelength grid with a fixed velocity interval size ($dv = 35.0$ km s$^{-1}$), and interpolating the smoothed normalized flux onto this new distance grid. Then we determine the proximity zone size by identifying the distance where the flux drops below $10\%$ of the continuum level for seven consecutive pixels, corresponding to $\gtrsim 205$ km s$^{-1}$.
The choice of a velocity interval $dv = 35.0$ km s$^{-1}$ arises from a compromise between having a finer wavelength grid for the spectra with the coarsest spectral resolution (to decrease the interval between two consecutive pixels) and keeping the number of consecutive pixels required to set the condition reasonably low.

% An example case of how we measure proximity zone sizes is shown in Fig. \ref{fig:prox_zone_example} with J1120$+$0641.
The measurements for all sources in this work are listed in Table \ref{tab:sample}, and reported in Fig. \ref{fig:proxmes_enigma}, \ref{fig:proxmes_exqr30}, and \ref{fig:proxmes_fire}.
They span the range $0.19 \leq R_{\mathrm{p}}$ [pMpc] $\leq 7.81$, with a median size of $R_{\mathrm{p}} = 2.76$ pMpc.

\subsubsection{Errors}\label{subsub:err}
We discuss the two main sources of uncertainty affecting our final results. The first one considered is the error on the redshift measurements, arising from the fact that the emission lines used to measure the quasar's redshift (i.e., [\ion{C}{II}], [\ion{O}{III}], and \ion{Mg}{II}) have different associated velocity errors (see Section \ref{sec:redshift}). It influences the calculation of the physical distance from the quasar, impacting the proximity zone size. The typical errors correspond to $\sigma_{\mathrm{z}} \simeq 0.12$ pMpc in the case of $z$ measurements coming from [\ion{C}{II}] and [\ion{O}{III}] emission lines, and $\sigma_{\mathrm{z}} \simeq 0.48$ pMpc if coming from \ion{Mg}{II}.

%% JFH Your section describing the Davies et al. continuum fitting procedure says nothing about the errors or how they are derived.
%% SO Fixed
The second source of uncertainty is the error on the continuum fit, which is used to normalize the quasar spectrum (see Section \ref{subsec:pca}). The PCA fit provides a model of the quasar continuum; any error in this fit affects the normalized flux and thus the determination of $R_{\mathrm{p}}$.
To estimate this uncertainty, we use the set of PCA continuum realizations. For each of them, we measure the proximity zone sizes using the same method as for the main measurements (see \ref{subsub:Rp}).
This results in a distribution of $R_{\mathrm{p}}$ values, one for each continuum realization.
The $16^{\text{th}}$ and $84^{\text{th}}$ percentiles of this distribution are taken as the lower and upper bounds of the uncertainty due to the PCA fit ($\sigma_{\mathrm{PCA}}$).
%% JFH it is unclear how you actually did this. I think you used the realizations provided by the code, computed the proximity zone sizes for each realization, and then took the percentiles of the proximity zone sizes that you measured. This would be the correct way to do it (I hope you did it this way). Anyway, you need to explain how you did this. If you did not do it this way, let's discuss. 
%% SO Done
They are generally small in a relative sense, with typical values of a few percent. Most quasars exhibit relative errors $\sigma_{\mathrm{PCA}}/R_{\mathrm{p}} \lesssim 0.03$, though the most extreme case (i.e., J2033$-$2738) reaches up to $\sim 75\%$.
%% JFH I think a relative error estimate is more relevant here than an absolute one. 
%% SO Done
%% JFH Well rebinning data to a finer grid does not increase your resolution. I think you are making a statement here about the resolution of the spectrograph and the pixel sampling of the spectra. How much this matters when you smooth things is unclear. Anyway, I would clean up this description because your reasoning for why it can be neglected, based on the rebinning spectra to a finer grid, does not really make sense to me. Maybe just omit this entirely since you ignore it.
%% SO Done
% In principle, there would be a third source of uncertainty arising from the different pixel sizes in the sample, but they get uniform and small across all the instruments after the re-binning procedure described above. Thus, we estimate that the impact of this error on the final measurements is negligible (as $\sigma_{\mathrm{pixel}} \simeq 0.04$ pMpc) and decide to ignore it.
Finally, we use the square root of the sum of the squares of all the individual errors to provide a total uncertainty on the proximity zone size ($\sigma_{\mathrm{up}}$ and $\sigma_{\mathrm{low}}$ in Table \ref{tab:sample}).
%% JFH Note that a better procedure would have been to do a Monte Carlo, i.e. grab a redshift from the error distribution, grab a realization from the continuum distribution, and perform the analysis many times and then take percentiles. I think the quadrature sum that you adopted is fine and what I describe is probably overkill, but it is good for you to understand and think about the "right" way to do it. My point is that these errors are actually probably correlated.
% SO I see, thanks.

% The third and last source of uncertainty is the error due to the pixel size of the instrument, arising from the finite resolution of the wavelength grid used in the analysis. $R_{\mathrm{p}}$ is determined by identifying the point where the flux drops below the threshold for consecutive pixels, and the exact position of this drop can vary within a pixel.
% The pixel size error is calculated as the difference between the proximity zone size and the distance of the consecutive pixel where the flux drop is identified.
% As already described above, we made the pixel size uniform across the different instruments to consistently apply the definition of proximity zone, by re-binning the spectrum to a fixed wavelength grid with a homogeneous velocity interval. Thus, the error on the pixel size is always the same for every instrument in this work, and it is $\sigma_{\mathrm{pixel}} \simeq 0.04$ pMpc.

To assess the impact of small SNR on our proximity zone sizes, we perform Monte Carlo noise-degradation tests on a representative subset of high-SNR quasars. For each object, the spectrum is artificially degraded to SNR $\simeq 5$ by adding Gaussian noise, and the full analysis, including PCA continuum refitting and proximity zone size measurement, is repeated for $100$ realizations.
We find that the median $R_{\mathrm{p}}$ remains unchanged in all tested cases, with the scatter induced by spectral noise being generally small ($\lesssim 5\%$). We therefore conclude that uncertainties due to finite SNR are subdominant compared to those arising from redshift measurements and continuum fitting, and can be safely neglected in the overall error budget.

To assess the impact of the chosen number of consecutive pixels on our measurements, we repeat the analysis using alternative conditions of five ($\gtrsim 135$ km s$^{-1}$) and nine ($\gtrsim 275$ km s$^{-1}$) consecutive pixels.
In the vast majority of cases, the measured proximity zone sizes remain unchanged. However, we identify a few objects (see Table \ref{tab:Rp_change}) for which the measurements exhibit differences, though they remain consistent within the error bars. Despite these minor variations, the overall statistical trends and conclusions remain robust, confirming that our choice of seven pixels is a reliable criterion for defining $R_{\mathrm{p}}$. 

\begin{table}
    \caption{Proximity zone size measurements for the objects where differences are observed when varying the number of consecutive pixels required for setting the condition. The table lists: the number of consecutive pixels considered, the quasar name, its redshift, and the measured $R_{\mathrm{p}}$, along with the associated uncertainties.}
    \centering
    \begin{tabular}{c|ccccc}
    \hline
     $N_{\mathrm{pixels}}$ & Name & $z$ & $R_{\mathrm{p}}$ [pMpc] & $\sigma_{\mathrm{up}}$ & $\sigma_{\mathrm{low}}$\\
    \hline
     \multirow{2}{1em}{5} & J2232+2930 & 6.666 & 2.04 & 0.13 & 0.12 \\
     & J2033$-$2738 & 5.803 & 3.23 & 0.25 & 0.16 \\ \hline
     \multirow{4}{1em}{9} & J0411$-$0907 & 6.8260 & 4.44 & 0.12 & 1.22 \\
     & J2232+2930 & 6.666 & 5.02 & 0.13 & 0.12 \\
     & J0713$+$0855 & 5.9647 & 2.51 & 4.43 & 0.14 \\
     & J0836+0054 & 5.773 & 10.10 & 0.58 & 7.36 \\
     \hline
    \end{tabular}
    \label{tab:Rp_change}
\end{table}

%%%%%% PROX ZONES ENIGMA
\begin{figure*}
    \centering
    \includegraphics[width=0.89\linewidth]{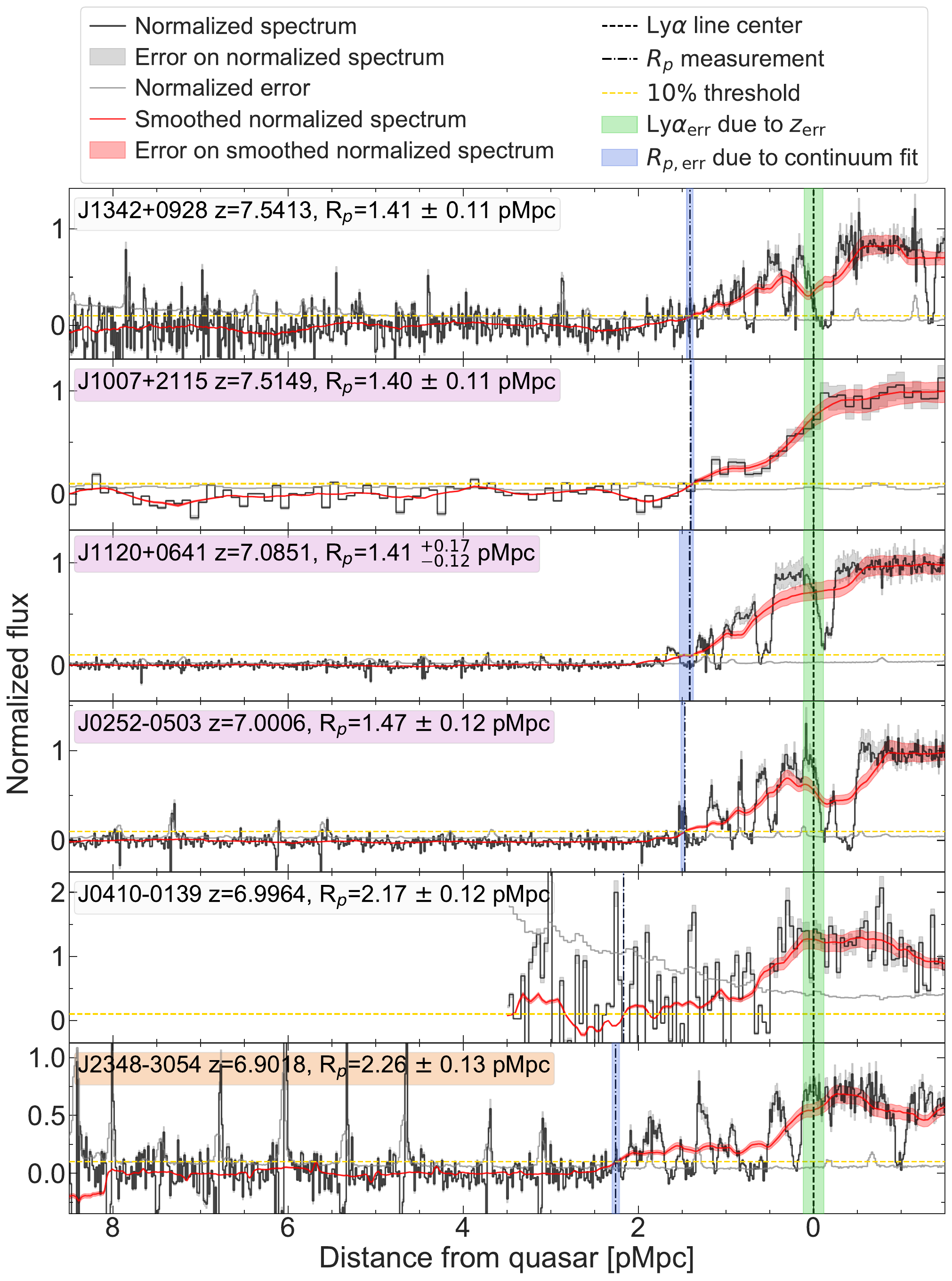}
     \caption{Proximity zone size measurements of the $35$ quasars in the ENIGMA sub-sample, sorted by decreasing $z$. The continuum-normalized flux, obtained by dividing the observed flux by the PCA reconstruction, is shown in black and its associated error due to the error in the PCA continuum fit is shown as a grey band. The normalized error on the flux is shown as a grey line. The smoothed normalized flux and its uncertainties are plotted as a red curve with a shaded region. The black dashed line sets the location of the quasar, while the black dash-dotted line marks the distance where the smoothed normalized flux crosses the $10\%$ threshold (dashed yellow line) setting the $R_{\mathrm{p}}$ measurement. The uncertainty on the position of the Ly$\alpha$ line due to the error on the redshift measurement is shown in green. The uncertainty on the $R_{\mathrm{p}}$ measurement due to the continuum prediction is reported in blue. The box containing name, redshift, and $R_{\mathrm{p}}$ is colored in orange to mark the BALs and in pink to highlight the small proximity zones defined in Section \ref{subsec:small}}.
     \label{fig:proxmes_enigma}
\end{figure*}

\begin{figure*}
    \addtocounter{figure}{-1}
    \centering
    \includegraphics[width=0.89\linewidth]{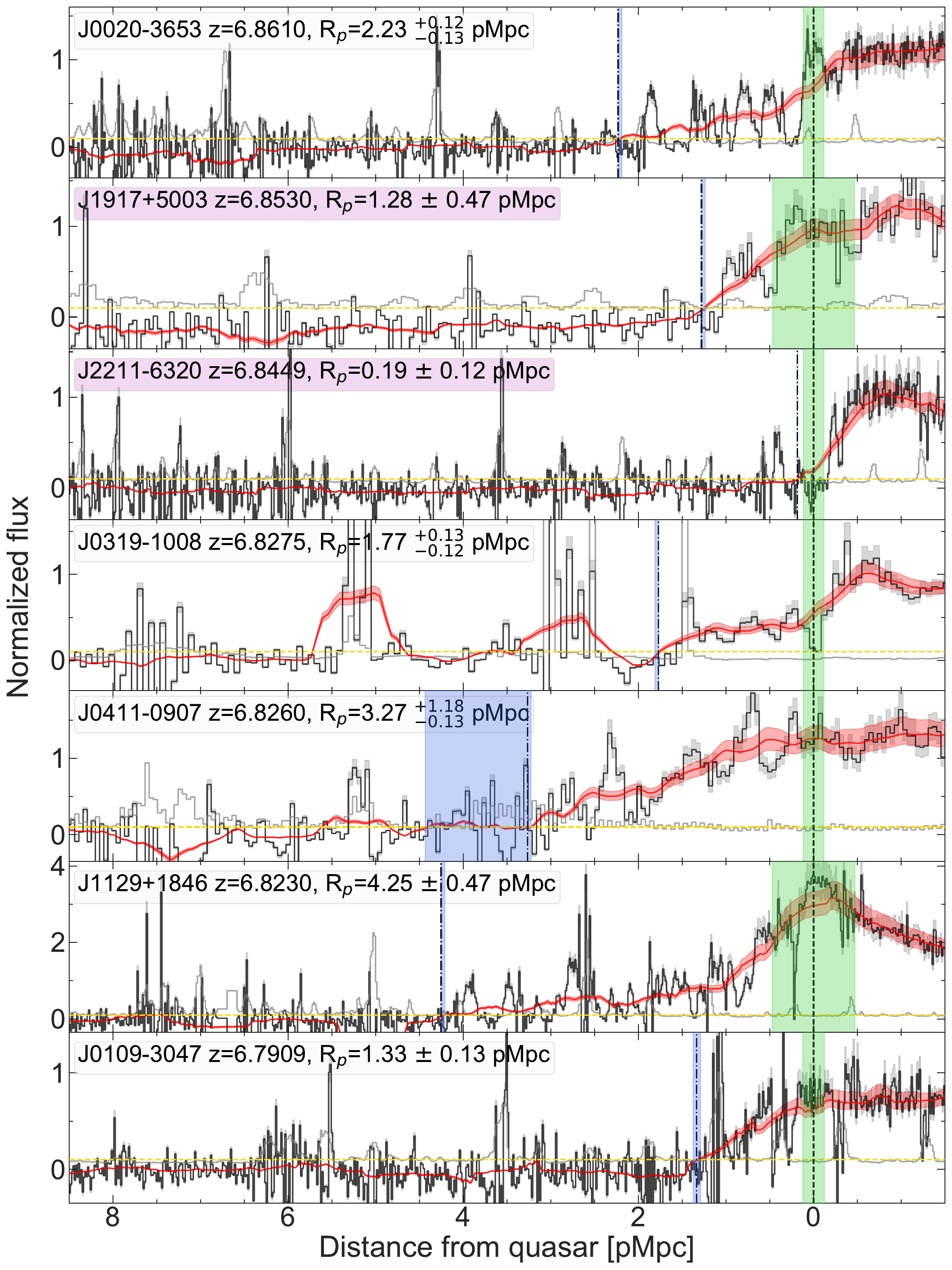}
     \caption{(Continued)}
\end{figure*}

\begin{figure*}
    \addtocounter{figure}{-1}
    \centering
    \includegraphics[width=0.89\linewidth]{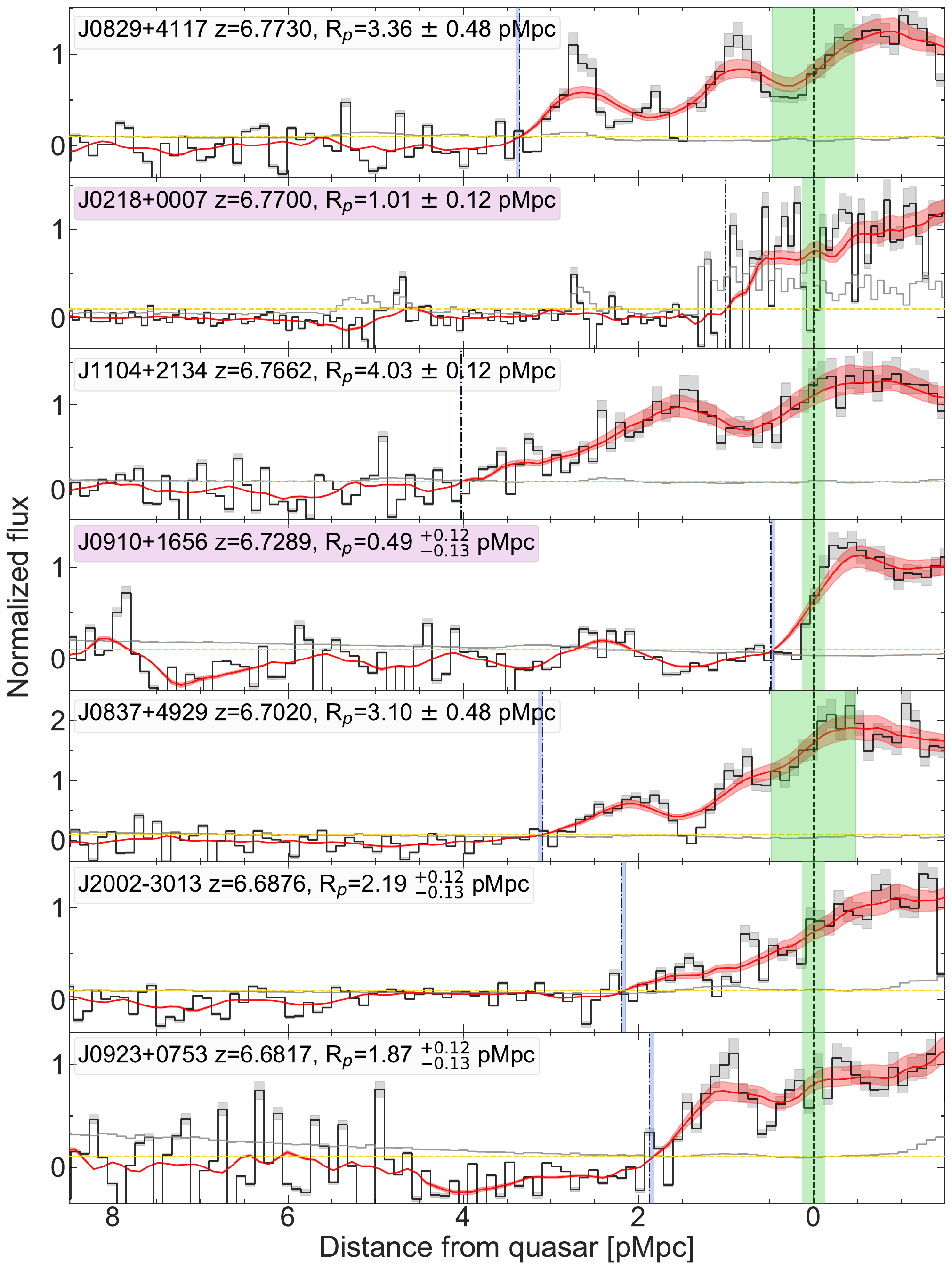}
     \caption{(Continued)}
\end{figure*}

\begin{figure*}
    \addtocounter{figure}{-1}
    \centering
    \includegraphics[width=0.89\linewidth]{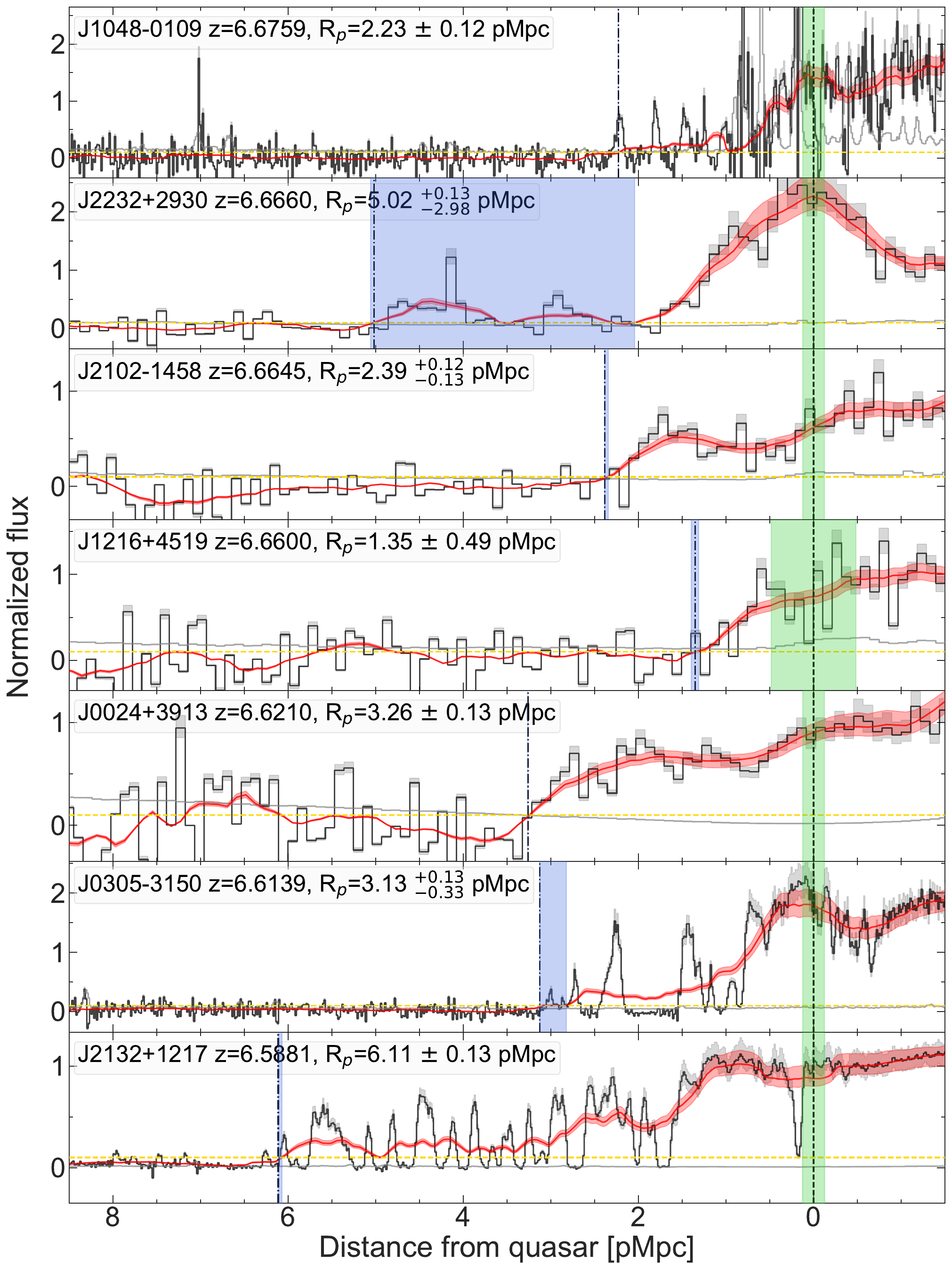}
     \caption{(Continued)}
\end{figure*}

\begin{figure*}
    \addtocounter{figure}{-1}
    \centering
    \includegraphics[width=0.86\linewidth]{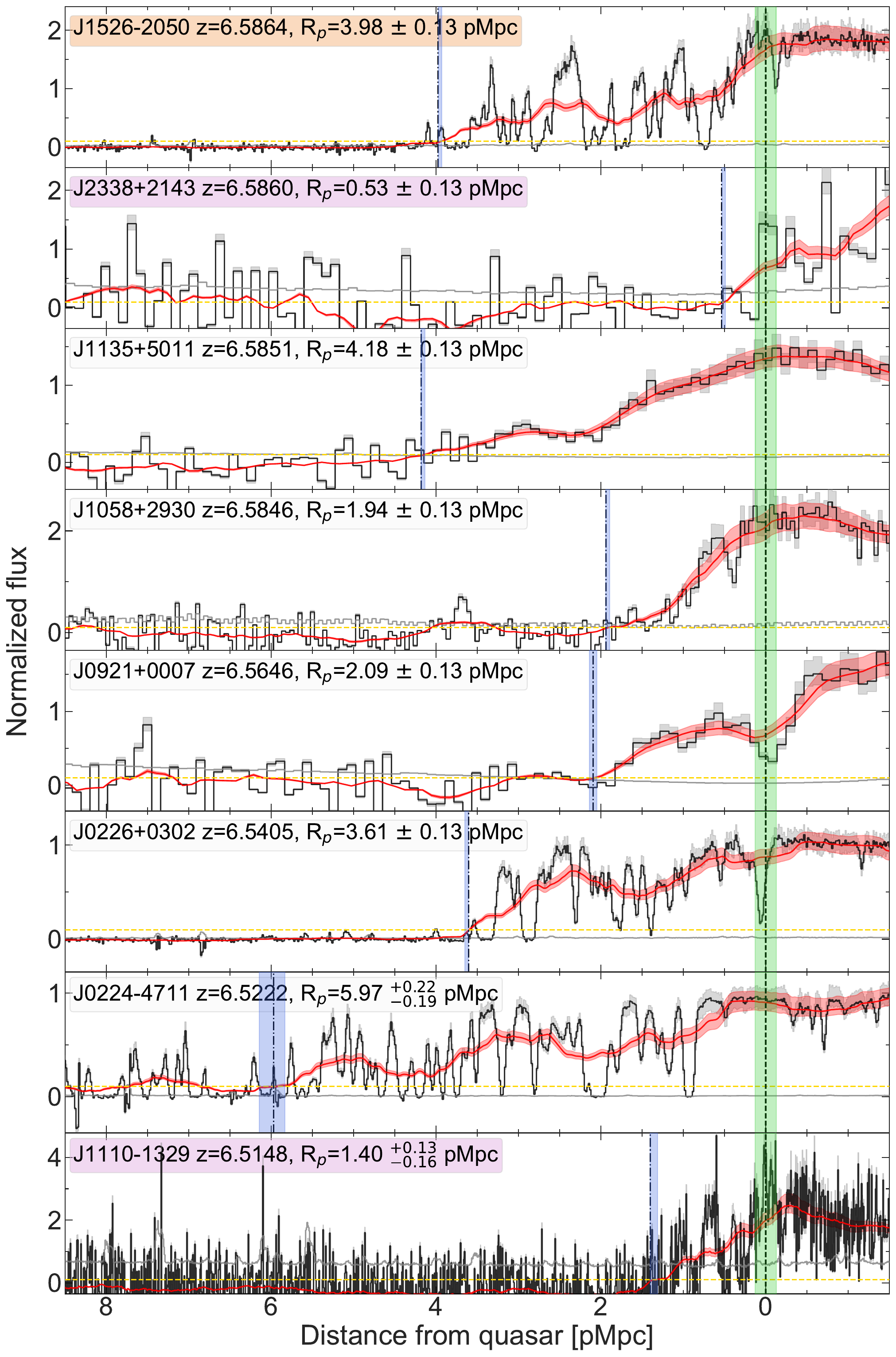}
     \caption{(Continued)}
\end{figure*}

%%%%%% PROX ZONES EXQR30
\begin{figure*}
    \centering
    \includegraphics[width=0.89\linewidth]{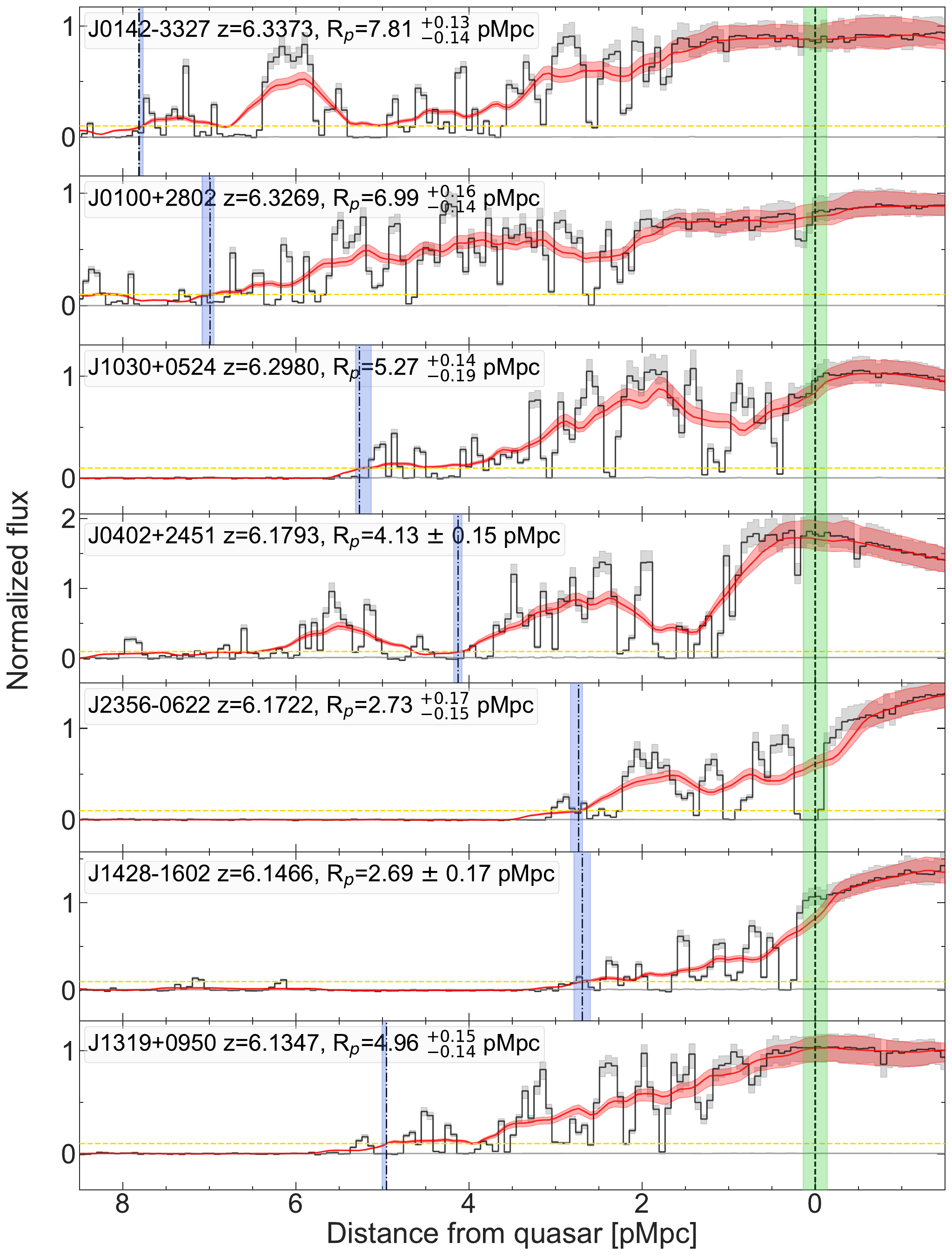}
     \caption{Proximity zone size measurements of the $21$ quasars in the E-XQR-30 sub-sample, sorted by decreasing $z$, as described in Fig. \ref{fig:proxmes_enigma}.}
     \label{fig:proxmes_exqr30}
\end{figure*}

\begin{figure*}
    \addtocounter{figure}{-1}
    \centering
    \includegraphics[width=0.89\linewidth]{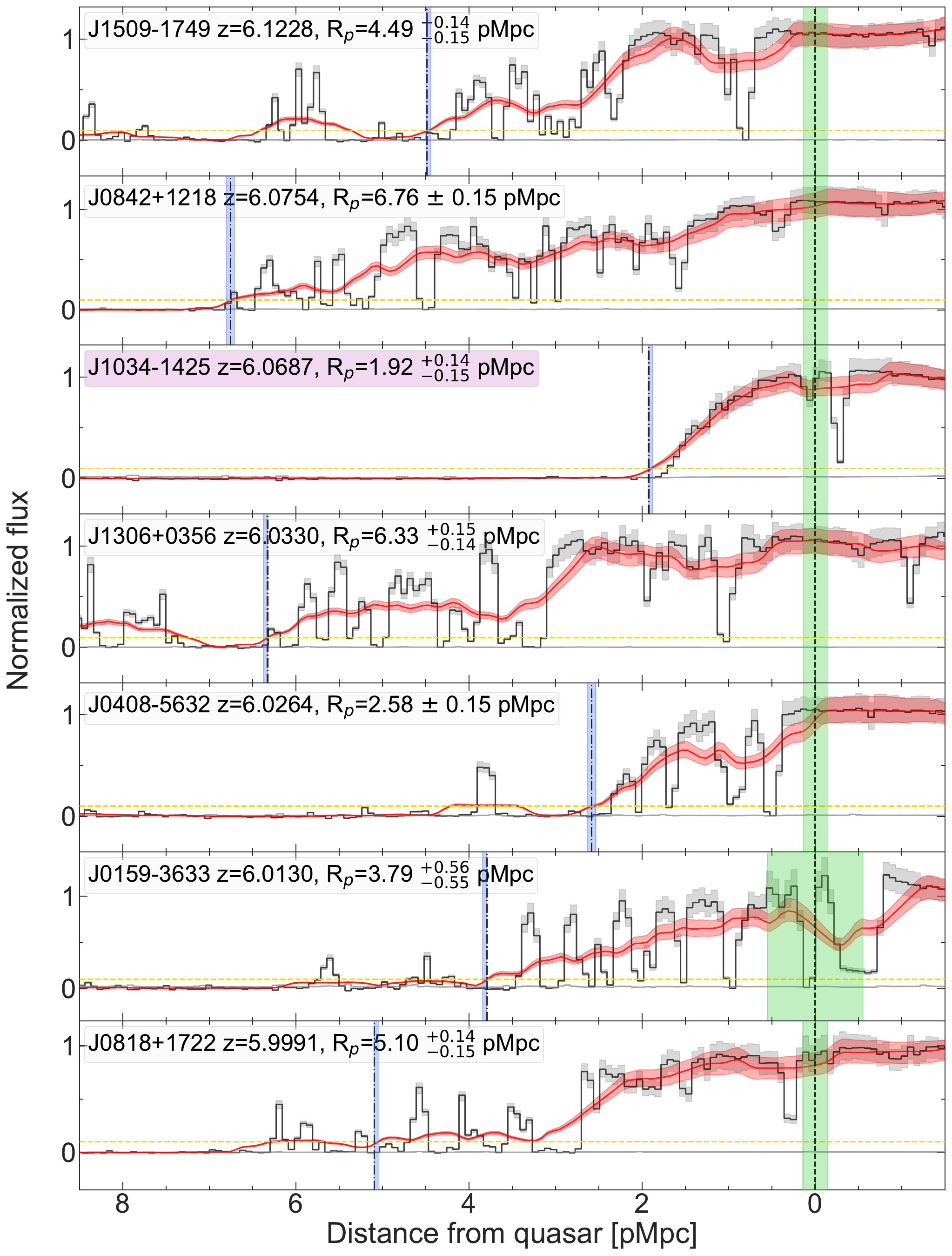}
     \caption{(Continued)}
\end{figure*}

\begin{figure*}
    \addtocounter{figure}{-1}
    \centering
    \includegraphics[width=0.89\linewidth]{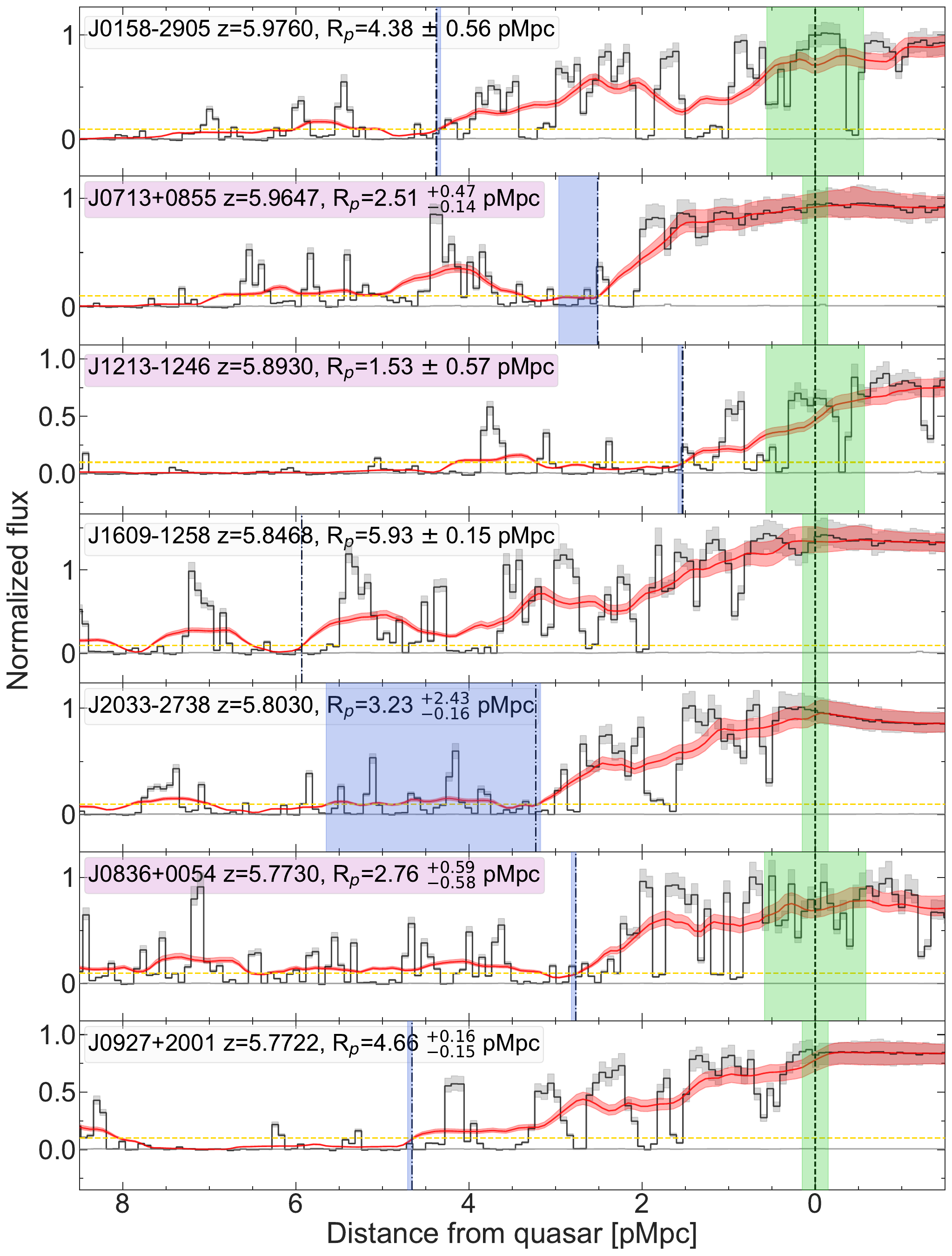}
     \caption{(Continued)}
\end{figure*}

%%%%%% PROX ZONES FIRE
\begin{figure*}
    \centering
    \includegraphics[width=0.89\linewidth]{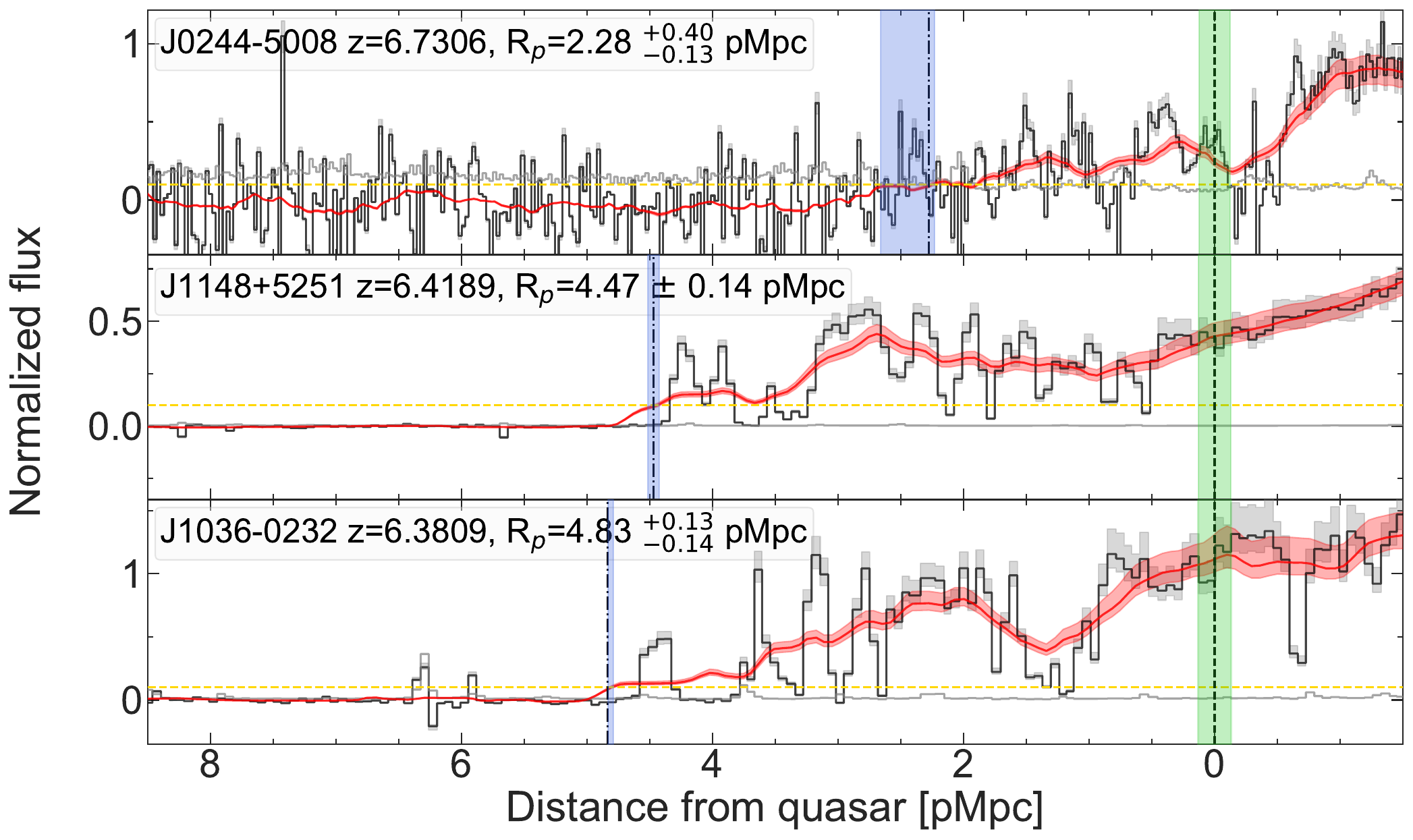}
     \caption{Proximity zone size measurements of the three quasars in the FIRE sub-sample, sorted by decreasing $z$, as described in Fig. \ref{fig:proxmes_enigma}.}
     \label{fig:proxmes_fire}
\end{figure*}

\section{Results}\label{sec:results}
\subsection{Comparison with the literature}\label{subsec:comp}
\begin{figure}
    \includegraphics[width=\linewidth]{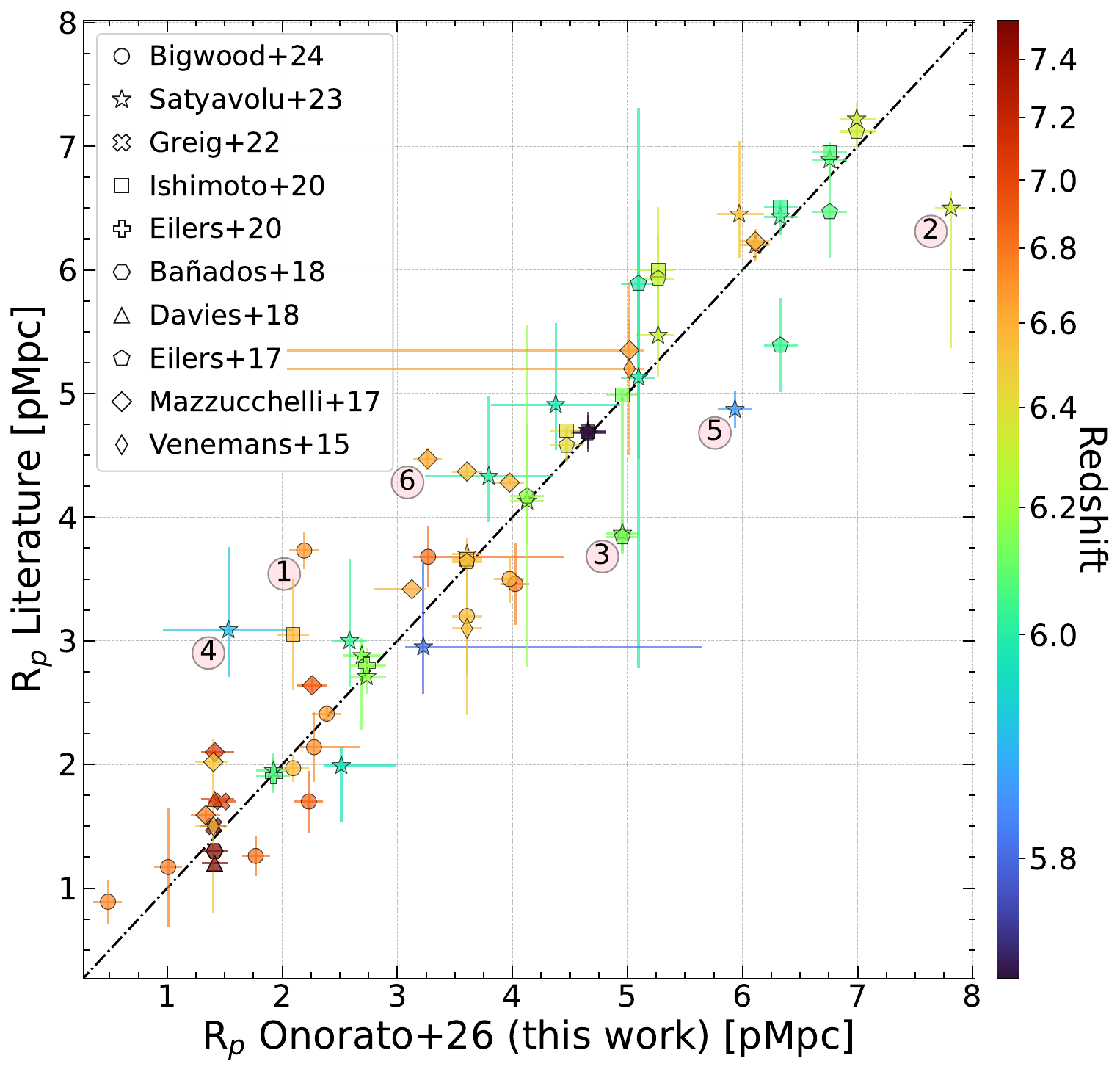}
    \caption{$R_{\mathrm{p}}$ measured in this work compared with those from the literature. The different markers represent data from various studies, while their color is determined by the redshift of the quasars, according to the colormap to the right. The error bars are the uncertainties on $R_{\mathrm{p}}$ estimated from the corresponding study. The 1:1 dash-dotted line indicates where the proximity zone sizes from this work and the literature would be equal. There is good agreement between $R_{\mathrm{p}}$ measured in this work and others from the literature. The six quasars where the scatter between our measurements and literature ones is $\Delta R_{\mathrm{p}} \gtrsim 1$ pMpc are marked with a number in a pink circle, corresponding to the number N in Table \ref{tab:scatter}, with their description.}
    \label{fig:Rp-scatter}
\end{figure}

We compare the proximity zone sizes obtained in this work with the literature on other proximity zone size analyses, such as \cite{Bigwood2024, Satyavolu2023, Greig2022, Ishimoto2020, Eilers2020,Eilers2017, Banados2018, Davies2018a, Mazzucchelli2017, Reed2017,Reed2015, Venemans2015}. The most important properties of the samples are summarized in Table \ref{tab:sample_compar}. Fig. \ref{fig:Rp-scatter} shows a scatter plot that compares the size of the proximity zones measured in this work with those reported in the literature. Overall, there is a good agreement between our measurements and those from the literature. The few sources showing the largest scatter ($\Delta R_{\mathrm{p}} \gtrsim 1$ pMpc) are listed in Table \ref{tab:scatter}, with numbers on the first column (N) corresponding to the labels in Fig. \ref{fig:Rp-scatter}.

\begin{table*}
    \caption{Main properties of the final sample in this paper, compared with those of other samples from the literature in which proximity zone analysis has also been performed. The columns show, respectively: the reference of the sample, the redshift range, the $M_{1450}$ range, the total number of quasars in the sample (with the objects in common with this work), and the instruments used to take the spectra.}
    \label{tab:sample_compar}
    \begin{threeparttable}
    \begin{tabular}{lcccc}
        \hline
        Sample & $z$ range & $M_{1450}$ range & $\rm{N_{QSOs}}$ (in common) & Instruments\\
        \hline
        Onorato et al. (this work) & $5.77-7.54$ & $[-29.13,-25.20]$ & $59$ & GNIRS/NIRES/X-Shooter/GMOS/LRIS/MODS/\\ & & & & FIRE/MOSFIRE/ESI\\
        \cite{Bigwood2024} & $6.17-6.86$ & $[-27.19,-25.32]$ & $19^{\rm{a}}$ ($14$) & FIRE\\
        \cite{Satyavolu2023} & $5.77-6.59$ & $[-29.13,-26.41]$ & $22$ ($22^{\rm{b}}$) & X-Shooter\\
        \cite{Greig2022} & $7.00-7.51$ & $[-26.82,-26.63]$ & $2$ ($2$) & NIRES/GNIRS/GMOS\\
        \cite{Ishimoto2020} (faint) & $5.93-6.56$ & $[-25.32,-23.10]$ & $11$ ($1$) & FOCAS/OSIRIS$^{\rm{c}}$\\
        \cite{Ishimoto2020} (bright) & $5.77-6.54$ & $[-27.75,-24.53]$ & $26$ ($7$) & Spectra from \texttt{igmspec}$^{\rm{d}}$ database\\
        \cite{Eilers2020} & $5.82-6.49$ & $[-27.46,-24.78]$ & $13$ ($2$) & X-Shooter/DEIMOS$^{\rm{e}}$\\
        \cite{Banados2018} & $7.54$ & $-26.34$ & $1$ ($1$) & GNIRS/FIRE\\
        \cite{Davies2018a} & $7.09-7.54$ & $[-26.57, -26.34]$ & $2$ ($2$) & FORS2$^{\rm{f}}$/GNIRS/FIRE\\
        \cite{Eilers2017} & $5.77-6.54$ & $[-29.13,-24.53]$ & $34^{\rm{g}}$ ($10$) & ESI/LRIS\\
        \cite{Mazzucchelli2017} & $6.42-7.08$ & $[-27.19,-24.89]$ & $15^{\rm{h}}$ ($10$) & X-Shooter/LRIS/FIRE/GNIRS/MODS\\ & & & & FORS2/DBSP$^{\rm{i}}$/Red Channel$^{\rm{j}}$\\
        \cite{Reed2017,Reed2015} & $6.07-6.25$ & $[-26.42,-26.02]$ & $9^{\rm{k}}$ & EFOSC2$^{\rm{l}}$/GMOS/MagE$^{\rm{m}}$\\
        \cite{Venemans2015} & $6.51-6.67$ & $[-27.19, -25.83]$ & $3$ ($3$) & FORS2/FIRE/EFOSC2/LRIS/Red Channel/MODS/LUCI$^{\rm{n}}$\\
        \hline
    \end{tabular}
    \begin{tablenotes}
    \item[a] Only $15$ out of $19$ quasars have $R_{\mathrm{p}}$ measurements.
    \item[b] We measure $R_{\mathrm{p}}$ for all the E-XQR-30 sources in \cite{Satyavolu2023}, and we include two more from \cite{D'Odorico2023}.
    \item[c] Acronyms for Faint Object Camera and Spectrograph \citep{Kashikawa2002}, and Optical System for Imaging and low-intermediate-Resolution Integrated Spectroscopy \citep{Cepa2000}, respectively.
    \item[d] \url{https://specdb.readthedocs.io/en/latest/igmspec.html}.
    \item[e] Acronym for DEep Imaging Multi-Object Spectrograph \citep{Faber2003}.
    \item[f] Acronym for FOcal Reducer/low dispersion Spectrograph $2$ \citep{Appenzeller1998}.
    \item[g] Only $31$ out of $34$ quasars have $R_{\mathrm{p}}$ measurements. Also, this sample significantly overlaps with \cite{Ishimoto2020}; thus, we favor comparisons with measurements from the latter being the most recent analysis with updated $z$ measurements.
    \item[h] Only $11$ out of $15$ quasars have $R_{\mathrm{p}}$ measurements.
    \item[i] Acronym for DouBle SPectrograph \citep{Oke1982}.
    \item[j] \cite{Schmidt1989}.
    \item[k] Only $5$ out of $9$ quasars have $R_{\mathrm{p}}$ measurements.
    \item[l] Acronym for ESO Faint Object Spectrograph and Camera (v.2; \citealt{Buzzoni1984}).
    \item[m] Acronym for Magellan Echellette Spectrograph \citep{Marshall2008}.
    \item[n] Formerly known as LUCIFER \citep{Seifert2003}.
    \end{tablenotes}
    \end{threeparttable}
\end{table*}

\begin{table}
    \caption{Quasars for which the proximity zone size measured in this work differs by $\Delta R_{\mathrm{p}} \gtrsim 1$ pMpc from literature values. The columns show, respectively: the number N corresponding to the quasar label in Fig. \ref{fig:Rp-scatter}, its name, its redshift, and the $R_{\mathrm{p}}$ measurements with errors, from both this work and the literature. The letters correspond to the literature samples from which the measurements are taken, listed below.}
    \centering
    \begin{threeparttable}
    \renewcommand{\arraystretch}{1.26}
    \begin{tabular}{lcc|ll|}
    \cline{4-5}
     &  &  & \multicolumn{2}{c|}{$R_{\mathrm{p}}$ [pMpc]} \\ 
    \hline
     \multicolumn{1}{c|}{N} & Name & $z$ & This work & Literature \\
    \hline
     \multicolumn{1}{c|}{1} & J2002$-$3013 & 6.6876 & $2.19^{+0.12}_{-0.13}$ & (a) $3.73 \pm 0.15$\\ \hline
     \multicolumn{1}{c|}{2} & J0142$-$3327 & 6.3373 & $7.81^{+0.13}_{-0.14}$ & (b) $6.50^{+0.14}_{-1.13}$\\ \hline
     \multicolumn{1}{c|}{\multirow{3}{*}{3}} & \multirow{3}{*}{J1319$+$0950} & \multirow{3}{*}{6.1347} & \multirow{3}{*}{$4.96^{+0.15}_{-0.14}$} & (b) $3.87^{+1.08}_{-0.14}$\\
     \multicolumn{1}{c|}{} &&&& (c) $3.84 \pm 0.14$ \\
     \multicolumn{1}{c|}{} &&&& (d) $4.99 \pm 0.04^{*}$ \\ \hline
     \multicolumn{1}{c|}{4} & J1213$-$1246 & 5.893 & $1.53 \pm 0.57$ & (b) $3.09^{+0.67}_{-0.38}$ \\ \hline
     \multicolumn{1}{c|}{5} & J1609$-$1258 & 5.8468 & $5.93 \pm 0.15$ & (b) $4.87 \pm 0.15$ \\ \hline
     \multicolumn{1}{c|}{6} & J0024+3913 & 6.6210 & $3.26 \pm 0.13$ & (e) $4.47 \pm 0.09$ \\
     \hline
    \end{tabular}
    \begin{tablenotes}
    \item Ref: a - \cite{Bigwood2024}; b - \cite{Satyavolu2023}; c - \cite{Eilers2017}; d - \cite{Ishimoto2020}; e - \cite{Mazzucchelli2017}.
    \item[*] Similar to our measurement.
    \end{tablenotes}
    \end{threeparttable}
    \label{tab:scatter}
\end{table}

From their normalized spectra (see Fig. \ref{fig:proxmes_enigma} and \ref{fig:proxmes_exqr30}), we observe that the smoothed flux remains close to the $10\%$ threshold. This suggests that minor variations in data quality, continuum normalization, or the specific criteria used to define $R_{\mathrm{p}}$ could slightly affect the final measurement.
For the cases labeled as 2, 4, and 5, the discrepancies with \citet{Satyavolu2023} seem to arise from different adopted reference wavelengths for the Ly$\alpha$ emission, despite identical reported emission redshifts.
In the case of J0024+3913, we suspect the discrepancy may stem from the fact that \cite{Mazzucchelli2017} re-binned their spectra, changing the SNR per re-binned pixel, and altering the effective resolution.
%% JFH rephrase. I don't think rebinning spectra alters the data quality. The data quality is an intrinsic property of the data. rebinning can change the S/N per rebinned pixel, alter the effective resolution, but it does not alter "data quality". 
%% SO Done

%% JFH It is hard to interpret this if you don't also quote your own measurement!
%% SO Done, I've reported everything in a table
% \begin{enumerate}[label=\arabic*., leftmargin=*, align=left]
%     \item J2002$-$3013: $R_{\mathrm{p}} = 3.73 \pm 0.15$ pMpc in \cite{Bigwood2024}.
%     \item J0142$-$3327: $R_{\mathrm{p}} = 6.50^{+0.14}_{-1.13}$ pMpc in \cite{Satyavolu2023}.
%     \item J1319$+$0950: $R_{\mathrm{p}} = 3.87^{+1.08}_{-0.14}$ pMpc in \cite{Satyavolu2023}, and $R_{\mathrm{p}} = 3.84 \pm 0.14$ pMpc in \cite{Eilers2017}, but $R_{\mathrm{p}} = 4.99 \pm 0.04$ pMpc in \cite{Ishimoto2020}, which is similar to our own measurement.
%     \item J1213$-$1246: $R_{\mathrm{p}} = 3.09^{+0.67}_{-0.38}$ pMpc in \cite{Satyavolu2023}.
%     \item J1609$-$1258: $R_{\mathrm{p}} = 4.87 \pm 0.15$ pMpc in \cite{Satyavolu2023}.
%     \item J0024+3913: $R_{\mathrm{p}} = 4.47$ pMpc in \cite{Mazzucchelli2017}.
% \end{enumerate}

%% JFH this title is inappropriate. I think the main point is that you perform a fit in L and z with a large sample. The fact that you use Bayesian inference, which is very standard in astronomy, is not the point.
%% SO Done
\subsection{Bivariate power-law fit of quasar proximity zones}\label{subsec:doublePL}
Observational and theoretical studies have shown that the proximity zone size depends on both the quasar’s UV luminosity and redshift (e.g., \citealt{Eilers2017, Davies2020}). However, the exact nature of these dependencies remains uncertain, particularly at $z > 6.5$, where constraints on quasar lifetimes and IGM neutral fractions are still debated.

In this section, we fit the dependence of $R_{\mathrm{p}}$ on $M_{1450}$ and $z$, using a Bayesian framework to robustly infer the scaling relations while accounting for measurement uncertainties. Specifically, we fit a bivariate power-law model to the observed $R_{\mathrm{p}}$ values and infer the posterior distributions of the parameters via Markov Chain Monte Carlo (MCMC) sampling. This approach allows us to quantify uncertainties and assess the robustness of our results.
We adopt the following parametric functional form 
%% JFH model --functional form. I would reconsider usage of model here which typically implies a physical model
%% SO Done
for $R_{\mathrm{p}}$:
\begin{equation}\label{eq:doublePL}
R_{\mathrm{p}}(M_{1450},z;\theta) = \mathrm{C} \cdot \left(\frac{1+z}{7}\right)^{\beta} \cdot 10^{-0.4(M_{1450} + 27)/\alpha},
\end{equation}
where $\theta = (\mathrm{s}, \mathrm{C}, \beta, \alpha)$ 
%% JFH Theta is always the whole parameter vector, but you have excluded sigma?
%% SO Done
represents the parameters to be inferred. This form simultaneously accounts for the evolution of proximity zones with redshift and luminosity and is therefore well-suited for fitting the data.
Given our data set $D=\{M_{1450,i},z_i,R_{{\mathrm{p},i}},\sigma_i\}$ with measured proximity zone sizes $R_{{\mathrm{p},i}}$ and associated uncertainties $\sigma_i$, we introduce an intrinsic scatter term, $s$, to account for stochastic effects (i.e., cosmic variance due to the IGM, variations in quasar lifetimes, and spatial fluctuations in $x_{\mathrm{HI}}$ due to the reionization).
% quasars with small and, likely, underestimated uncertainties. 
%% JFH s does not represent "uncertainties". This is very confusing. The additional scatter is "cosmic variance" due to the IGM, plus possibly a distribution of quasar lifetimes, plus possibly fluctuations in xHI due to reionization etc. It is "intrinsic scatter" i.e. physical not associated with measurement error. Don't call it "uncertainty". It is stochasticity.
%% SO Done
We assume Gaussian-distributed deviations 
%% JFH errors should not be used in this context see above.
%% SO Done
and define the log-likelihood function as:
\begin{equation}\label{eq:loglike}
\ln{\mathcal{L}(D|\theta)} = -\frac{1}{2} \sum_{i} \left[ \frac{(R_{{\mathrm{p},i}}-R_{{\mathrm{p}}}(M_{1450,i},z_i;\theta))^2}{\sigma_{i}^{2} + s^2} + \ln(2\pi(\sigma_{i}^{2} + s^2)) \right].
\end{equation}
This formulation accounts for both the measured uncertainties and the additional scatter parameter $s$, ensuring a more realistic assessment of observational errors.
We impose uniform priors on the parameters: $0 < s < 10$, $0 < C < 10$, $-10 < \beta < 10$, and $0 < \alpha < 10$.
The posterior probability is given by Bayes' theorem: $P(\theta|D) \propto P(D|\theta) P(\theta)$; which in logarithmic form becomes: $\ln P(\theta|D) = \ln \mathcal{L}(D|\theta) + \ln P(\theta)$.

To sample the posterior, we use the affine-invariant MCMC ensemble sampler implemented in the Python package \texttt{emcee} \citep{Foreman-Mackey2013}, initializing $32$ walkers around an optimized starting point. This initial estimate is obtained via optimization, finding a plausible parameter set before MCMC sampling.
%% JFH don't say differential evolution. Just say optimization. Differential evolution is the optimizer you used which is a technical detail.
%% SO Done
We run the chains for $50000$ 
%% JFH this is a huge number of steps??
%% SO Yeah
steps per walker and discard the first $5000$ steps ($10\%$) of each chain as burn-in.
%% JFH if you really did 50,000 steps a burn in of 100 is pretty tiny!! It probably does not matter. I would probably do 5000 steps and a 10% burnin.
%% SO I'm doing 50000 steps and 10% burn-in.
We monitor the acceptance fraction of the MCMC chains to ensure proper mixing and convergence.

We run the fit for the first time on a large ensemble of $105$ quasars, made by both this work and the literature, after removing the two quasars in our sample classified as BALs (i.e., J2348$-$3054 and J1526$-$2050), as they could introduce systematic effects in our $R_{\mathrm{p}}$ measurements. The resulting constraints on the model parameters are reported in Table \ref{tab:param} (No BALs) as median values with $1\sigma$ uncertainties ($16^{\text{th}}$ and $84^{\text{th}}$ percentiles).
We evaluate our result by computing the residuals ($\chi$) of the fit between the data (observed $R_{\mathrm{p}}$) and the model predictions. Specifically, we define:
\begin{equation}\label{eq:chi}
    \chi = \frac{\mathrm{data}-\mathrm{model}}{\sigma_\mathrm{{eff}}}
\end{equation}
where $\sigma_\mathrm{{eff}}=\sqrt{\sigma^2 + s^2}$ represents the effective uncertainty on the data, combining the real errors on $R_{\mathrm{p}}$ and the scatter parameter inferred from the MCMC. The distribution of $\chi$ is shown in Fig. \ref{fig:residuals}. In Section \ref{subsec:small}, we use these residuals to define a threshold ($\chi \leq -0.75$) to rigorously identify quasars with small $R_{\mathrm{p}}$ and investigate them for the presence of possible associated absorbers prematurely truncating the proximity zone.

%% JFH A few points about this section. 
%% JFH
%% 1. Currently you define small according to R_p, without any correction. I'm not sure what's been done in the literature before, but maybe since you have a 2d fit now, it makes more sense to define small as a significant outliers from your fit. 
%% 2. I think figures illustrating the bona fide small-zone objects should be in the main text. Discussions of things contaminated by metal-line absorbers should just be put in the appendix. 
%% 2. J1342 J1120 and J0252 all have iGM DW analyses and lifetime estimates and a detailed metal line analyses has been performed for each. You don't cite those and then you redo this yourself. You need to mention that past work. 
%% 3. The main result here appears to be the small zone quasars in your sample that are new. I would highlight that, or anything that you can newly classify as being small given your fit. 
%% I have a t_Q estimate for J0410 that is not that small, but we won't mention that. Anyway, it is important to understand whether it's zone is really small relative to your fit or not. As you know these things depend on redshift and luminosity. 

\subsubsection{Possible contaminants in proximity zones}\label{subsec:small}
Small proximity zone sizes (typically $R_{\mathrm{p}} \lesssim 2$ pMpc) can be interpreted as signatures of short quasar lifetimes ($t_{\mathrm{Q}} \lesssim 10^4$ yr). This interpretation depends on the ionization state of the surrounding IGM. If it is significantly neutral, a small proximity zone arises because the quasar’s ionization front has not had sufficient time to propagate outward and fully ionize the surrounding gas \citep{Eilers2017,Eilers2020}. Conversely, if the IGM is already highly ionized, $R_{\mathrm{p}}$ is sensitive to the quasar lifetime only when $t_{\mathrm{Q}} \lesssim t_{\mathrm{eq}} \sim 3 \times 10^4$ yr (see Section \ref{sec:intro}).
%, since the ionization front driven by the quasar has not had sufficient time to propagate outward and ionize the ambient IGM \citep{Eilers2017,Eilers2020}. 
%% JFH This explanation is true if the IGM is neutral. What if the IGM is already ionized around the quasar? You need to understand this distinction.
%% SO Done.
These measurements are critical for constraining the growth timescales of SMBHs at early cosmic epochs, which must accrete substantial mass within $\lesssim 1$ Gyr of the Big Bang \citep{Volonteri2010, Volonteri2012}.

However, the interpretation of a small $R_{\mathrm{p}}$ as evidence for a young quasar can be confounded by the presence of absorption systems that truncate the transmitted Ly$\alpha$ flux. 
These systems can absorb ionizing photons or Ly$\alpha$ photons and thus artificially shorten the observed proximity zone, independent of the quasar's age.
A rigorous classification of such contaminants is necessary to avoid misidentification. Absorbers can be broadly divided into two categories:
\begin{itemize}
    \item Intervening absorbers, which lie along the line of sight at redshifts lower than the quasar, and hence are not physically associated with its environment. These typically have limited impact on proximity zone measurements unless they are exceptionally strong metal absorption systems, with line equivalent width EW $> 1$ {\angstrom}.
    % unusually strong (e.g., a foreground DLA).
    %% JFH I would just say an exceptionally strong metal absorption system possibly quote an EW > 1 A or something like that.
    %% SO Done
    \item Associated absorbers, which reside at or near the quasar redshift, often within its host halo or immediate surroundings. This is the most relevant class of contaminants for proximity zone analyses.
\end{itemize}

Both types of absorbers can, in principle, affect proximity zone measurements, but only \textit{associated}, optically thick absorbers are considered true contaminants in the context of $R_{\mathrm{p}}$ interpretation. Below, we outline several key scenarios, previously discussed in \cite{Eilers2017, Eilers2020}:

\begin{enumerate}[label=\arabic*., leftmargin=*, align=left]
    \item BALs (and possibly mini BALs): these are intrinsic to the quasar and arise from high-velocity outflows, manifesting as broad, blue-shifted absorption features. They can absorb Ly$\alpha$ and associated metal-line photons, leading to suppression of flux near the systemic redshift. As they alter the continuum and transmitted flux in the proximity zone region, BAL quasars are usually excluded from proximity zone studies.
    
    \item Optically thick associated absorbers (Lyman Limit Systems - LLSs and pDLAs): these include systems with column densities $N_{\mathrm{HI}} \simeq 10^{17-20}$ cm$^{-2}$ (LLSs) and $N_{\mathrm{HI}} \gtrsim 10^{20}$ cm$^{-2}$ (DLAs), located within a few thousand km s$^{-1}$ of the quasar redshift. These absorbers, which likely reside in the quasar's circumgalactic medium (CGM), are self-shielded and optically thick to both ionizing and Ly$\alpha$ photons, and they can truncate the proximity zone by absorbing the entire ionizing flux at their location. From a modeling standpoint, these systems break the assumptions of Ly$\alpha$ forest simulations, which assume a low-density optically thin IGM, and therefore must be excluded from samples used to infer quasar ages.
    %% JFH Fit into here somehow that these are CGM absorbers
    %% SO Done
    Identifying these systems requires detection of saturated Ly$\alpha$ absorption (i.e., line-black regions) in combination with strong low-ionization metal lines (e.g., \ion{O}{I}, \ion{Si}{II}, \ion{C}{II}). We note, however, that at high-$z$ and low metallicity, optically thick associated absorbers may also be traced by weak low-ionization metal lines \citep[e.g.,][]{Sodini2024}. As a result, the SNR of the spectra plays a crucial role in their identification, and in low-SNR data it becomes increasingly challenging to robustly rule out the presence of such absorbers.
    
    \item Optically thin associated absorbers: these exhibit narrow high-ionization metal-line systems (e.g., \ion{C}{IV}, \ion{Si}{IV}) which are detected without associated saturated Ly$\alpha$ absorption, and are optically thin to ionizing photons. Quasars showing such systems are often retained in proximity zone analyses, as their impact on the observed $R_{\mathrm{p}}$ is limited. However, some caution is warranted: such systems may trace previously optically thick gas that has been photoionized away by the quasar. If so, the absorber may still alter the ionization state of the gas in a manner not captured by optically thin Ly$\alpha$ forest models. While these subtleties are usually neglected in practice, they highlight the importance of carefully evaluating the spectral context of each system.

\end{enumerate}

It is therefore imperative to conduct a thorough assessment of each quasar with a small proximity zone to exclude the presence of these contaminants.
Unlike previous studies, where small proximity zones were usually defined as $R_{\mathrm{p}} \lesssim 2$ pMpc (see \citealt{Eilers2017, Eilers2020, Satyavolu2023}), here we take advantage of our bivariate power-law fit described in Section \ref{subsec:doublePL} and identify such objects as those with residuals $\chi \leq -0.75$ (see Fig. \ref{fig:residuals}). They are highlighted in Fig. \ref{fig:proxmes_enigma} and \ref{fig:proxmes_exqr30} by coloring in pink the box containing their details.
 
\begin{figure}
    \centering
    \includegraphics[width=\linewidth]{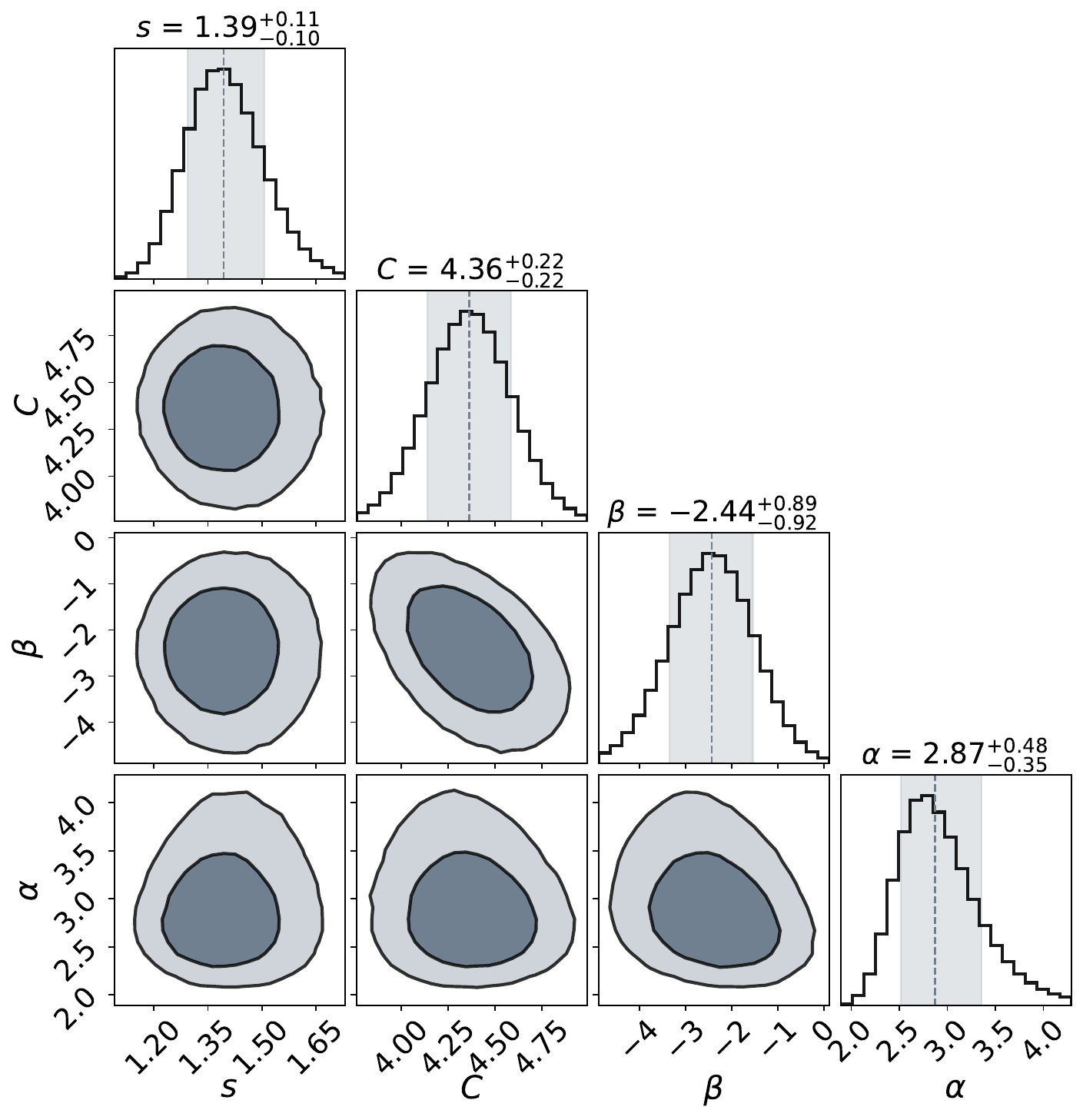}
     \caption{Corner plots of the 4D posterior distribution for the fit described in Section \ref{subsec:doublePL}. They represent our best-fit, obtained after excluding all the contaminated sources identified in Section \ref{subsec:small} (i.e., BALs, pDLAs, quasars whose $R_{\mathrm{p}}$ is affected by metal absorbers). The contours in the 2D histograms highlight the $1 \sigma$, and $2 \sigma$ regions, whereas the dashed lines in the 1D histograms represent the median values of the parameters with the $1 \sigma$ errors shown as shaded regions. These values are reported on top of each histogram.}
     \label{fig:corner_plot}
\end{figure}

We perform a detailed metal absorbers analysis of these quasars (see Appendix \ref{app:int_abs}), and we summarize our main results in Table \ref{tab:smallzone}. We find that the proximity zone size of the quasar J2211$-$6320 (see Fig. \ref{fig:J2211}) may be truncated by the presence of an associated absorber, 
%% JFH explain what kind?
%% SO I would like to know. We detect C IV only, without low-ion lines. Is it an Optically thin associated absorbers?
and we exclude it from our fit, along with the two BALs in our sample already excluded 
%% JFH included in the sample is confusing. Say in our sample
%% SO Done
(i.e., J2348$-$3054 and J1526$-$2050) and the other identified four contaminated sources from the literature that exhibit BAL (J1558$-$0724, J1743$+$4124), pDLA (J0346$-$1628) features, or $R_{\mathrm{p}}$ truncated by metal absorbers (J0840+5624; see Table \ref{tab:smallzone}). 
%% JFH Can you explain what kind of absorbers these contaminated literature sources are??
%% SO Done
Our best-fit parameters are those obtained by excluding these seven contaminated quasars, and we report them in Table \ref{tab:param} and Fig. \ref{fig:corner_plot}.
In Fig. \ref{fig:Rp-mag} and \ref{fig:Rp-z}, these objects excluded from the fit are marked with a black cross.

%% JFH You need to distinguish here between associated and intervening absorbers. Furthermore, the physics behind why this is a problem is not being explained, and it needs to be. There are two scenarios
%% 1. Something like a BAL, mini-BAL whereby either the HI possibly associated with the absorber, or blueshift metals from the absorber will clobber the proximity zone in the region of interest. 
%% 2. A strong associated low-ionization metal absorption line system that is likely to be optically thick to ionizing radiation. For this kind of proximate LLS, or DLA the proximity zone is going to terminate (the HI goes line black) at the absorber redshift. The issue with such objects is that we cannot model them easily with the types of Ly-a forest simulations used to study proximity zones. Or stated another way, the zones are small because of the absorber not because the quasar is young. 
%% 3. If you see optically thin absorbers, like say CIV, but the HI does not go line black, we usually keep these things in the sample. However, this is a bit subtle since it could be that they were once LLSs or DLAs that got ionized away. As such, using Ly-a forest simulations to treat these can be suspect. I would not go into details about this possibility, but you need to understand it. 

\begin{table*}
    \centering
    \begin{threeparttable}
    \caption{Quasars with small $R_{\mathrm{p}}$, identified as those with $\chi \leq -0.75$ from the residuals of the fit in Section \ref{subsec:doublePL}. For both this work and the literature, we list the quasar name, its redshift, the proximity zone size measurement with associated uncertainties in pMpc, and notes. In the last column, we indicate whether metal absorption systems (abs) have been identified in the quasar spectrum, whether these are likely to affect $R_{\mathrm{p}}$, or if the spectrum suffers from other limitations (e.g., low SNR, poor continuum fit, BAL or pDLA features).}
    \label{tab:smallzone}
    \begin{tabular}{l|lc|ccc|l}
    \hline
     Sample & Name & $z$ & $R_{\mathrm{p}}$ [pMpc] & $\sigma_{\mathrm{up}}$ & $\sigma_{\mathrm{low}}$ & Notes\\
    \hline
     \multirow{13}{13em}{Onorato et al. (this work)} & J1007$+$2115 & 7.5149 & 1.40 & 0.11 & 0.11 & No abs identified. \\
     & J1120$+$0641 & 7.0851 & 1.41 & 0.17 & 0.12 & Many abs, not affecting $R_{\mathrm{p}}$ (see 5). \\
     & J0252$-$0503 & 7.0006 & 1.47 & 0.12 & 0.12 & Many abs, not affecting $R_{\mathrm{p}}$ (see 6). \\
     & J1917$+$5003$^{*}$ & 6.853 & 1.28 & 0.47 & 0.47 & Abs at $z_{\mathrm{abs}} \simeq 3.687$, not affecting $R_{\mathrm{p}}$. \\
     & J2211$-$6320$^{*,\mathrm{a}}$ & 6.8449 & 0.19 & 0.12 & 0.12 & Abs at $z_{\mathrm{abs}} \simeq 6.8447$, affecting $R_{\mathrm{p}}$. \\
     & J0218$+$0007 & 6.7700 & 1.01 & 0.12 & 0.12 & No abs identified. \\
     & J0910$+$1656 & 6.7289 & 0.49 & 0.12 & 0.13 & No abs identified. \\
     & J2338$+$2143$^{*}$ & 6.586 & 0.53 & 0.13 & 0.13 & Low SNR to find abs. \\
     & J1110$-$1329 & 6.5148 & 1.40 & 0.13 & 0.16 & Low SNR to find abs. \\
     & J1034$-$1425 & 6.0687 & 1.92 & 0.14 & 0.15 & Many abs, not affecting $R_{\mathrm{p}}$ (see 1, 3, 7). \\
     & J0713$+$0855 & 5.9647 & 2.51 & 0.47 & 0.14 & Many abs, not affecting $R_{\mathrm{p}}$ (see 1, 7). \\
     & J1213$-$1246 & 5.893 & 1.53 & 0.57 & 0.57 & Many abs, not affecting $R_{\mathrm{p}}$ (see 7). \\
     & J0836$+$0054$^{*}$ & 5.773 & 2.76 & 0.59 & 0.58 & Many abs, not affecting $R_{\mathrm{p}}$ (see 7). \\ \hline
    \multirow{10}{13em}{Literature} & J1406$-$0116 & 6.292 & 0.14 & 0.05 & 0.05 & (2, faint): noisy spectrum, poor PCA fit. \\
     & J0330$-$4025 & 6.249 & 1.68 & 0.62 & 0.38 & (3): no abs identified. \\
     & J2229+1457 & 6.1517 & 0.48 & 0.04 & 0.04 & (2, bright; 3): low SNR to find abs. \\
     & J1558$-$0724$^{\mathrm{a}}$ & 6.1097 & 1.32 & 0.14 & 0.14 & (3): BAL, might affect $R_{\mathrm{p}}$. \\
     & J2100$-$1715 & 6.0806 & 0.37 & 0.14 & 0.14 & (3): no abs identified. \\
     & J1743+4124$^{\mathrm{a}}$ & 6.0263 & 1.04 & 0.14 & 0.14 & (3): BAL, might affect $R_{\mathrm{p}}$. \\
     & J0346$-$1628$^{\mathrm{a}}$ & 5.967 & 0.79 & 0.14 & 0.14 & (3): pDLA, might affect $R_{\mathrm{p}}$. \\
     & J1335+3533 & 5.9012 & 0.70 & 0.10 & 0.10 & (2, bright; 4): no abs identified. \\
     & J0840+5624$^{\mathrm{a}}$ & 5.8441 & 0.88 & 0.15 & 0.15 & (4): Many abs, affecting $R_{\mathrm{p}}$. \\
     & J0017+1705 & 5.8165 & 1.16 & 0.15 & 0.15 & (3): abs at $z_{\mathrm{abs}} \simeq 5.8165$, not affecting $R_{\mathrm{p}}$. \\ \hline
    \end{tabular}
    \begin{tablenotes}
    % \scriptsize
    \item Ref: 1 - \cite{Satyavolu2023}; 2 - \cite{Ishimoto2020}; 3 - \cite{Eilers2020}; 4 - \cite{Eilers2017}; 5 - \cite{Bosman2017}; 6 - \cite{Wang2020}; 7 - \cite{Davies2023}.
    \item[*] New $R_{\mathrm{p}}$ measurement.
    \item[a] Excluded from the bivariate power-law fit.
    \end{tablenotes}
    \end{threeparttable}
\end{table*}

\begin{table}
    \caption{Constraints (median values, $16^{\text{th}}$, and $84^{\text{th}}$ percentiles) on the model parameters of the bivariate power-law fit. The first row (No BALs) is obtained after excluding the two BAL quasars in the sample (i.e., J2348$-$3054 and J1526$-$2050). The second row is our best-fit, obtained after excluding all the contaminated sources identified in Section \ref{subsec:small} (i.e., BALs, pDLAs, quasars whose $R_{\mathrm{p}}$ is affected by metal absorbers). These are based on the corner plots in Fig. \ref{fig:corner_plot}.}
    \centering
    \renewcommand{\arraystretch}{1.26}
    \begin{tabular}{l|cccc}
    \hline
     & s & C & $\beta$ & $\alpha$\\
    \hline
    No BALs & $1.52^{+0.12}_{-0.11}$ & $4.06 \pm 0.23$ & $ -1.92^{+0.97}_{-1.01}$ & $2.95^{+0.60}_{-0.42}$\\ \hline
    Best-fit & $1.39^{+0.11}_{-0.10}$ & $4.36\pm 0.22$ & $ -2.44^{+0.89}_{-0.92}$ & $2.87^{+0.48}_{-0.35}$\\
    \hline
    \end{tabular}
    \label{tab:param}
\end{table}

\subsubsection{Correlation with luminosity}\label{subsec:corrL}
The evolution of proximity zone sizes with the luminosity could depend on different factors, one of them being the ionization state of the surrounding IGM. From Equation \ref{eq:prox_zone} we see that the size of the ionization front expanding into a neutral IGM evolves as $R_{\mathrm{p}} \propto \dot{N}_\gamma^{1/3}$. However, the analytical model from \cite{Bolton2007b} shows that $R_{\mathrm{p}} \propto \dot{N}_\gamma^{1/2}$ in a highly ionized medium. Given the patchy topology of the IGM at these redshifts (e.g., \citealt{Davies2018a}), the expected scaling should fall between the two limits.
We also consider the predicted evolutions inferred by the simulations in \cite{Eilers2017} for a mostly neutral (Equation \ref{eq:neutr}) and ionized (Equation \ref{eq:ion}) IGM, respectively, described by the following power-laws:
\begin{align}
R_{\mathrm{p}} \approx 5.03 \cdot 10^{-0.4(M_{1450} + 27)/2.45}, \label{eq:neutr}\\
R_{\mathrm{p}} \approx 5.57 \cdot 10^{-0.4(M_{1450} + 27)/2.35}, \label{eq:ion}
\end{align}
implying that $ R_{\mathrm{p}} \propto 10^{-0.4 M/\alpha} \propto L^{1/\alpha} \propto \dot{N}_\gamma^{1/\alpha}$, with $\alpha = 2.45$ and $2.35$, from Equations \ref{eq:neutr} and \ref{eq:ion}, respectively.
%% JFH Maybe say that these imply N^alpha, i.e. quote the number 1/2.45 and 1/2.35 to make it more clear
%% SO Done

\begin{figure}
    \includegraphics[width=\linewidth]{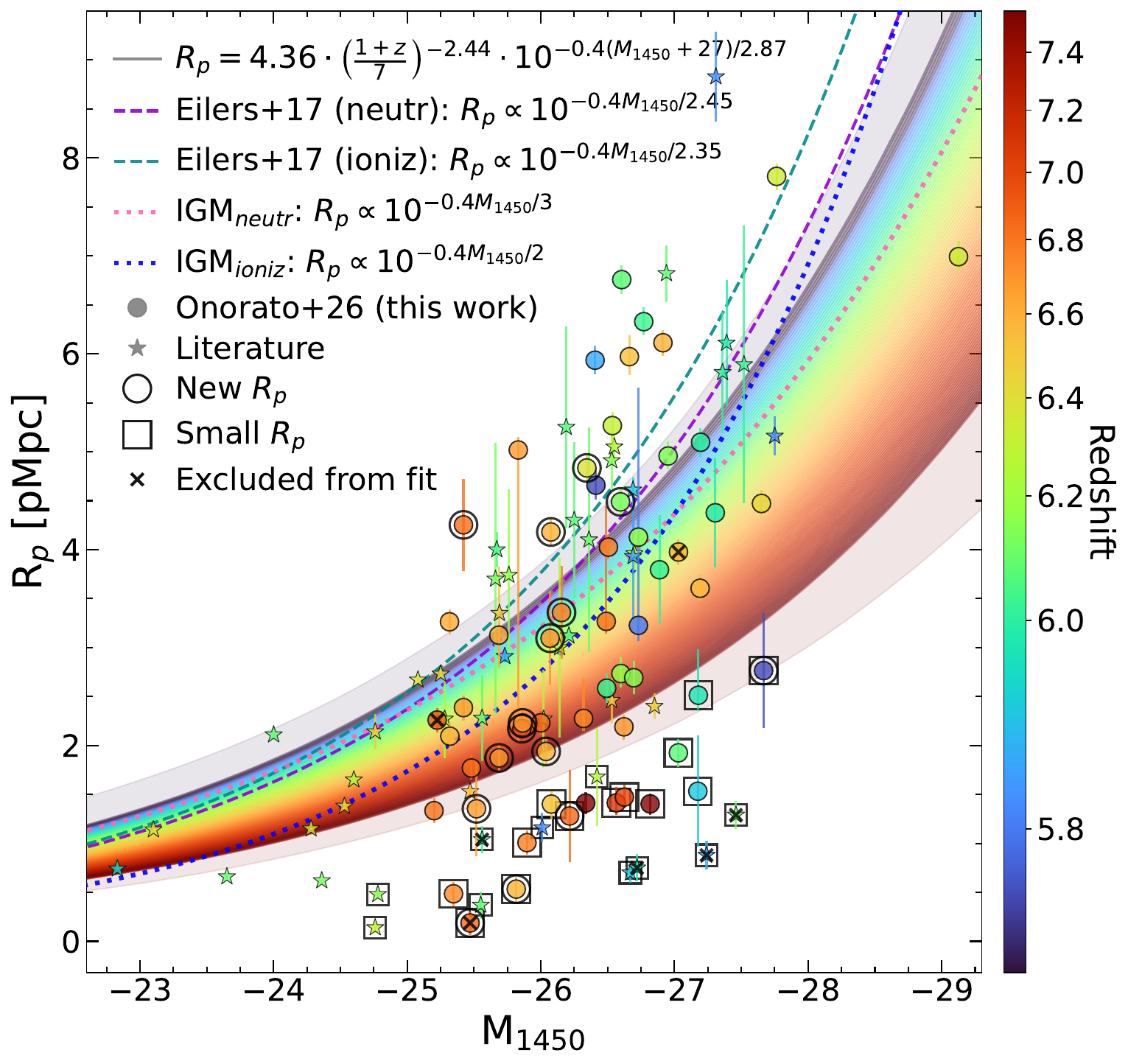}
    \caption{Distribution of $R_{\mathrm{p}}$ as a function of $M_{1450}$ for the quasars included in this study (points) and for others from the literature (stars), with the corresponding error bars. The color of the symbols is determined by the redshift of the quasars, according to the colormap to the right. The big black circles surrounding some points highlight the new $R_{\mathrm{p}}$. The black squares mark the objects with small $R_{\mathrm{p}}$, identified from the residuals of the fit in Section \ref{subsec:doublePL} ($\chi \leq -0.75$). The quasars excluded from the plotted final fit (i.e., BALs, pDLAs, $R_{\mathrm{p}}$ truncated by metal absorbers) are marked with a black cross. We show a distribution of curves from our bivariate power-law fit, obtained by varying the redshift from $z=7.55$ (the darkest red curve) to $z=5.76$ (the darkest blue curve). Median values (solid lines) come from randomly sampling the MCMC for the posterior distribution $1000$ times. The $1\sigma$ uncertainty regions (shaded areas) are derived from the $16^{\text{th}}$ and $84^{\text{th}}$ percentiles of the distribution, but we only show the lower uncertainty bound for the highest-$z$ curve and the upper uncertainty bound for the lowest-$z$ curve.
    The dashed purple and green curves come from the radiative transfer simulations for a mostly neutral and ionized IGM by \citet{Eilers2017}, respectively. The dotted pink and blue curves show the theoretical expectations for a mostly neutral IGM surrounding the quasar and the expected evolution of $R_{\mathrm{p}}$ from the analytical model by \citet{Bolton2007b} in a highly ionized IGM, respectively. Our fit falls between the models' predictions, and we find many quasars with small $R_{\mathrm{p}}$, that, hence, might have short lifetimes ($t_{\mathrm{Q}}\lesssim 10^4$ yr).}
    \label{fig:Rp-mag}
\end{figure}

In Fig. \ref{fig:Rp-mag}, we consider these predictions together with our data and the results obtained from our bivariate power-law fit, finalized in Section \ref{subsec:small}.
We present the distribution of $R_{\mathrm{p}}$ as a function of the quasars' magnitude, color-coded with their redshift. Given the large sample size of this work and the literature, the sources considered span in a wide magnitude range $-29.13 \leq M_{1450} \leq -22.83$, in which the distribution of proximity zone sizes shows significant variation and scatter. 
% Particularly, this scatter increases when moving from lower to higher luminosity quasars. 
%% JFH Not sure we need this below, and anyway it should only be presented after all the empirical results in a separate paragraph. 
% To explain this evidence, we echo \cite{Satyavolu2023}, as they attribute it to the different expansion rates of ionization fronts. In luminous objects, they rapidly expand, increasing the likelihood of finding neutral hydrogen islands along different lines of sight. In contrast, fainter quasars require more time for their ionization fronts to propagate to similar distances, limiting their interaction with neutral regions. As a result, the surrounding environment of fainter quasars appears more uniformly ionized, leading to a narrower distribution of $R_{\mathrm{p}}$.

We point out the presence of many outliers, highlighted by black squares. These objects show small $R_{\mathrm{p}}$ and possibly have short lifetimes ($t_{\mathrm{Q}}\lesssim 10^4$ yr), except for those also marked with a black cross, identified as contaminated quasars and, as such, excluded from our final fit (see Section \ref{subsec:small}).
% where one of them (i.e., J2211$-$6320) is explained by the presence of dense associated metal absorbers 
%% JFH Why are we discussing contaminated things here? Focus on the outliers which are the short lifetimes. 
%% SO Done
% that alter the shape of the proximity zone (see Section \ref{subsec:small}), while the others could be explained by short quasar lifetimes (see Section \ref{sec:intro}).
% A detailed study of the quasars in this sample that exhibit small $R_{\mathrm{p}}$, and hence possibly have short $t_{\mathrm{Q}}$, will be performed in future works.
%% JFH Omit reference to future work. 
%% SO Done
We plot a distribution of curves from our fit, obtained by varying the redshift from $z=7.55$ (the darkest red curve) to $z=5.76$ (the darkest blue curve). Median values (solid lines) are obtained by randomly sampling the Markov chains for the posterior distribution $1000$ times. The $1\sigma$ uncertainty regions (shaded areas) are derived from the $16^{\text{th}}$ and $84^{\text{th}}$ percentiles of the distribution; however, for clarity, we only show the lower uncertainty bound for the highest-$z$ curve and the upper uncertainty bound for the lowest-$z$ curve.

We compare them with the theoretical expectations for a mostly neutral IGM (dotted pink line), the analytical model described by \citealt{Bolton2007b} (dotted blue line) and the predictions obtained by \cite{Eilers2017} for a highly neutral and ionized medium, shown in Equations \ref{eq:neutr} and \ref{eq:ion} (dashed purple and green lines, respectively).
The normalization for the theoretical and analytical curves was arbitrarily chosen in \cite{Eilers2017}, so we change and set it to our derived parameter C, for visual and comparison purposes. We also quote that our normalization corresponds to a smaller value than those inferred in the simulations by \cite{Eilers2017}.
Our estimated luminosity dependence falls between the limits expected for neutral and highly ionized environments.

To conclude, we try to explain the visibly increasing scatter in the $R_{\mathrm{p}}$ distribution when moving from lower to higher luminosity quasars, echoing \cite{Satyavolu2023}. They attribute this evidence to the different expansion rates of ionization fronts. In luminous objects, they rapidly expand, increasing the likelihood of finding neutral hydrogen islands along different lines of sight. In contrast, fainter quasars require more time for their ionization fronts to propagate to similar distances, limiting their interaction with neutral regions. As a result, the surrounding environment of fainter quasars appears more uniformly ionized, leading to a narrower distribution of $R_{\mathrm{p}}$.

\subsubsection{Correlation with redshift}\label{subsec:corrz}
\begin{figure}
    \includegraphics[width=\linewidth]{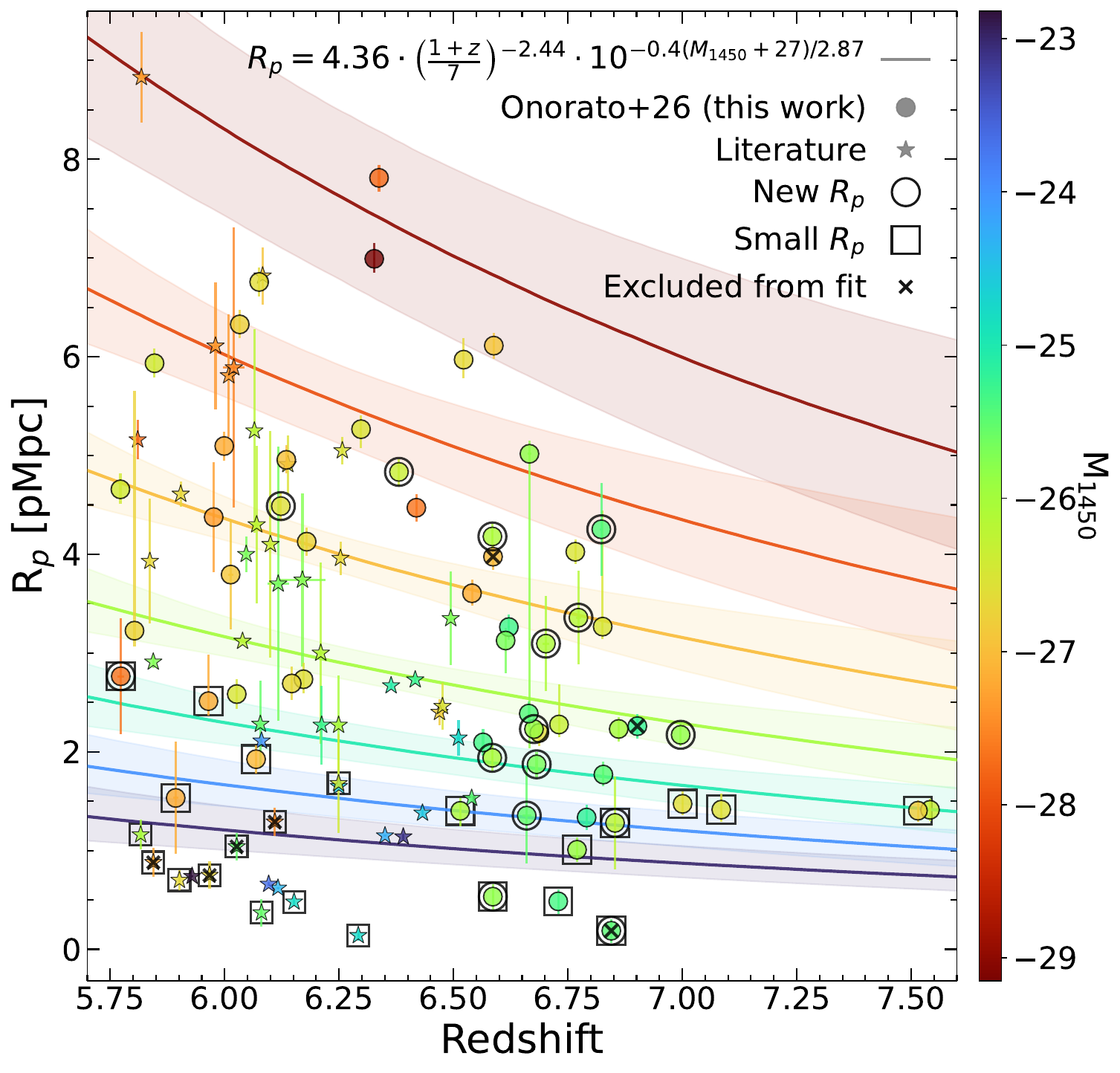}
    \caption{Distribution of $R_{\mathrm{p}}$ non-corrected for their luminosity as a function of $z$ for the quasars included in this study (points) and for others from the literature (stars), with the corresponding error bars. The color of the symbols is determined by the $M_{1450}$ of the quasars, according to the colormap to the right. The big black circles surrounding some points highlight the new $R_{\mathrm{p}}$. The black squares mark the objects with small $R_{\mathrm{p}}$, identified from the residuals of the fit in Section \ref{subsec:doublePL} ($\chi \leq -0.75$). The quasars excluded from the plotted final fit (i.e., BALs, pDLAs, $R_{\mathrm{p}}$ truncated by metal absorbers) are marked with a black cross. We show seven curves from our bivariate power-law fit, obtained by setting the magnitude to be $M_{1450}=-29$ (dark red), $M_{1450}=-28$ (light red), $M_{1450}=-27$ (dark yellow), $M_{1450}=-26$ (lime), $M_{1450}=-25$ (turquoise), $M_{1450}=-24$ (light blue) and $M_{1450}=-23$ (dark blue). Median values (solid lines) and $1\sigma$ uncertainty regions (shaded areas) come from randomly sampling the MCMC for the posterior distribution $1000$ times. They show a steeper redshift evolution than found in previous studies, suggesting that the IGM’s ionization state might play a role in shaping $R_{\mathrm{p}}$.}
    \label{fig:Rp-z}
\end{figure}

We study the variation in proximity zone sizes with redshift as it can provide valuable information on the reionization process. In this paper, we follow the same approach as adopted in \cite{Satyavolu2023} and do not apply luminosity corrections to standardize proximity zones to a single magnitude, as the relationship between $R_{\mathrm{p}}$ and $M_{1450}$ varies significantly with $z$, making a universal correction inapplicable. Thus, Fig. \ref{fig:Rp-z} shows the proximity zone sizes non-corrected for their luminosity, from both this work and the literature, as a function of redshift.
We plot seven curves from our bivariate power-law fit (Section \ref{subsec:doublePL}), obtained by fixing the $M_{1450}$ to be, in the order, $M_{1450}=-29$ (dark red), $M_{1450}=-28$ (light red), $M_{1450}=-27$ (dark yellow), $M_{1450}=-26$ (lime), $M_{1450}=-25$ (turquoise), $M_{1450}=-24$ (light blue) and $M_{1450}=-23$ (dark blue). Median values (solid lines) and $1\sigma$ uncertainty regions (shaded areas) are obtained by randomly re-sampling the Markov chains for the posterior distribution $1000$ times.

Our results indicate a steeper redshift dependence than found in previous studies (e.g., \citealt{Satyavolu2023}: $R_{\mathrm{p}} \propto (1+z)^{-0.89}$), suggesting that the evolution of the IGM’s ionization state may play a more significant role in shaping proximity zone sizes than previously assumed.
We conclude that at fixed $M_{1450}$, variations in $R_{\mathrm{p}}$ can arise either from $z$ evolution (i.e., if the quasars exhibit different redshifts in Fig. \ref{fig:Rp-mag}) or from differences in the quasar lifetimes. Further investigation is needed to disentangle these effects.

\section{Summary and conclusions}\label{sec:summary}
In this work, we have conducted a comprehensive analysis of quasar proximity zones using a sample of $59$ quasars spanning redshifts $5.77 \leq z \leq 7.54$. This data set includes new measurements for $15$ quasars that had not been previously analyzed. By leveraging good quality spectra from multiple instruments, we have obtained robust estimates of proximity zone sizes ($R_{\mathrm{p}}$) and explored their dependence on quasar luminosity and redshift. Our main findings can be summarized as follows:

\begin{itemize}
    \item Our measurements yield a range of $0.19 \leq R_{\mathrm{p}} \leq 7.81$ pMpc, with a median size of $R_{\mathrm{p}} = 2.76$ pMpc. The distribution exhibits significant scatter, reflecting potential variations in quasar lifetimes and IGM properties.
    
    \item We fit a bivariate power-law model to our data and literature. The inferred dependence on redshift and quasar luminosity follows the scaling $R_{\mathrm{p}} \propto (1+z)^{\beta} \cdot 10^{-0.4 M_{1450} / \alpha}$, with $\alpha = 2.87^{+0.48}_{-0.35}$ and $\beta = -2.44^{+0.89}_{-0.92}$, when we exclude contaminated sources (i.e., BALs, pDLAs, quasars with $R_{\mathrm{p}}$ truncated by metal absorbers). The evolution of the proximity zones with the luminosity falls between the model predictions for a highly neutral and ionized IGM. The redshift evolution is steeper than in other studies (e.g., \citealt{Satyavolu2023}), suggesting that the ionization state of the IGM might play a role in shaping proximity zone sizes.
    
    \item There are $13$ quasars that exhibit $R_{\mathrm{p}}$ values smaller than expected from model predictions. For all of them except J2211$-$6320, we rule out the presence of proximate dense absorbers as the cause, leading us to propose short quasar lifetimes ($t_\mathrm{Q} \lesssim 10^4$ yr) as a plausible explanation, and hence posing challenges on SMBH growth.

\end{itemize}

%% JFH maybe mention more your PDf paper in prep
%% SO Done
Our findings highlight the importance of considering both reionization and quasar lifetimes when interpreting proximity zone sizes. In a future work (Onorato et al., in prep), we will incorporate a new inference approach \citep{Hennawi2024,Kist2024} and hydrodynamical simulations of ionization front evolution. This will enable us to rigorously test the reliability of our models by comparing the Ly$\alpha$ transmission probability distribution function (PDF) derived from observed quasar spectra in our sample with that from mock spectra, generated under specific assumptions on the neutral hydrogen fraction $x_{\mathrm{HI}}$ and the quasar lifetime $t_{\mathrm{Q}}$.
% refining the constraints on both the neutral hydrogen fraction $x_{\mathrm{HI}}$ and the quasar lifetime $t_{\mathrm{Q}}$.

\section*{Acknowledgements}
We acknowledge \cite{D'Odorico2023} and \cite{Durovcikova2024} for their public data release, which made this work possible.
We acknowledge helpful conversations with the ENIGMA group at UC Santa Barbara and Leiden University.
SO is grateful to Timo Kist, Shane Bechtel, and Daming Yang for comments on an early version of the manuscript.
JFH acknowledges support from the European Research Council (ERC) under the European Union’s Horizon 2020 research and innovation program (grant agreement No 885301), from the National Science Foundation (NSF) under Grant No. 2307180, and from NASA under the Astrophysics Data Analysis Program (ADAP, Grant No. 80NSSC21K1568).

%%%%%%%%%%%%%%%%%%%%%%%%%%%%%%%%%%%%%%%%%%%%%%%%%%
\section*{Data Availability}
The data underlying this article will be shared on reasonable request to the corresponding author.



\bibliographystyle{mnras}
\bibliography{ref} % if your bibtex file is called example.bib

%\begin{thebibliography}{99}
%\end{thebibliography}

%%%%%%%%%%%%%%%%%%%%%%%%%%%%%%%%%%%%%%%%%%%%%%%%%%

%%%%%%%%%%%%%%%%% APPENDICES %%%%%%%%%%%%%%%%%%%%%

\appendix

\section{Continuum modeling for the quasars in the sample}
\begin{figure*}
    \centering
    \includegraphics[width=0.95\linewidth]{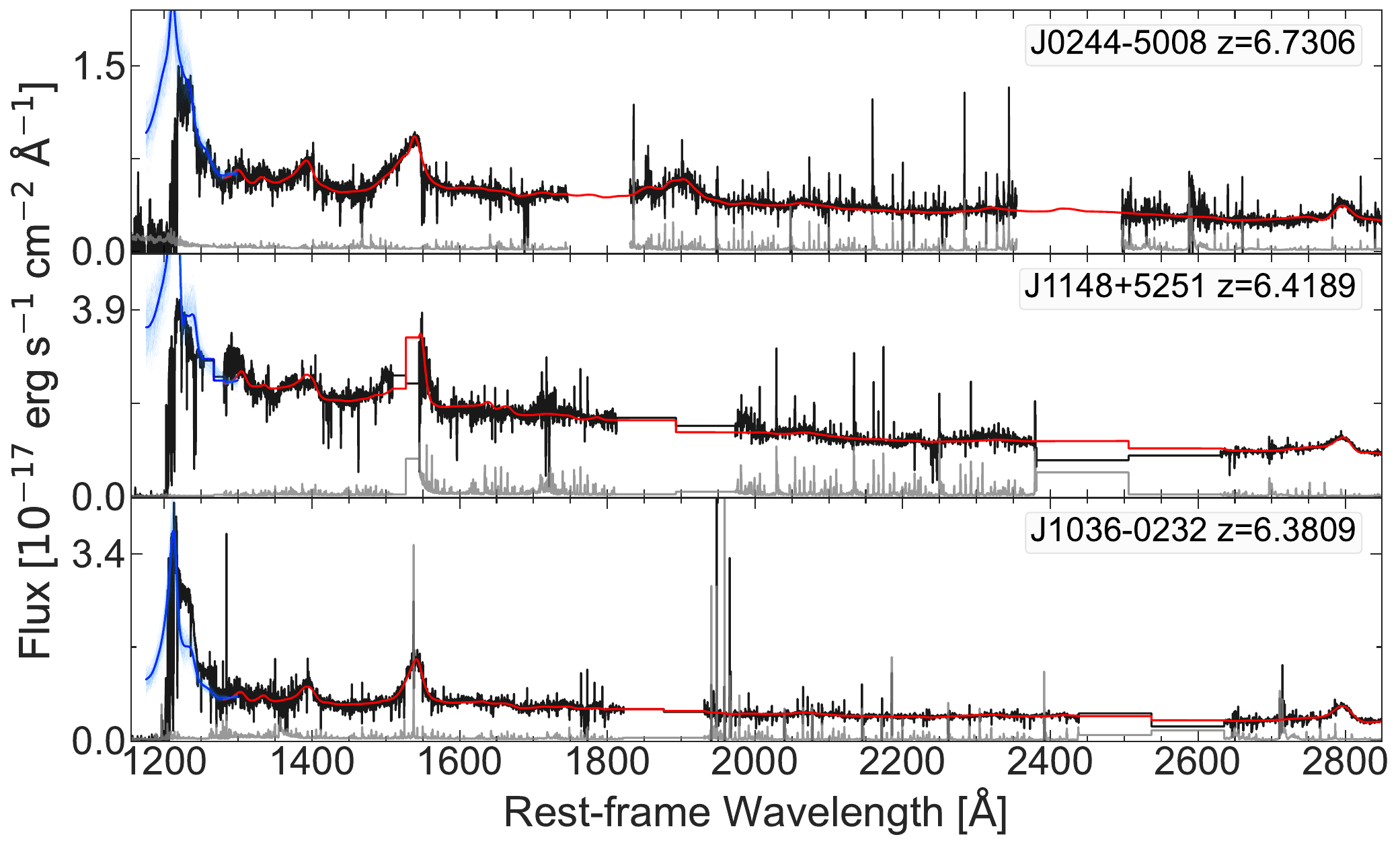}
     \caption{Spectra of the three quasars in the FIRE sub-sample, sorted by decreasing $z$, and their best-fitting continuum model using principal component analysis (PCA) as described in Fig. \ref{fig:pca}. All the spectra are smoothed for visual purposes and we have masked regions of strong telluric absorption.}
     \label{fig:pca_plot_fire}
\end{figure*}

Due to the large sample size, for displaying choices, we report here the continuum model reconstructions for the quasars in this work (see Fig. \ref{fig:pca_plot_enigma}, \ref{fig:pca_plot_exqr30}, and \ref{fig:pca_plot_fire}).
% , and their proximity zone size measurements (see Fig. \ref{fig:proxmes_enigma}, \ref{fig:proxmes_exqr30}, and \ref{fig:proxmes_fire}).
A full description of the method is provided in Section \ref{subsec:pca}.

%%%%%% PCA ENIGMA
\begin{figure*}
    \centering
    \includegraphics[width=0.95\linewidth]{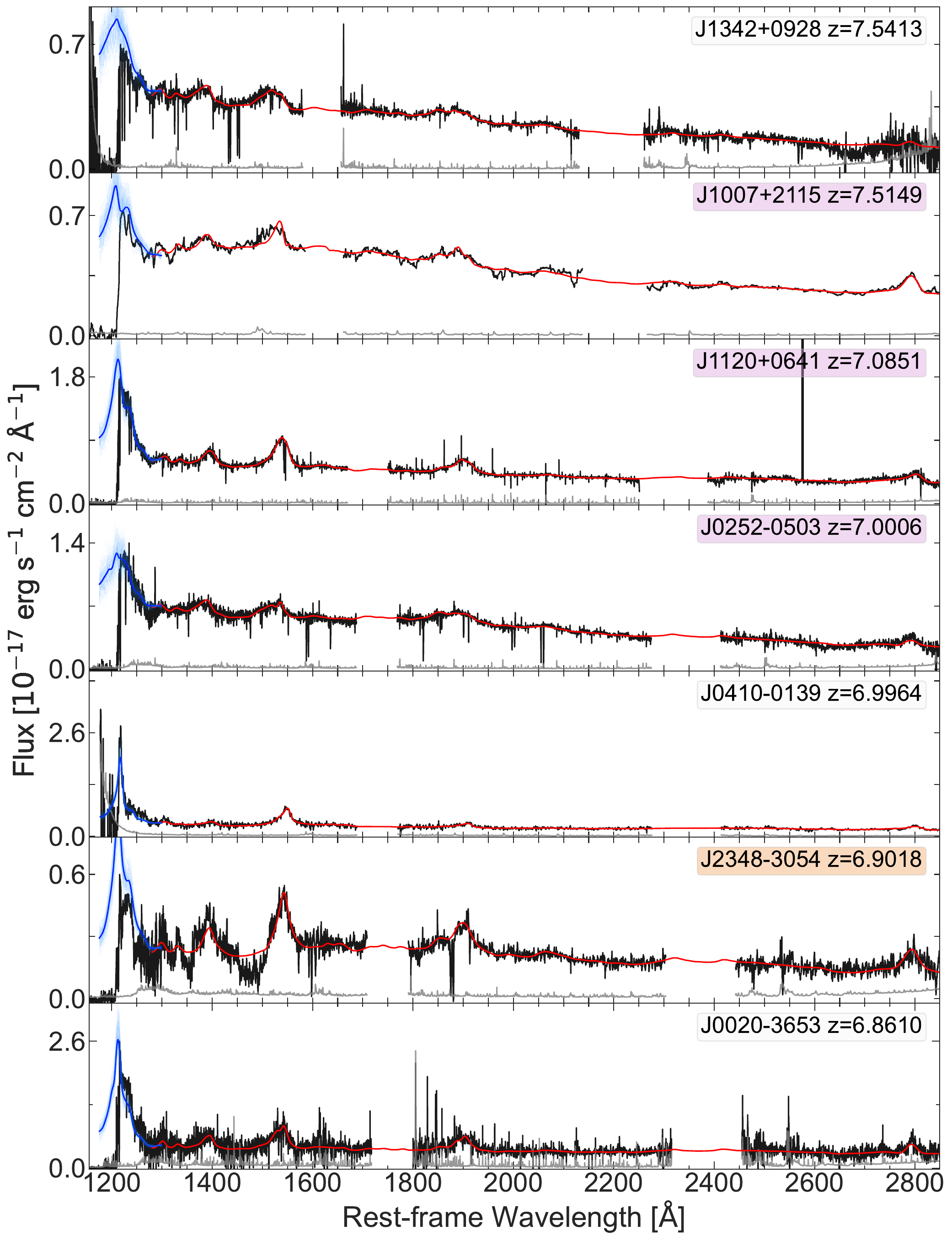}
     \caption{Spectra of the $35$ quasars in the ENIGMA sub-sample, sorted by decreasing $z$, and their best-fitting continuum model using principal component analysis (PCA) as described in Fig. \ref{fig:pca}. All the spectra are smoothed for visual purposes and we have masked regions of strong telluric absorption. The box containing the name and the redshift of the quasars is colored in orange to mark the BALs and in pink to highlight the small proximity zones defined in Section \ref{subsec:small}.}
     \label{fig:pca_plot_enigma}
\end{figure*}

\begin{figure*}
    \addtocounter{figure}{-1}
    \centering
    \includegraphics[width=0.95\linewidth]{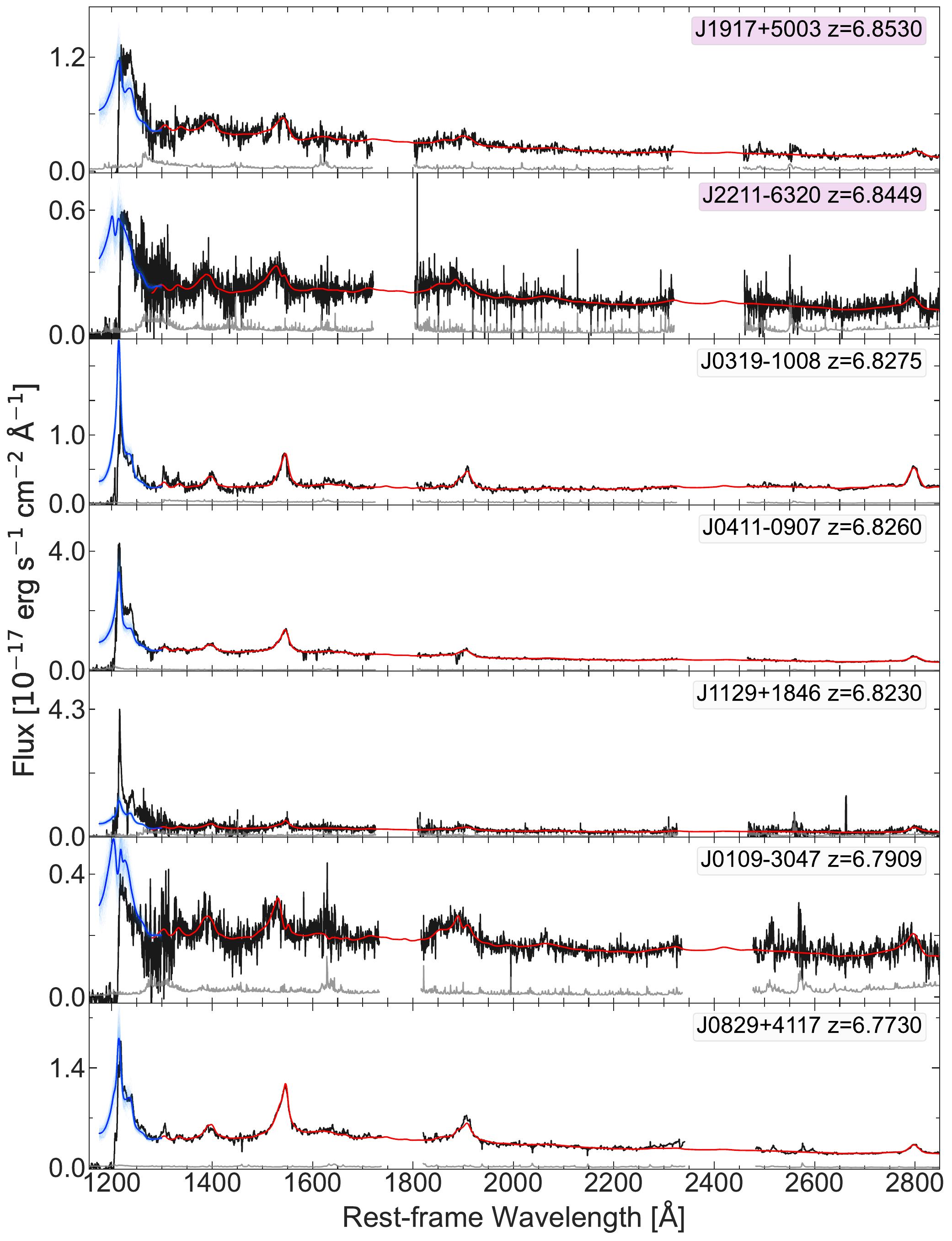}
     \caption{(Continued)}
\end{figure*}

\begin{figure*}
    \addtocounter{figure}{-1}
    \centering
    \includegraphics[width=0.95\linewidth]{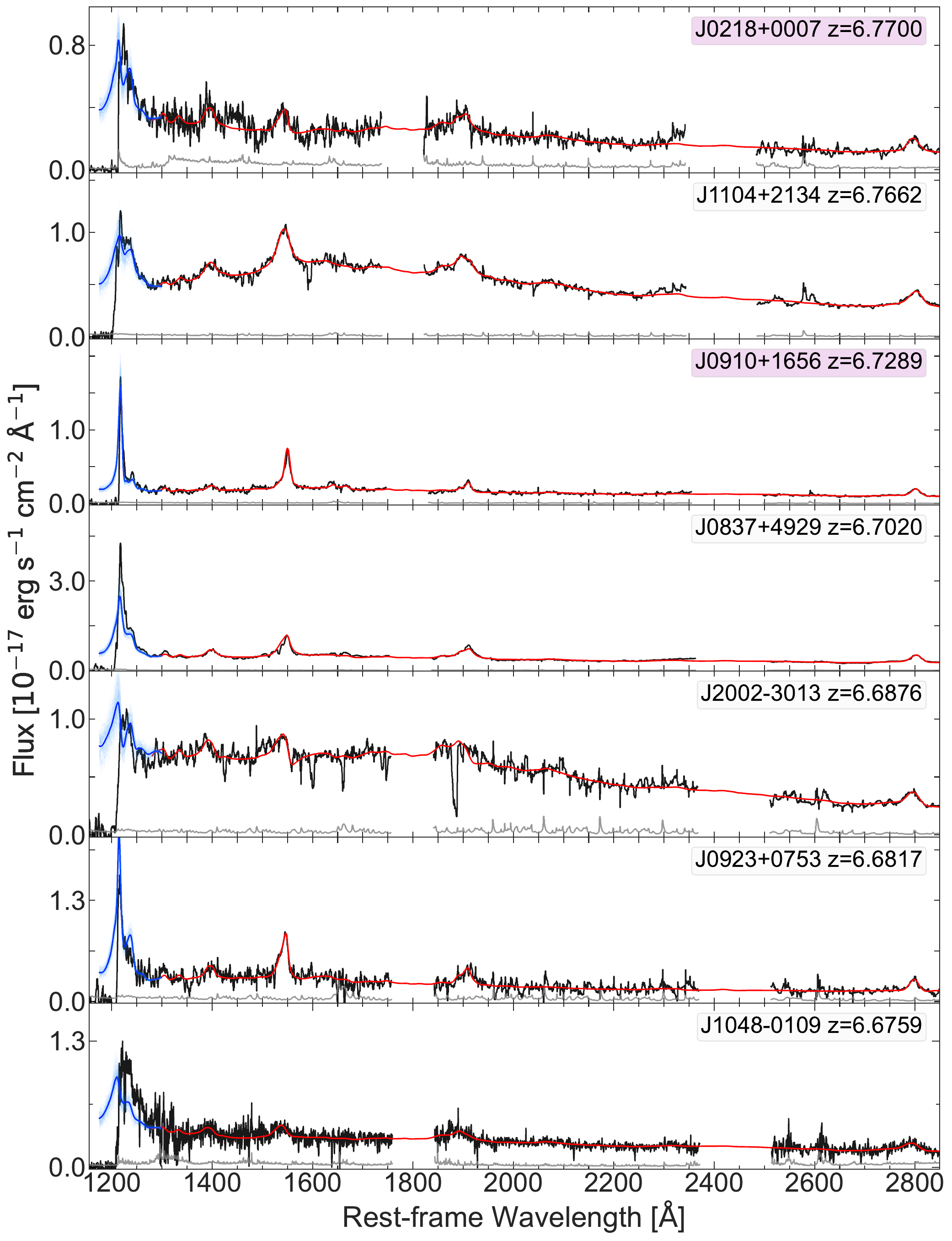}
     \caption{(Continued)}
\end{figure*}

\begin{figure*}
    \addtocounter{figure}{-1}
    \centering
    \includegraphics[width=0.95\linewidth]{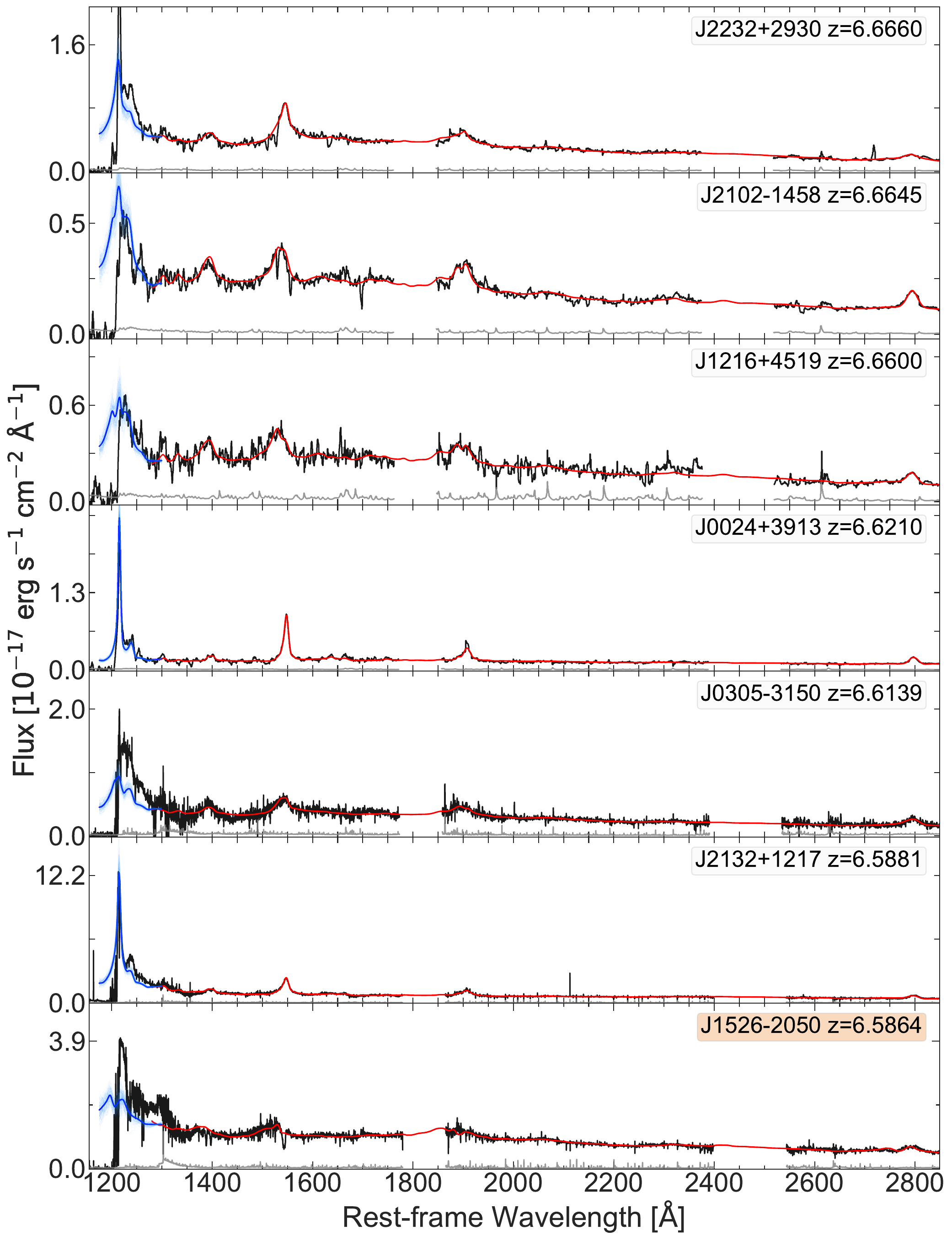}
     \caption{(Continued)}
\end{figure*}

\begin{figure*}
    \addtocounter{figure}{-1}
    \centering
    \includegraphics[width=0.95\linewidth]{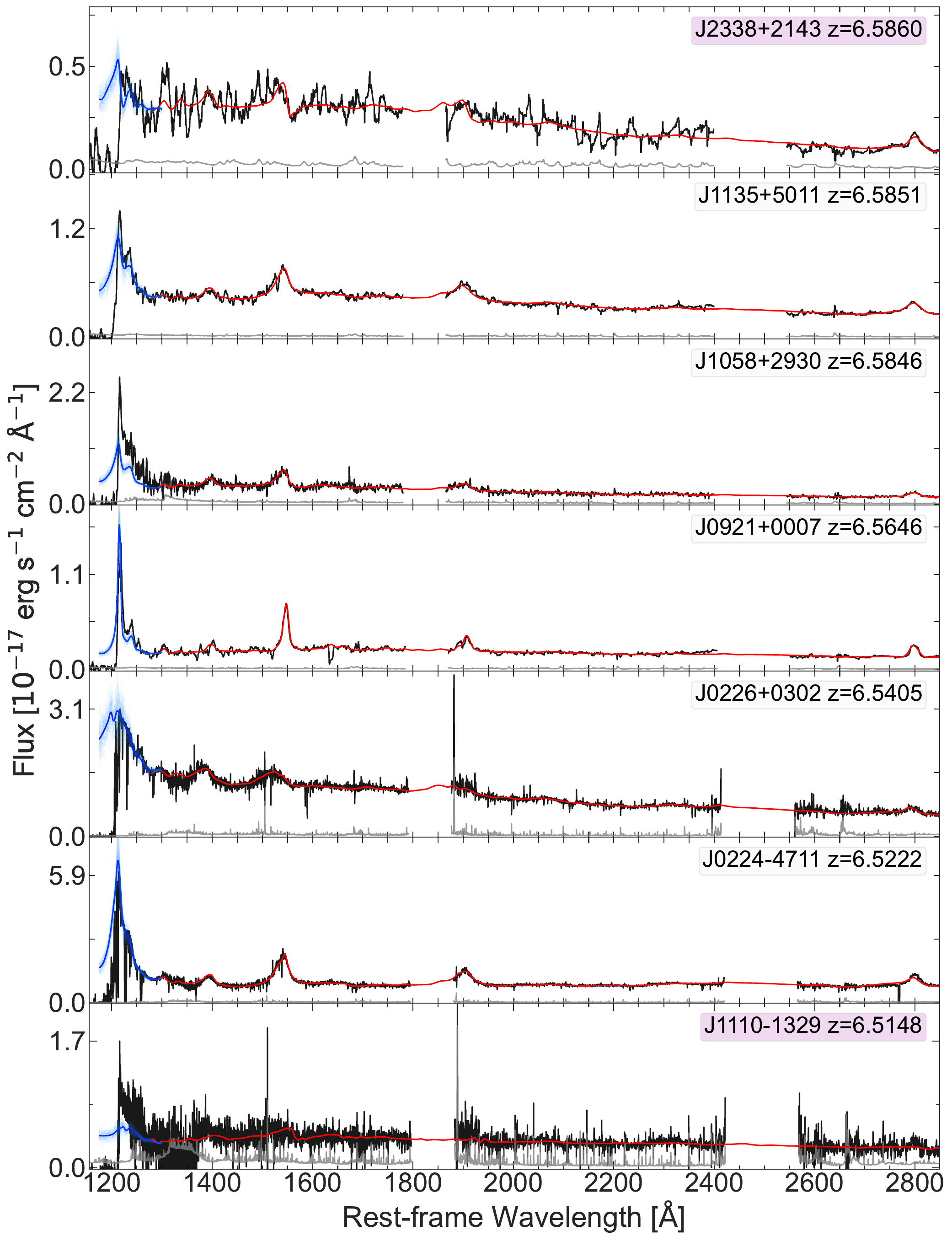}
     \caption{(Continued)}
\end{figure*}

%%%%%% PCA EXQR30
\begin{figure*}
    \centering
    \includegraphics[width=0.95\linewidth]{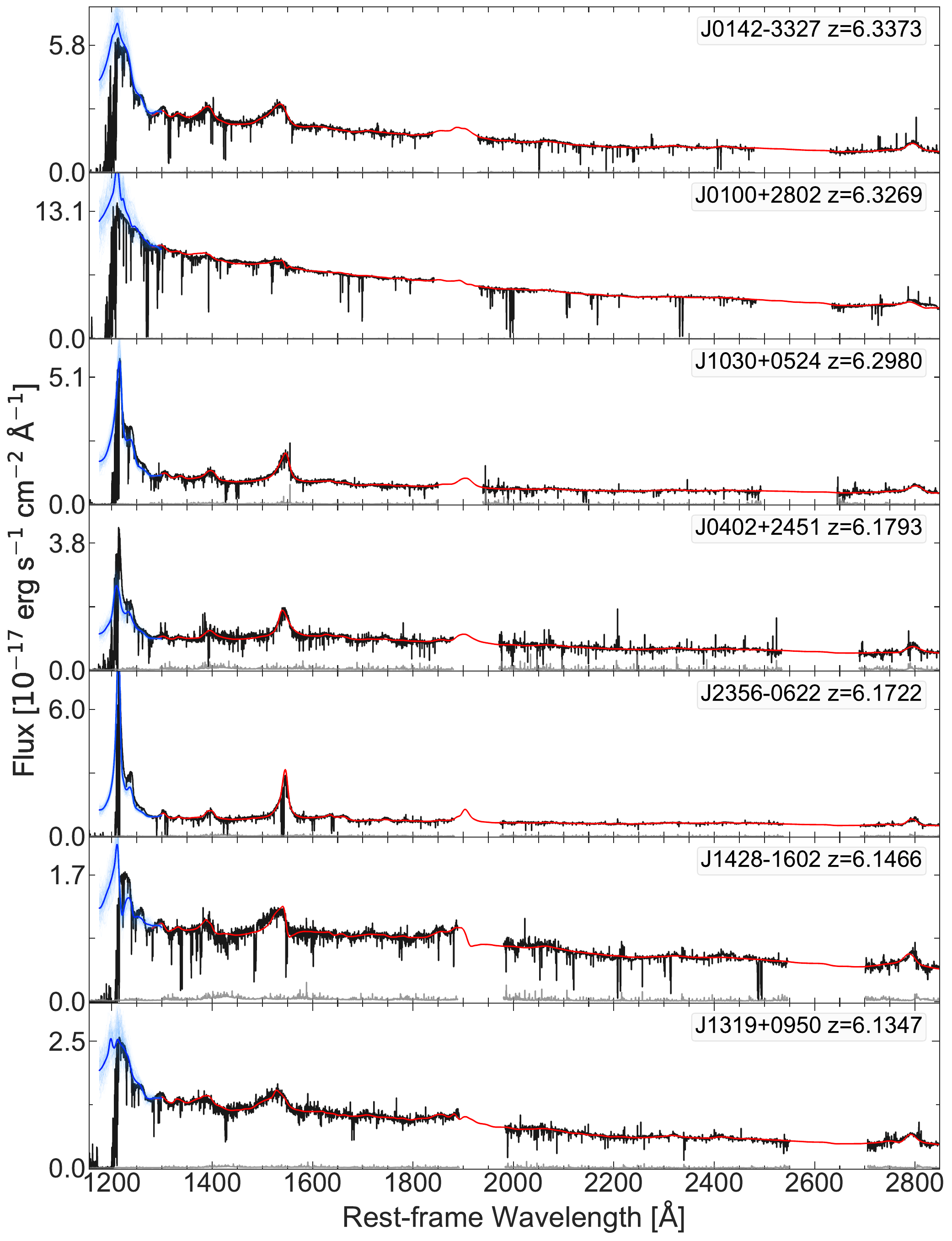}
     \caption{Spectra of the $21$ quasars in the E-XQR-30 sub-sample, sorted by decreasing $z$, and their best-fitting continuum model using principal component analysis (PCA) as described in Fig. \ref{fig:pca}. All the spectra are smoothed for visual purposes and we have masked regions of strong telluric absorption. The box containing the name and the redshift of the quasars is colored in pink to highlight the small proximity zones defined in Section \ref{subsec:small}.}
     \label{fig:pca_plot_exqr30}
\end{figure*}

\begin{figure*}
    \addtocounter{figure}{-1}
    \centering
    \includegraphics[width=0.95\linewidth]{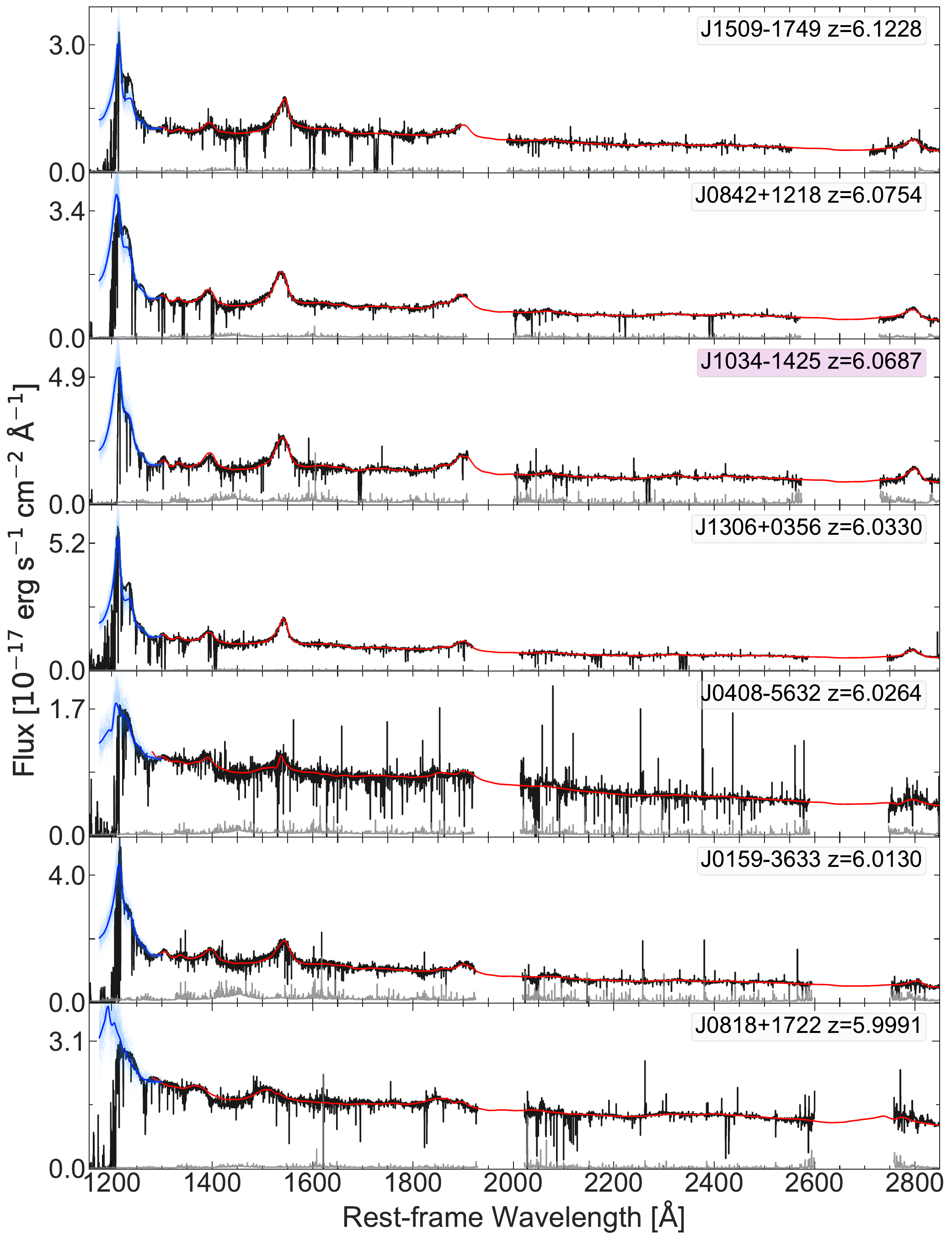}
     \caption{(Continued)}
\end{figure*}

\begin{figure*}
    \addtocounter{figure}{-1}
    \centering
    \includegraphics[width=0.95\linewidth]{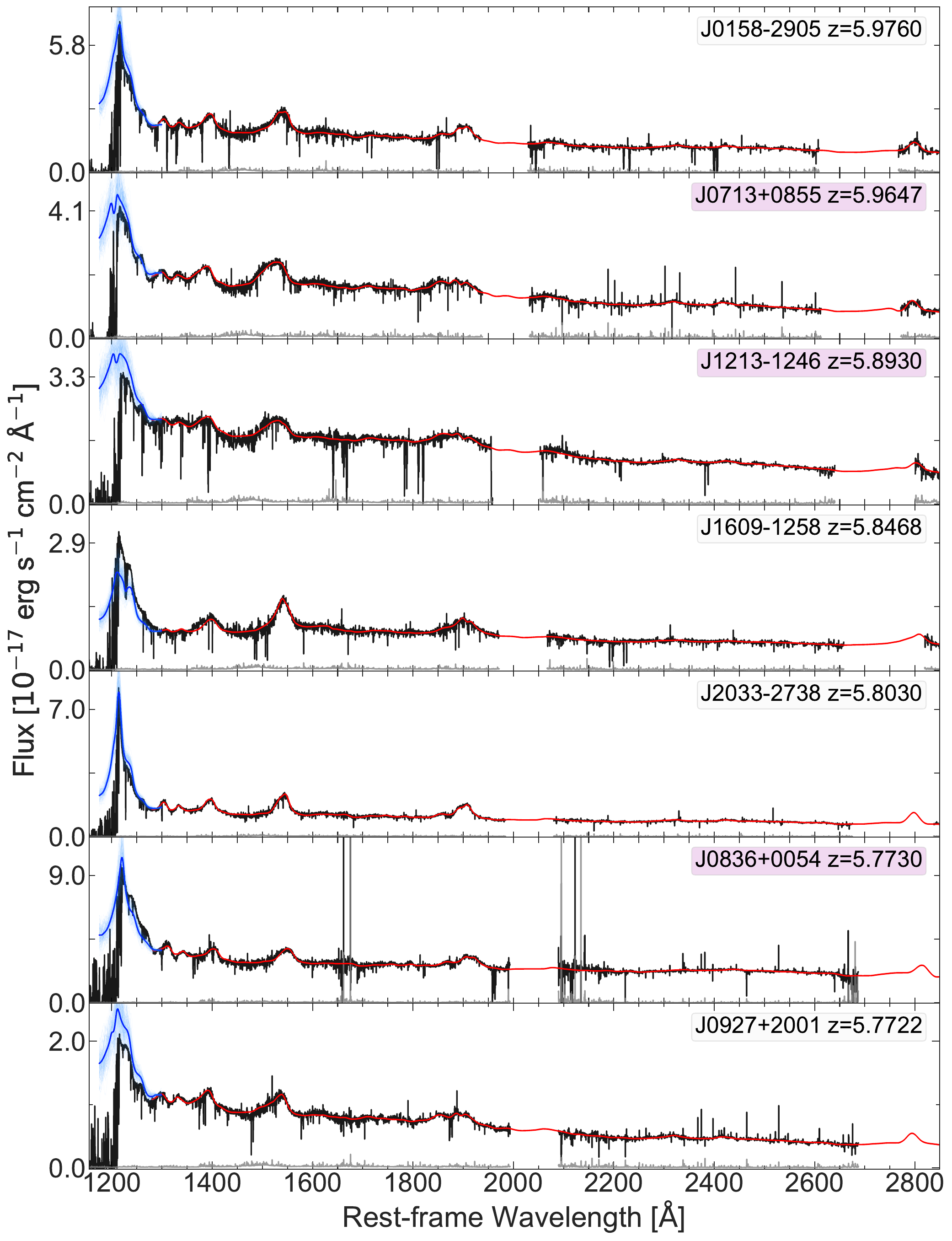}
     \caption{(Continued)}
\end{figure*}

%%%%%% PCA FIRE

\section{Additional results from the fit}
\begin{figure}
    \centering
    \includegraphics[width=\linewidth]{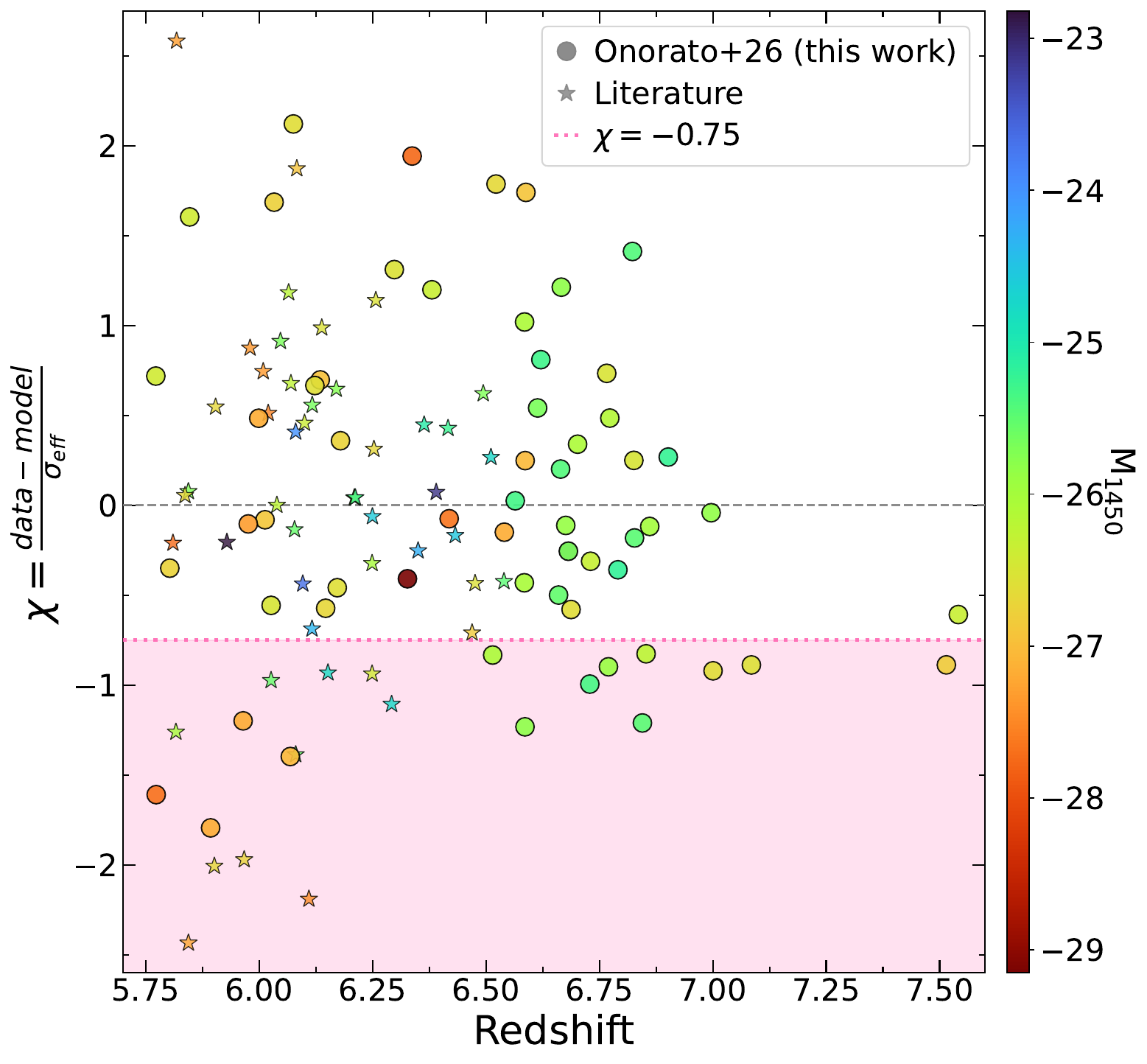}
     \caption{Residuals ($\chi$) of the bivariate power-law fit performed after excluding the two BAL quasars in the sample (i.e., J2348$-$3054 and J1526$-$2050) as a function of redshift. The symbols are the data from this work (points) and the literature (stars), and they are color-coded with $M_{1450}$, according to the colormap on the right. The dashed gray line indicates where the model perfectly matches the data. The dotted pink line marks the threshold of $\chi = -0.75$, and the shaded region below highlights the quasars with small $R_{\mathrm{p}}$ (see Section \ref{subsec:small}).}
     \label{fig:residuals}
\end{figure}

In Fig. \ref{fig:residuals}, we show the residuals of the bivariate power-law fit described in Section \ref{subsec:doublePL}, and the threshold used to identify the quasars with small proximity zone size ($\chi \leq -0.75$, see Section \ref{subsec:small}).

\begin{figure}
    \centering
    \includegraphics[width=\linewidth]{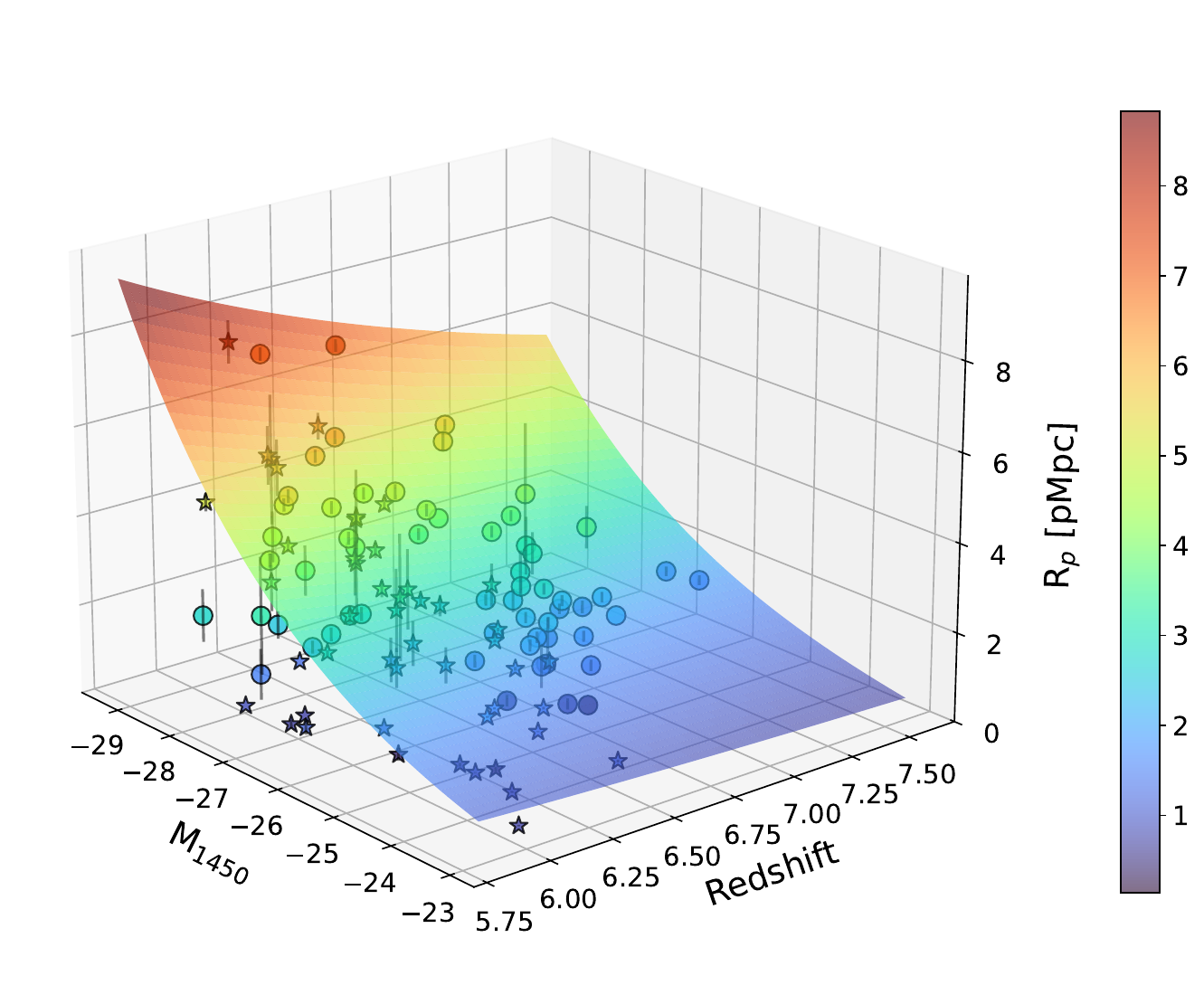}
     \caption{3D surface plot of the proximity zone size as a function of redshift and absolute magnitude. The surface represents the best-fit bivariate power-law model obtained from the MCMC analysis in Section \ref{subsec:doublePL}, with colors varying according to $R_{\mathrm{p}}$, following the colormap on the right. The symbols are the data from this work (points) and the literature (stars). Their color corresponds to the proximity zone size (also according to the colormap), with error bars indicating the uncertainties in the measurements. The plot provides a comprehensive visualization of the relationship between $R_{\mathrm{p}}$, $z$, and $M_{1450}$.}
     \label{fig:3D_plot}
\end{figure}

Furthermore, we compare our predictions with observational data in a three-dimensional (3D) space of ($M_{1450}, z, R_{\mathrm{p}}$). A surface plot is shown in Fig. \ref{fig:3D_plot} to visualize the functional dependence of $R_{\mathrm{p}}$ on $M_{1450}$ and $z$, overlaid with observational data points and their associated error bars.
% In Fig. \ref{fig:3D_plot}, we provide a 3D visualization of how the size of the proximity zone $R_{\mathrm{p}}$ varies as a function of the absolute magnitude $M_{1450}$ and the redshift $z$ (see Section \ref{subsec:doublePL}).

\section{Metal absorbers analysis}\label{app:int_abs}
\begin{figure*}
    \includegraphics[width=\linewidth]{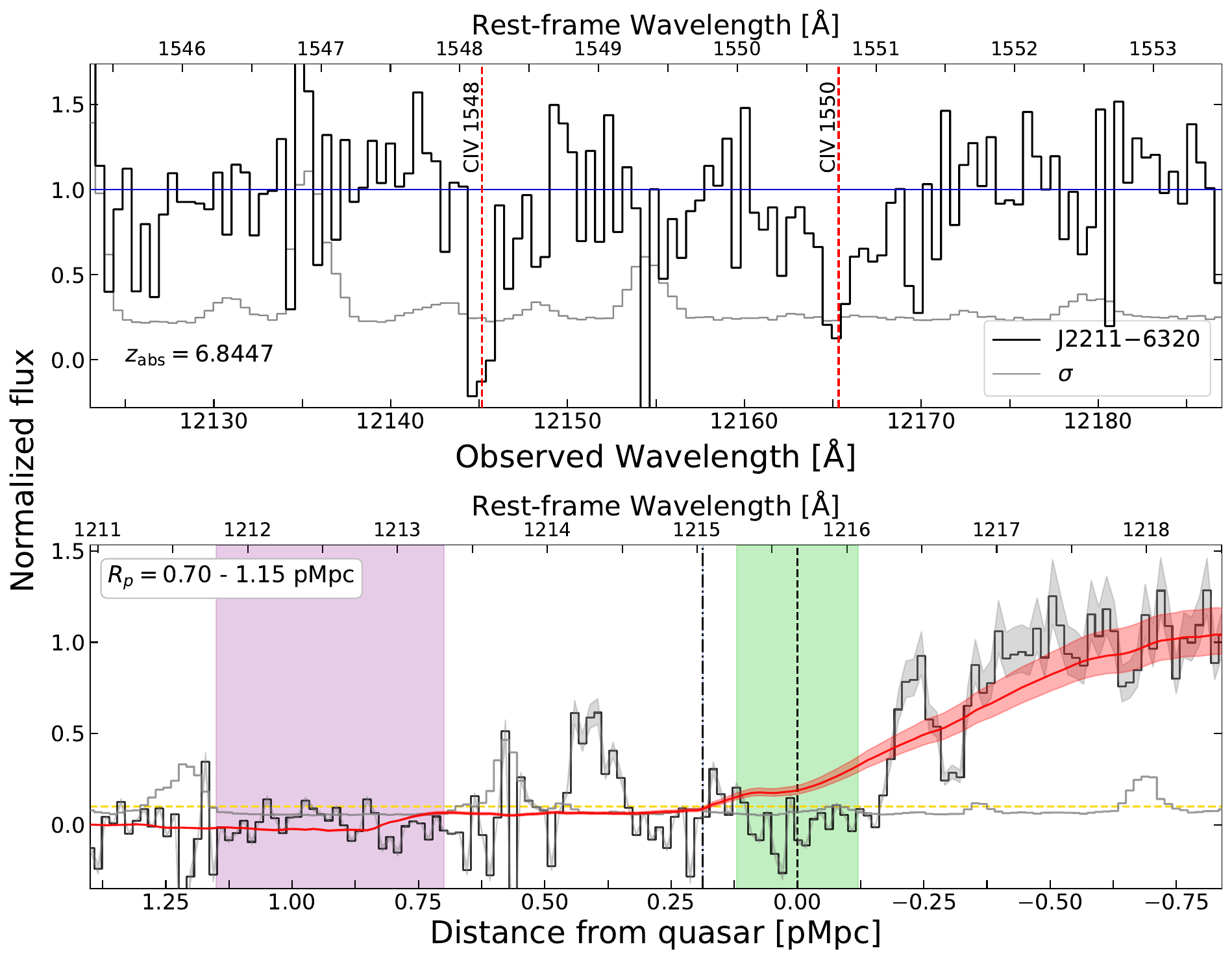}
    \caption{\textit{Top:} Zoom-in of the continuum-normalized spectrum (black) and noise vector (grey) of J2211$-$6320. We identify \ion{C}{IV} as the main absorption line system (dashed red lines) associated with an absorber at $z_{\mathrm{abs}} = 6.8447$, which is likely to affect the shape of the proximity zone of the quasar. \textit{Bottom:} Proximity zone size measurement of J2211$-$6320, as described in Fig. \ref{fig:proxmes_enigma}. In magenta, we show the area in which we suspect $R_{\mathrm{p}}$ of J2211$-$6320 would fall ($R_{\mathrm{p}}= 0.70 - 1.15$ pMpc) without the truncation effect due to the metal absorber.}
    \label{fig:J2211}
\end{figure*}

Here we describe our metal absorbers analysis for the objects showing small proximity zone size ($\chi \leq -0.75$, from the fit in Section \ref{subsec:doublePL}) performed via visual inspection of the spectra (see Section \ref{subsec:small}):

%% JFH I'm not sure why you focus on performing a full intervening absorber analysis here. It is the proximate absorbers that we care about. The intervening absorber catalog is helpful in that if you can ID all the lines, you can rule out lines near the quasar itself, but I'm not sure you need to include all this in the main text. Maybe in the appendix. 
%% SO Done
\begin{itemize}
    % \item J1342$+$0928: we find \ion{Mg}{II} doublets and \ion{Fe}{II} lines associated with at least three metal absorbers at $z_{\mathrm{abs}} \simeq 6.8435$, $3.4302$, and $3.3760$ (see also \citealt{Banados2018}). They imprint several absorption features redward the Ly$\alpha$ line, but should not affect the size of the measured proximity zone.
    %% JFH You are missing a very important and relevant paper by Davies et al. 2024 on this object that makes exactly this point. 

    \item J1120$+$0641: we recognize \ion{C}{IV}, \ion{Mg}{II} doublets and several metal lines associated with at least seven intervening absorbers at $z_{\mathrm{abs}} \simeq 7.0603$, $7.0555$, $7.0167$, $5.7950$, $4.4725$, $2.8097$, $2.7952$ (see also \citealt{Bosman2017}). They imprint several absorption features redward the Ly$\alpha$ line, but due to their distance from the quasar, the absorbers are unlikely to influence the Ly$\alpha$ region and the size of the proximity zone.
    %% JFH Here you should just be citing the very careful analysis of this quasar carried by out Bosman. She looked into all the metal absorbers here and cataloged them using the same data you are using. 
    %% SO Done

    \item J0252$-$0503: we distinguish \ion{C}{IV} doublet associated with a possible absorber at $z_{\mathrm{abs}} \simeq 6.9633$, and \ion{Mg}{II} doublet and \ion{Fe}{II} absorption lines associated with another absorber at $z_{\mathrm{abs}} \simeq 4.2094$ (see also \citealt{Wang2020}). They appear to imprint \ion{H}{I} and \ion{Al}{III} absorption lines in the Ly$\alpha$ region, respectively, but should not alter the size of the proximity zone. We also find a \ion{Mg}{II} doublet and many \ion{Fe}{II} absorption lines associated with four absorbers at $z_{\mathrm{abs}} \simeq 4.8790$, $4.7148$, $4.1904$, and $3.5426$. The first one imprints absorption features redward the Ly$\alpha$ line but does not affect the proximity zone, while all the others do not have any particular absorption lines in that portion of the spectrum.
    %% JFH I guess Feige Wang's IGM damping wing analysis of this quasar probably looked into this in detail. You need to be aware of these studies to prevent from spending extra time doing your own analysis. It is also problematic that three highly relevant papers to this sectoin (Davies 24, Bosman on 1120, and Wang on 0252 are not being cited. You are supposed to know the literature on your thesis topic better than your supervisor!
    %% SO Done

    % \item J0410$-$0139: we identify both \ion{C}{IV} and \ion{Mg}{II} doublet associated with a metal absorber at $z_{\mathrm{abs}} \simeq 6.40$. Due to its distance from the quasar, the absorber is unlikely to influence the Ly$\alpha$ region and the size of the proximity zone. However, we must note that the SNR of the spectrum is not high enough (i.e., $\langle \rm{SNR_{J}} \rangle = 6$, from \citeO25) to identify weak metal absorption features.

    \item J1917$+$5003: we recognize a \ion{Mg}{II} doublet associated with an intervening absorber at $z_{\mathrm{abs}} \simeq 3.687$. As before, it is unlikely that the absorber influences the size of the proximity zone.

    \item J2211$-$6320: we find a \ion{C}{IV} doublet at $z_{\mathrm{abs}} \simeq 6.8447$, likely associated with a metal absorber, which we suspect could be the quasar’s host galaxy. However, we do not see any low-ionization lines, probably due to the spectrum’s low SNR (i.e., $\langle \rm{SNR_{J}} \rangle = 9$, see \citeO25). We conservatively decide to exclude this object from our final fit. Because of the closeness of the absorber to the quasar, it likely influences the Ly$\alpha$ region and distorts the proximity zone. The top panel of Fig. \ref{fig:J2211} displays the identified \ion{C}{IV} absorption lines, while the bottom panel highlights in magenta the region where we estimate the true $R_{\mathrm{p}}$ to lie ($R_{\mathrm{p}}= 0.70 - 1.15$ pMpc). This estimate is based on visible fluctuations in the normalized flux up to $\simeq 0.70$ pMpc, which we interpret as remnants of a truncated proximity zone. 
    
    % \item J0109$-$3047: we identify N V, \ion{Si}{II}, \ion{Si}{IV}, \ion{C}{IV}, and \ion{Mg}{II} doublets associated with a metal absorber at $z_{\mathrm{abs}} \simeq 6.7882$. We also find a \ion{C}{IV} doublet and several \ion{Fe}{II} lines associated with another metal absorber at $z_{\mathrm{abs}} \simeq 6.6020$. The absorbers both appear to affect the Ly$\alpha$ region: the closest to the quasar with its \ion{H}{I} absorption line feature, while the furthest one with the \ion{N}{V} absorption line system (see Fig. \ref{fig:J0109}). However, we do not see any particular fluctuations in the normalized flux of the quasar (Fig. \ref{fig:proxmes_enigma}), so we think the proximity zone measurement should not suffer much from the presence of the absorbers, with a final expected value falling in the range $R_{\mathrm{p}}= 1.33 - 1.50$ pMpc. As in the previous case, this spectrum suffers from low SNR (i.e., $\langle \rm{SNR_{J}} \rangle = 6.7$, see \citeO25), making it difficult to distinguish additional absorption features.

    \item J1007$+$2115, J0218$+$0007, J0910$+$1656: we do not identify any metal absorber.
    % J0319$-$1008, J0923$+$0753, J1216$+$4519, J1058$+$2930

    \item J2338$+$2143: because of the spectrum's very low SNR ($\langle \rm{SNR_{J}} \rangle = 3.4$, from \citeO25), we cannot infer anything about the presence of metal absorbers.
    % J0109$-$3047, J1110$-$1329

    \item J1034$-$1425, J0713$+$0855, J1213$-$1246, J0836$+$0054: these quasars belong to the E-XQR-30 sample, and the presence of metal absorbers has already been widely studied by \cite{Davies2023}. We refer to them for a complete list of the absorbers identified in the whole catalog, while here we focus on the ones that might contaminate the shape of the proximity zone in the above-mentioned sources.
    
    For J1034$-$1425, the closest metal absorber is located at $z_{\mathrm{abs}} \simeq 5.8987$. We do not see any clear low-ionization lines or fluctuations in the normalized flux of this object (see Fig. \ref{fig:proxmes_exqr30}), and \cite{Satyavolu2023, Eilers2020} have excluded the presence of an absorber ahead of the quasar. However, \cite{Durovcikova2025} quote the presence of a proximate Ly$\alpha$ emitter (LAE) that might truncate $R_{\mathrm{p}}$. We do not recognize any metal absorption lines on the quasar spectrum associated with it, except for a possible DW feature redward of the Ly$\alpha$ line. We decide to keep this object in our fit.

    For J0713$+$0855, we find \ion{Mg}{II} and \ion{Fe}{II} doublets associated with an intervening absorber at $z_{\mathrm{abs}} \simeq 3.5085$ that imprints \ion{Al}{III} absorption lines in the Ly$\alpha$ region, but due to its distance from the quasar, it does not contaminate the proximity zone. Furthermore, \cite{Satyavolu2023, Davies2023} quote a high-ionization associated absorber at $z_{\mathrm{abs}} \simeq 5.9238$ that might truncate $R_{\mathrm{p}}$, but we do not see any clear \ion{Mg}{II} or \ion{C}{IV} absorption lines associated with it, and \ion{H}{I} does not go line-black. We conclude that the $R_{\mathrm{p}}$ of this object should not be affected.
    
    For J1213$-$1246, the absorbers at $z_{\mathrm{abs}} \simeq 5.6024$ and $4.8710$ imprint \ion{Si}{II} and \ion{Si}{IV} absorption lines in the Ly$\alpha$ region, while the one at $z_{\mathrm{abs}} \simeq 3.8254$ imprints an absorption feature redward of the Ly$\alpha$ region. However, due to their distance from the quasar, they do not affect $R_{\mathrm{p}}$.

    For J0836$+$0054, we identify several intervening absorbers at $z_{\mathrm{abs}} \simeq 5.1256$, $4.9965$, $3.7439$, and $2.2990$, but they do not contaminate $R_{\mathrm{p}}$.
\end{itemize}

% \begin{figure*}
%     \includegraphics[width=\linewidth]{figures/J0109-3047_small_2abs.pdf}
%     \caption{\textit{Top:} Zoom-in of the spectrum (black) and noise vector (grey) of J0109$-$3047. We show N V, \ion{Si}{II}, \ion{Si}{IV}, and \ion{C}{IV} absorption line system (dashed red lines) associated with an absorber at $z_{\mathrm{abs1}} = 6.7882$, which seems to affect the shape of the proximity zone of the quasar. The spectrum is smoothed for display purposes. \textit{Bottom:} Same as in the top panel, but for an absorber at $z_{\mathrm{abs2}} = 6.6020$, which affects the proximity zone with its \ion{N}{V} absorption line system. We conclude that, as we do not see strong fluctuations in the normalized flux of the quasar (Fig. \ref{fig:proxmes_enigma}), we think the proximity zone should not be altered much from the presence of the absorbers, with a plausible expected value in the range $R_{\mathrm{p}}= 1.33 - 1.50$ pMpc.}
%     \label{fig:J0109}
% \end{figure*}

%%%%%%%%%%%%%%%%%%%%%%%%%%%%%%%%%%%%%%%%%%%%%%%%%%

% Don't change these lines
\bsp	% typesetting comment
\label{lastpage}
\end{document}